\documentclass[graybox]{svmult}

\usepackage{type1cm}        
\usepackage{makeidx}         
\usepackage{graphicx}        
\usepackage{multicol}        
\usepackage[bottom]{footmisc}

\usepackage{newtxtext}       %
\usepackage[varvw]{newtxmath}       

\usepackage[utf8]{inputenc}
\usepackage[english]{babel}
\usepackage{amsmath,amsthm,amsfonts}
\usepackage{multirow}
\usepackage{tikz}
\usepackage{verbatim}
\usepackage{todonotes}
\usepackage{tocloft} 
\usepackage{isotope}
\usepackage{braket}
\usepackage{siunitx}
\DeclareSIUnit\torr{Torr} 
\DeclareSIUnit\gauss{G}

\usepackage{hyperref}
\hypersetup{
    colorlinks=true,
    linkcolor=blue,
    filecolor=blue,      
    urlcolor=cyan,
    citecolor=blue,
    pdftitle={Fundamentals of Trapped Ions and Quantum Simulation of Chemical Dynamics},
    pdfpagemode=FullScreen,
    }

\usepackage{wrapfig}
\usepackage{caption}

\newcommand{\Be}{$^{9}$Be$^+$\,}
\newcommand{\Mg}{$^{25}$Mg$^+$\,}

\newcommand{\Baf}{$^{137}$Ba$^+$\,}
\newcommand{\Ca}{$^{40}$Ca$^+$\,}
\newcommand{\Caf}{$^{43}$Ca$^+$\,}
\newcommand{\Sr}{$^{88}$Sr$^+$\,}

\newcommand{\Yb}{$^{171}$Yb$^+$\,}

\makeindex             

\begin{document}

\title*{Fundamentals of Trapped Ions and Quantum Simulation of Chemical Dynamics}
\author{Guido Pagano\orcidID{0000-0002-6760-4015},\\ 
Wojciech Adamczyk\orcidID{0000-0002-6732-2413} and \\
Visal So\orcidID{0009-0001-1326-3120}
}
\institute{Guido Pagano \at Department of Physics and Astronomy and Smalley-Curl Institute, Rice University, 6100 Main Street, Houston, TX 77005, USA \email{pagano@rice.edu}
\and Wojciech Adamczyk \at Institute for Quantum Electronics, ETH Zürich, Zürich, Switzerland \email{wadamczyk@phys.ethz.ch}
\and Visal So \at Department of Physics and Astronomy and Smalley-Curl Institute, Rice University, 6100 Main Street, Houston, TX 77005, USA
\email{vs39@rice.edu}}
%
%
\maketitle

\abstract{Trapped atomic ions are among the most advanced platforms for quantum simulation, computation, and metrology, offering long coherence times and precise, individual control over both internal and motional degrees of freedom. In this review, we present a pedagogical introduction to trapped-ion systems, covering the physics of ion trapping, qubit encodings, and laser-ion interactions. We explain how spin-dependent forces generated by light fields enable both analog and digital quantum simulations of spin and spin–boson models, as well as high-fidelity quantum logic gates. We then highlight an emerging frontier in the simulation of chemical dynamics, summarizing recent experiments that demonstrate the capability of trapped ions to simulate vibronic models and excitation-transfer processes. Finally, we outline future directions in quantum simulation and discuss open challenges in scaling up trapped-ion architectures.}


\section{Introduction}

Early proposals by Feynman~\cite{Feynman1982, Feynman1986} envisioned leveraging quantum-mechanical systems to perform quantum computations and simulations of complex many-body problems that are beyond the reach of classical computers. Among the various quantum platforms explored to realize this vision, trapped atomic ions provide exceptional qubit quality and a high level of controllability~\cite{Wineland1978, Wineland1998, Blatt2008, Bruzewicz2019, monroe2021programmable}. These features stem from two key factors. First, ions are suspended in a vacuum chamber by electromagnetic fields, removing much of the influence of the decohering environment that typically affects solid-state quantum systems. Second, the internal electronic states of ions offer naturally defined qubits, and their motional degrees of freedom provide a “quantum bus” to mediate entangling interactions between distant ions via electromagnetic radiation~\cite{Cirac1995, Leibfried2003}.

These features allowed trapped-ion research to advance significantly over the past three decades in quantum simulation, quantum computation, and metrology. Since the first single-ion trapping and laser cooling experiments~\cite{Wineland1978, Wineland1998}, trapped ions have enabled precise tests of quantum mechanics and led to the development of the trapped-ion optical atomic clocks~\cite{Wineland2013Nobel}. Trapped ions have achieved record coherence times, demonstrating $T_2^*=50$ s~\cite{Harty2014} and $T_2^*$ exceeding 1 hour~\cite{Wang2021Single}. High-fidelity ($>99.9\%$) two-qubit gates have been demonstrated with both lasers~\cite{Leibfried2003, Ballance2016, Gaebler2016, Moses2023} and microwave radiation~\cite{Srinivas2021, loschnauer2024scalablehighfidelityallelectroniccontrol}.
In parallel with these advances, microfabricated “chip” traps have dramatically improved scalability prospects by enabling complex electrode geometries with carefully controlled potentials~\cite{Stick2006, Pino2021, Jain2024}. Efforts are ongoing to integrate photonic interconnects and new ion species to address the challenges of mid-circuit measurement, sympathetic cooling, and high-speed ion shuttling~\cite{Bruzewicz2019}.

On the quantum simulation side, it is possible to engineer long-range spin interactions governed by the phonon modes of the ion crystal, enabling analog simulations of Ising-like and spin-boson Hamiltonians~\cite{Porras2004, Porras2008, Blatt2012}. 
Quantum simulations of spin models and spin-boson models~\cite{monroe2021programmable, Schneider2012, Safavi2018, Kang2024} have been realized in experiments with several tens of ions in 1D crystals~\cite{ Zhang2017a, Pagano2020, tan2021domain, joshi2022observing}, and even hundreds of ions in 2D crystals~\cite{Britton2012, Guo2024}. Hybrid digital-analog approaches also extend the capabilities of trapped-ion quantum simulators to non-native models, such as Heisenberg Hamiltonians, lattice gauge theories, and chemical dynamics, by breaking the target Hamiltonian into small, non-commuting Trotter evolution steps~\cite{lanyon2011universal, Martinez2016real, Kokail2019, Whitlow2023, Kranzl2023observation}. 

This review offers a pedagogical overview of the trapped-ion platform with an emphasis on quantum simulation and computing. 
In Section~\ref{sec_Motion}, we begin by discussing the physics of ion trapping, covering both the classical and quantum treatments of single-ion and multi-ion motion. 
Section~\ref{sec_qubits} covers the most popular qubit encodings, discusses their respective strengths and weaknesses, and reviews the state of the art in state preparation and measurement errors (SPAM) as well as coherence times.
Section~\ref{sec_laser-ion_interactions} then details how laser-ion interactions give rise to both single-qubit operations and the spin-dependent forces that underlie entangling gates and quantum simulations of Ising and XY spin models and spin-boson models. 
Following this general framework, in Section~\ref{sec_progress}, we outline a broad set of applications, ranging from quantum magnetism to high-energy and chemical dynamics simulations, focusing specifically on the latter.
We conclude in Section~\ref{sec_outlook} with an outlook on open challenges and emerging directions, including integrated photonic technologies, the pursuit of larger-scale ion processors, and more advanced quantum simulations of high-energy physics and complex chemical processes.

We hope this review will serve as a useful tutorial to the foundational aspects of trapped-ion quantum technology and guide readers to the frontiers of this vibrant field.

\section{Ion Trapping and Motion} \label{sec_Motion}


Confinement of charged particles in three dimensions using only static electric fields is fundamentally impossible since Gauss's law ($\nabla \cdot \mathbf{E} = 0$) implies that the electrostatic potential must have negative curvature in at least one direction. By contrast, time-varying electric fields can circumvent this restriction, as demonstrated by Wolfgang Paul~\cite{Paul1958Forsch, paul1990electromagnetic} and Hans Dehmelt~\cite{Dehmelt1990Nobel}. The Paul trap relies on a simple principle: a time-varying and spatially-inhomogeneous electric field can give rise to a time-averaged restoring force for trapped particles. Building on this idea, Dehmelt's vision~\cite{DEHMELT196853} of 
\begin{quotation}
isolated atomic systems floating at rest in free space for unlimited periods and free from any undesired outside perturbations
\end{quotation}
is now routinely realized, at least to a large extent, in many laboratories worldwide.

\begin{figure}[b!]
    \centering
\includegraphics[width=\columnwidth]{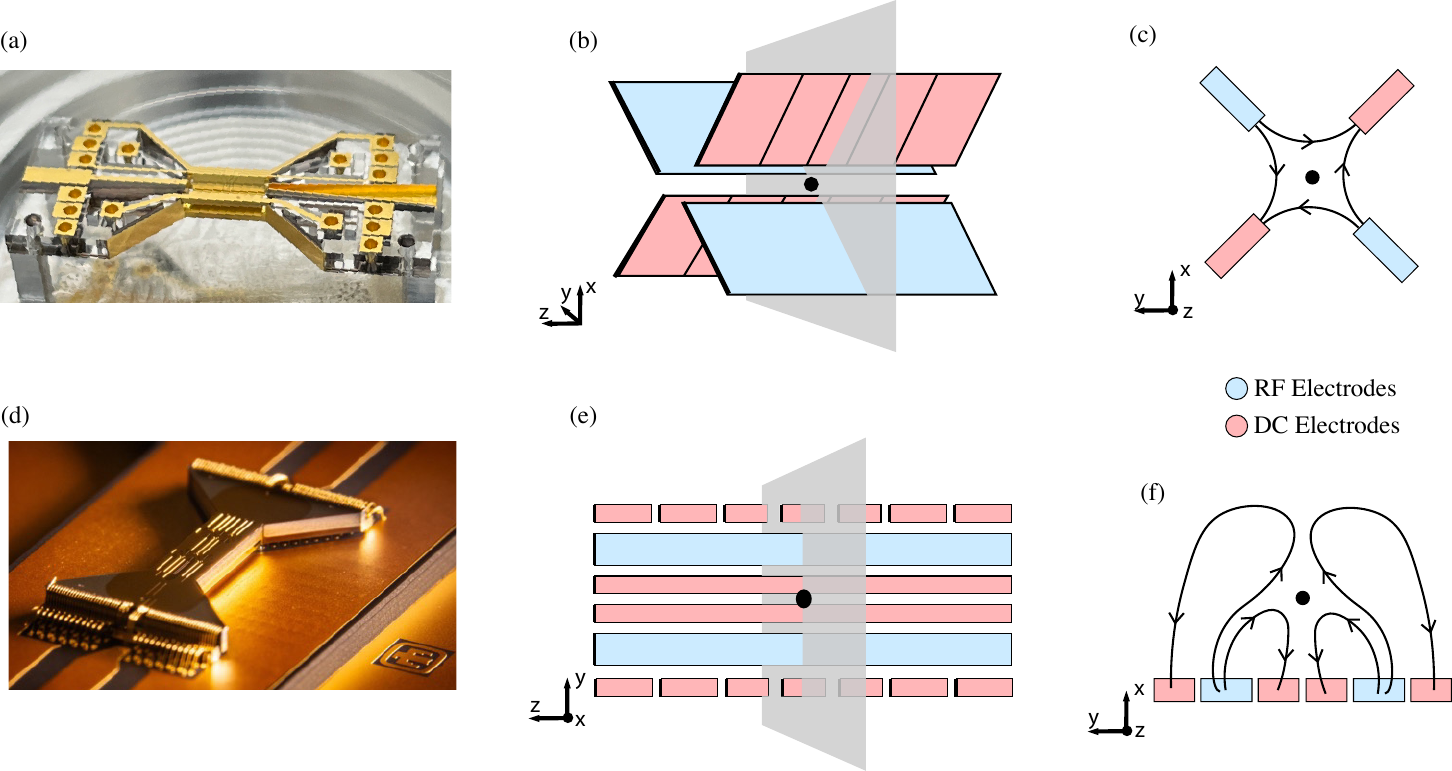}
	\caption {{\bf Trap architectures of a segmented three-dimensional (3D) trap (a-c) and a segmented planar two-dimensional (2D) trap (d-f).} Panels (a) and (d) show photographs of a high optical access 3D monolithic blade trap \cite{Menon2026} and of a high optical access (HOA \cite{osti_1237003}) microfabricated chip trap, respectively. Panels (b,c,e,f) present schematic diagrams of the trap electrodes in the two cases along different directions. Light blue segments represent RF electrodes, and light red segments represent DC electrodes. The light gray plane corresponds to the cross-sections depicted in (c) and (f), illustrating the electric-field-line geometry for both trap types. Notably, the planar 2D trap can be viewed as a projection of the 3D trap onto a single plane.}
	 \label{fig_1} 
\end{figure}

While a Paul trap can be built using different electrode configurations (see Ref.~\cite{Siverns2017} for a non-exhaustive list), they can be grouped into two types of traps: (a) three-dimensional traps, where electrodes are arranged in three dimensions around the ions, and (b) planar ``chip'' traps, where electrodes are arranged in two dimensions (see Fig.~\ref{fig_1}). In this review, we will focus on the more symmetric trapping potential generated by three-dimensional traps, but this treatment can be easily adapted to two-dimensional chip traps.

The simplest Paul trap is built out of four electrodes arranged symmetrically around the ion. By applying a time-varying voltage $V_{\rm rf}$ oscillating at frequency $\omega_{\rm rf}$ to two opposing electrodes and grounding the other pair, a quadrupole potential is generated in the transverse plane. This oscillation periodically inverts the quadrupole field's orientation, alternating the configuration of the saddle point. Considering a single direction, a charged particle of charge $e$ at first glance would seem to be confined and deconfined half of the time. So why is the average force on the particle different from zero? The reason is that the quadrupole field has a finite spatial gradient, so for small displacements, this gives rise to an average restoring force $\bar{F}(x)\sim e^2(\partial E (x)/\partial x) (E(x)/2 {\rm m} \omega^2_{\rm rf})$, where $\rm m$ is the particle mass. As we will see in the following, this results in an effective confining potential (sometimes referred to as pseudopotential) defined as $\psi(x)=e^2 E(x)^2/(4 {\rm m} \omega_{\rm rf}^2)$~\cite{DEHMELT196853}.


\subsection{Classical Motion}
Knowing that the curvature of the potential is the fundamental ingredient for the confinement of charged particles, we can write the most general  quadratic time-dependent potential near the trap center, including both static ($V_{\rm dc}$) and time-varying ($V_{\rm rf}$) parts, as:
\begin{equation}
\label{eq_pot}
\varphi(x,y,z,t)= \frac{V_{\rm dc}}{2} (\alpha_x x^2 + \alpha_y y^2 + \alpha_z z^2) + \frac{V_{\rm rf}}{2}(\alpha'_x x^2 + \alpha'_y y^2 + \alpha'_z z^2)\cos(\omega_{\rm rf} t).
\end{equation}
Although this equation has 6 degrees of freedom, Laplace's equation ($\nabla^2 \varphi=0$) imposes two boundary conditions, leaving only four free parameters:
\begin{equation}
\label{eq_alphas_Laplace}
\begin{aligned}
\alpha_x + \alpha_y +  \alpha_z &= 0, \\
\alpha'_x + \alpha'_y  + \alpha'_z &= 0, 
\end{aligned}
\end{equation}
The fact that the problem is underconstrained suggests that there are many possible electrode configurations. For simplicity, we will focus our discussion on the linear traps, where the dc-confinement is only along the $z$-direction (dc deconfinement along $x$- and $y$-directions). The rf potential is opposite in amplitude along the $x$- and $y$-directions at any given time and is negligible along the $z$-direction, although the trap imperfections can sometimes lead to axial rf fields.
The coefficients $\alpha_i$ defined by this symmetry choice obey the following constraints:
\begin{equation}
\label{eq_linear_trap_constraints}
\begin{aligned}
\alpha_x + \alpha_y &= - \alpha_z, \,\,\, \alpha_z >0, \\
\alpha'_x &= - \alpha'_y, \,\,\,\, \alpha'_z = 0.
\end{aligned}
\end{equation}
It is convenient to redefine the geometric coefficients $\alpha_i$ in terms of the characteristic dimensions of the trap:
\begin{eqnarray}
\label{eq_1}
\alpha_i &=&\kappa_i/z_0^2,\nonumber\\
\alpha'_x &=&\kappa'_r/ r_0^2=-\alpha'_y,
\end{eqnarray}
where $r_0$ and $z_0$ are the ion-electrode distances along the radial and axial directions, respectively. The geometric dimensionless factor $\kappa'_r$, ranging between 0 and 1, indicates how much the electrode configuration departs from the ideal hyperbolic electrode shape. Hyperbolic electrodes follow the equipotential lines of a perfect quadrupole and correspond to $\kappa'_r=1$. However, this electrode configuration is not experimentally practical due to its limited optical access. Typical traps usually feature ion-electrode distances in the few tens to hundreds of $\rm \mu m$ ($r_0,z_0\sim 20 - 500\,\rm \mu m$) with rf voltages in the $V_{\rm rf}\sim 100-500 \,\rm V$ range and dc voltages in the $V_{\rm dc} \sim 1-50 \, \rm V$ range, depending on the desired confinement.

The potential in Eq.~\eqref{eq_pot} gives rise to the following classical equation of motion\footnote{Here, we only consider one dimension, but the treatment can be straightforwardly extended to three dimensions.}:
\begin{equation}
\label{eq_motion}
 \frac{d^2 x}{dt^2} =-\frac{e}{\rm m}\frac{\partial\varphi}{\partial x} =-\frac{e}{\rm m} \left[ V_{\rm dc} \alpha_x + V_{\rm rf} \alpha'_x \cos(\omega_{\rm rf} t)\right]x.
\end{equation}
Rescaling the time to a dimensionless parameter $\xi=\omega_{\rm rf} t/2$, we can rewrite Eq.~\eqref{eq_motion} as a Mathieu equation~\cite{Leibfried2003}:
\begin{equation}
\label{eq_mathieu_equation_ion}
 \frac{d^2 x}{d\xi^2} + \left[ a_x  + 2 q_x \cos( 2 \xi )\right]x=0,
\end{equation}
where $a_x$ and $q_x$ are defined as:
\begin{eqnarray}
a_x &=& \frac{4 e V_{\rm dc} \alpha_x}{{\rm m} \omega_{\rm rf}^2} = 
\left(\frac{ e V_{\rm dc} }{ \frac{1}{2} {\rm m} \omega_{\rm rf}^2 z_0^2}  \right) 2 \kappa_x, \label{eq_ax}\\
q_x &=& \frac{2 e V_{\rm rf} \alpha'_x}{{\rm m} \omega_{\rm rf}^2} = \left(\frac{e V_{\rm rf} }{\frac{1}{2} {\rm m} \omega_{\rm rf}^2 r_0^2 }\right)  \kappa'_r \label{eq_qx}.
\end{eqnarray}
The dimensionless Mathieu coefficients $a_x (q_x)$ are proportional to the ratio between the electrostatic energy due to the $V_{\rm dc} (V_{\rm rf})$ voltages seen by the ion and the potential energy associated with the rf frequency and the trap geometry, weighted by their respective geometric factors. In the special, symmetric configuration with $\alpha_x=\alpha_y=-\alpha_z/2<0$, $a_z=\left({ e V_{\rm dc} / \frac{1}{2} {\rm m} \omega_{\rm rf}^2 z_0^2}  \right) 2\kappa_z>0$ depends only on the static confinement along the axial $z$ direction~\cite{Wineland1998}. 

\begin{figure}[t]
\centering
\includegraphics[width=0.9\columnwidth]{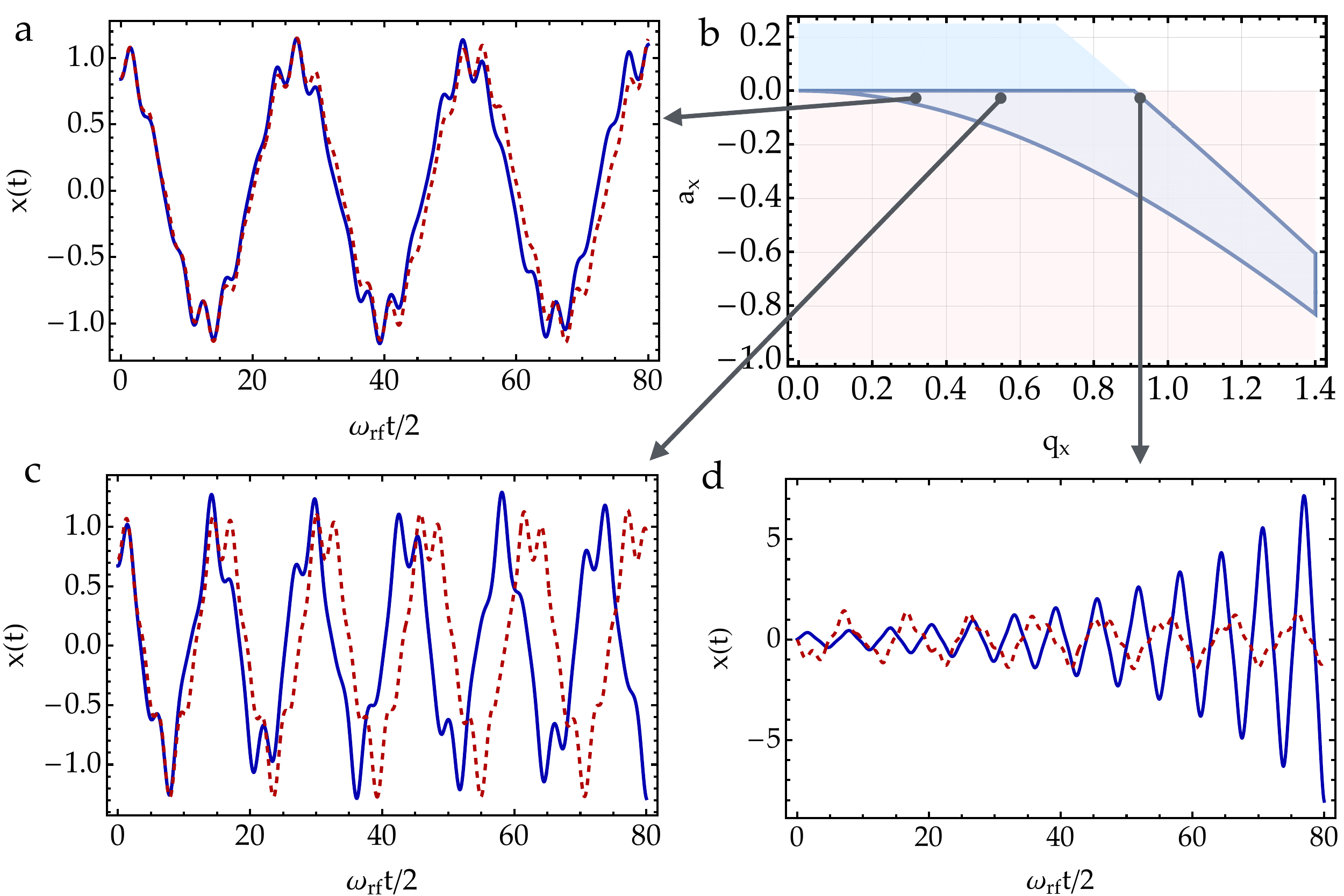}
\caption{{\bf Ion motion along one dimension and stability region}. Exact solution (dark blue) and lowest-order approximation (dark red) for $a_x=-0.01$ and (a) $q_x=0.3$, (c) 0.55, and (d) 0.92. When the ion is confined, the slow secular motion oscillating at $\omega$ is modulated by the micromotion oscillating at frequency $\omega_{\rm rf}$. (b) Stability region in the $(a_x,q_x)$ plane of Eq.~\eqref{eq_mathieu_equation_ion} in the case of a linear Paul trap.
    }
	 \label{fig:2} 
\end{figure}

The Mathieu equation can be solved analytically by inserting as an ansatz a sum of functions oscillating with different harmonics of $\xi$:
\begin{equation}
x(\xi)=A e^{i\beta_x \xi} \sum_{n=-\infty}^\infty C_{2n}e^{i2n \xi} + B e^{-i\beta_x \xi} \sum_{n=-\infty}^\infty C_{2n}e^{-i2n \xi},
\label{eq_sol_motion}
\end{equation}
where $\beta_x$ and $C_{2n}$ are functions of the Mathieu coefficients $a_x$ and $q_x$ (see Ref.~\cite{Leibfried2003} for the exact expressions). 
Assuming  $\left|a_x\right|, q_x^2 \ll 1$, and thus $\beta_x\simeq\sqrt{a_x + q_x^2/2}$, the solution for the ion motion can be approximated to the lowest order to:
\begin{equation}
\label{eq_low_order}
x(t)\propto \underbrace{\cos\left[\left(\sqrt{a_x + \frac{q_x^2}{2}}\right) \frac{\omega_{\rm rf}t}{2}\right]}_{\text{secular motion}}\underbrace{\left[1 - \frac{q_x}{2} \cos(\omega_{\rm rf} t)\right]}_{\text{micromotion}}.
\end{equation}
This solution shows how the ion, when confined, undergoes oscillations with a slow secular frequency $\omega=\beta_x \omega_{\rm rf}/2$ modulated by a fast micromotion oscillation at $\omega_{\rm rf}$ (see Fig.~\ref{fig:2}a,c). 

In typical linear traps, the radial confinement is much tighter than the axial confinement ($V_{\rm rf}\gg V_{\rm dc}$), therefore we can assume $q_x^2 \gg a_x$. Hence
the frequency of the secular motion in the lowest order approximation in Eq.~\eqref{eq_mathieu_equation_ion} can be expressed as~\cite{Berkeland1998}:
\begin{equation}
\omega=\sqrt{a_x + \frac{q_x^2}{2}}\frac{\omega_{\rm rf}}{2} \approx \frac{q_x}{2\sqrt{2}} \omega_{\rm rf}= \frac{1}{\sqrt{2}}\left(\frac{e}{{\rm m}}\right) \left( \frac{V_{\rm rf}}{\omega_{\rm rf}} \right) \left(\frac{\kappa'_r}{r_0^2}\right).
\label{eq_omega}
\end{equation}
Equation~\eqref{eq_omega} is a central relation in ion trap design, as it explicitly connects the trap frequency to all the relevant physical parameters. In particular, the secular frequency is directly proportional to the ion’s charge-to-mass ratio, linearly proportional to the applied rf voltage $V_{\rm rf}$, and inversely proportional to the rf drive frequency $\omega_{\rm rf}$. The third ratio in Eq.~\eqref{eq_omega} accounts for the trap geometry through the ion-electrode distance and the scaling factor $\kappa'_r$. 
In quantum simulation experiments, larger radial secular frequencies offer several advantages: they allow for storing longer linear ion crystals, facilitate Doppler and ground-state cooling~\cite{Monroe1995}, and typically reduce the heating rate of the motional degrees of freedom (see also Section~\ref{sec_Motion}), which typically scales $\sim S_E(\omega)/\omega$, where $S_{E}(\omega)$ is the electric field noise density. In the absence of technical sources of noise with a specific spectrum, it has been observed in multiple experiments that $S_E(\omega)$ typically falls with $1/\omega$, leading to a heating rate scaling of $\dot{\bar{n}}\sim 1/\omega^{2}$~\cite{Brownnutt2015}. 

The most direct way to increase the trap frequency is to raise the applied rf voltage, but this has both physical and practical limitations. Simply applying higher voltage will increase the micromotion amplitude $q_x$ (see Eq.~\eqref{eq_qx} and Fig.~\ref{fig:2}a,c). 
As explained in Section~\ref{sec_laser-ion_interactions}, a larger micromotion decreases the laser-ion interaction. 
Furthermore, the Mathieu equations define a stability region (see Fig.~\ref{fig:2}b) in the $(a_x,q_x)$ plane, and, in the case of a linear Paul trap, for $q_x\gtrsim 0.9$, the ion motion will not be confined as the solution will exponentially diverge (see Fig.~\ref{fig:2}d). 
Achieving a large trap frequency $\omega$ with a small micromotion amplitude $q_x$ implies using both high voltages and rf frequencies, which are limited by the maximum amount of power that can be safely applied to the trap and the other electrical components of the system. Since in most cases of interest the loss tangent of the substrate is negligible, the power dissipated on the trap can be estimated by calculating the ohmic contribution as $P_{\rm diss}= R_t C^2 \omega_{\rm rf}^2 V_{\rm rf}^2$/2~\cite{Stick2006}, where $R_t$ and $C$ are the effective series resistance of the electrodes and the net capacitance of the trap’s RF circuit, respectively. Hence, the operating voltage and frequency are usually the result of a trade-off between obtaining large trap frequencies and small micromotion amplitudes and limiting the total power supplied to the trap and the other electrical components, such as the feedthroughs and connection cables.


\subsection{From Trapped Ions to Optical Lattices}

Interestingly, it is possible to make a mathematical analogy between trapped-ion motion (a potential periodic in time) and optical lattices (a potential periodic in space). Let us consider an optical lattice in 1D, which is formed by potential $V(x)=V_0 \sin ^2(k_L x)$, leading to the single-particle Hamiltonian:
\begin{equation}
\hat{H}=\frac{\hat{p}^2}{2 {\rm m}}+V_0 \sin ^2(k_L \hat{x}).
\end{equation}
The eigenvalue equation $\hat{H} \psi(x) = E \psi(x)$ with $\psi(x) = \left<x|\psi\right>$ can be cast in a dimensionless form, similarly to the ion case, by defining dimensionless length $w=k_L x$, potential $s = V_0/E_r$ and energy $\tilde{E} = E/E_r$, with $E_r=\hbar^2 k_L^2/2{\rm m}$ the recoil energy:
\begin{equation}
-\frac{d^2\psi(w)}{d w}+\frac{s}{2}[1-\cos (2 w)] \psi(w)=\tilde{E} \psi(w).
\end{equation}
After rearranging the terms, we get:
\begin{equation}
\frac{d^2\psi(w)}{d w^2}+\left[\underbrace{\left(\tilde{E} -\frac{s}{2} \right)}_{a_x}+\underbrace{\left(-\frac{s}{2}\right)}_{2q_x}\cos (2 w)\right] \psi(w)=0.
\end{equation}
This equation is mathematically equivalent to the Mathieu equation that describes the motion of a trapped ion, with the key distinction being that the potential oscillates spatially rather than temporally. 
\begin{figure}[t]
\centering
\includegraphics[width=\columnwidth]{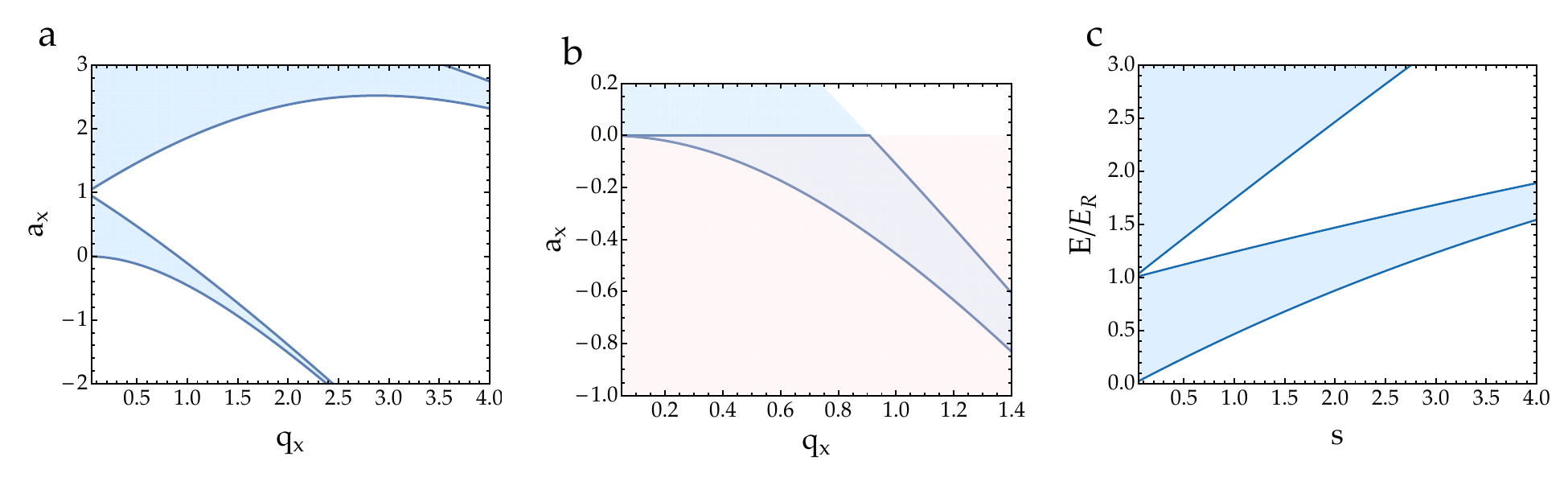}
	\caption {{\bf Stability diagram for (a) Mathieu equation, (b) trapped-ion in a linear Paul trap, (c) band structure of a wavefunction in an optical lattice.} 
    In panel (b), the light blue region indicates stable ion motion in the $x$-$y$ plane under the conditions $a_x = a_y$ and $q_x = -q_y$. The light red region shows stability along the $z$-direction, given by $a_x = -\tfrac{1}{2} a_z$ and $q_z = 0$. Since no rf-field is applied in the $z$-direction, $a_z$ must be positive to ensure trapping, which implies $a_x < 0$. The dark blue region then denotes the parameter space for which the ion remains trapped simultaneously along all three directions.}
	 \label{fig_stability} 
\end{figure}
Therefore, the Mathieu stability region (see Fig.~\ref{fig_stability}a) determines the dynamics of the system in both cases. In the case of optical lattices, allowed energy bands and forbidden band gaps arise (see Fig.~\ref{fig_stability}c), analogous to the stable and unstable solutions of the Mathieu equation. In this analogy, band gaps correspond to parameter regions where the Mathieu exponent becomes real, analogous to unstable (exponentially growing) solutions in the ion case.
The parameters $s$ and $\tilde{E}$ are related to $a_x$ and $q_x$ through $s = -4q_x$ and $\tilde{E} = a_x - 2q_x$. This relationship implies that understanding the stability of the wavefunction solutions in an optical lattice potential can be directly related to the stability analysis of Mathieu equations governing the ion dynamics.

However, there is an important distinction. In ion traps, the stability region is determined by the stability of motion in all three spatial directions. Consequently, increasing $a_x$ to access another stability band in a single direction might not be feasible, as it would also affect the parameters $a_y$ and $a_z$. Therefore, special care has to be taken such that, when changing the trap parameters, the motion is still stable in all directions. Depending on the relationship between $\alpha_x$, $\alpha_y$, and $\alpha_z$ and their rf counterparts, the stability region will have a different shape. In Fig.~\ref{fig_stability}b, we plot the stability regions of a linear Paul trap with $\alpha_x=\alpha_y=-\alpha_z/2$, and quadrupole rf field in the $x$-$y$ plane. In this case, one has to consider the intersection of the stability regions to ensure confinement in all directions. For linear Paul traps, since ideally there is no rf field in the $z$-direction, the stability in that direction is purely determined by the dc confinement. Due to Gauss's law, the static confinement along $z$-direction implies radial dc deconfinement. 


Interestingly, we can use this analogy to better understand the solutions to the motion of the trapped ion. Optical lattices, governed by the same equation class, are typically analyzed via Bloch's theorem. In a system with translational symmetry, described by a potential $V(\mathbf{r}+\mathbf{R})=V(\mathbf{r})$ for $\forall \,\mathbf{R}$ in a Bravais lattice, the eigenstates of the corresponding Schr\"{o}dinger equation can be expressed as:
\begin{equation}
\psi_{\boldsymbol{k}}^{(j)}(\boldsymbol{r})=\mathrm{e}^{\mathrm{i} \boldsymbol{k} \cdot \boldsymbol{r}} v_{\boldsymbol{k}}^{(j)}(\mathbf{r}), \text{ with} \quad  v_{\mathbf{k}}^{(j)}(\mathbf{r}+\mathbf{R})=v_{\mathbf{k}}^{(j)}(\mathbf{r}), \, \forall \,\mathbf{R}.
\end{equation}
Here, each eigenstate factorizes into a plane wave with wavevector $\boldsymbol{k}$, the so-called Bloch function $v^{(j)}_{\boldsymbol{k}}(\mathbf{r})$ that shares the periodicity of the lattice. Superscript $j$ corresponds to the band index, which differentiates different eigenstates with the same symmetry, labeled by $\mathbf{k}$. The Bloch function can be expanded as a superposition of all reciprocal lattice vectors~\cite{Ashcroft76}. For a one-dimensional system, a general Bloch function takes the form:
\begin{equation}
    \psi_k^{(j)}(x)=e^{i k x} \sum_{n\in \mathbb{Z}} C^{(j)}_{n, k} e^{i 2 n k_L x},
    \label{eq_bloch_sol}
\end{equation}
where $C^{(j)}_{n, k}$ are Fourier coefficients of the Bloch's function $v_{k}^{(j)}(r)$. \\

Analogous to the optical lattice case, we can also use Bloch's theorem to solve the trapped-ion Mathieu equation in Eq.~\eqref{eq_sol_motion}. However, because the $x_\omega^{(j)}(t)$ is an observable, it has to be real. Therefore, simply plugging the same eigenfunctions as for optical lattices would not be sufficient. To make the observable real, it is necessary to pair each base function with the counter-rotating term and constrain the coefficients as below, where $T={2\pi/\omega_{\rm rf}}$ is a period of the electric field oscillation:
\begin{equation}
    x_\omega^{(j)}(t) \;=\; e^{i \omega t}\,v_{\omega}^{(j)}(t) \;+\; e^{-i \omega t}\,v_{-\omega}^{(j)}(t),
    \quad
    \text{with}
    \quad
    \begin{cases}
        \big[v_{\omega}^{(j)}(t)\bigr]^* \;=\; v_{-\omega}^{(j)}(t),\\[5pt]
        v_{\omega}^{(j)}(t + T)\;=\;v_{\omega}^{(j)}(t),
    \end{cases}
    \quad \forall\,t.
    \label{bloch_function_position}
\end{equation}
Because of the periodicity constraint of $v_{\omega}^{(j)}(t)$, we can expand the Bloch's function in terms of the Fourier series:
\begin{equation}
    v_{\omega}^{(j)}(t) = \sum_{n\in \mathbb{Z}} \tilde{C}^{(j)}_{n, \omega} e^{i n \omega_{\rm rf} t}.
\end{equation}
Inserting Eq.~\eqref{bloch_function_position} into Eq.~\eqref{eq_mathieu_equation_ion}, we get a series of continued fractions, which can be used to relabel the ansatz into the more traditional form:
\begin{equation}
x_\omega^{(j)}(t)= A e^{i \omega t} \left(\sum_{n\in \mathbb{Z}} C^{(j)}_{2n, \omega} e^{i n \omega_{\rm rf} t}\right)+Be^{-i \omega t}  \left(\sum_{n\in \mathbb{Z}} C^{(j)}_{2n, \omega}e^{-i n \omega_{\rm rf} t}\right),
\end{equation}
with $\tilde{C}_{n, \omega}^{(j)} = AC_{2n, \omega}^{(j)}$, and $\left(\tilde{C}_{n, \omega}^{(j)}\right)^* = BC_{2n, \omega}^{(j)}$. We can use these coefficients to define the solution to the trapped-ion Mathieu equation in Eq.~\eqref{eq_sol_motion} with the time analog of the Bloch function:
\begin{equation}
\begin{aligned}
u(\xi) \equiv e^{i\beta_x \xi} v(\xi) &= e^{i\omega t} \sum_{n\in \mathbb{Z}} C_{2n}e^{i2n \omega_{\rm rf} t/2} \\&=  e^{i\beta_x \omega_{\rm rf} t/2} \sum_{n\in \mathbb{Z}} C_{2n}e^{i2n \omega_{\rm rf} t/2} \\&= e^{i\beta_x \xi} \sum_{n\in \mathbb{Z}} C_{2n}e^{i2n \xi}.
\label{eq_u(xi)}
\end{aligned}
\end{equation}
As shown in Eqs.~\eqref{eq_bloch_sol} and~\eqref{eq_u(xi)}, the two solutions have the same mathematical shape and can therefore be written using the same 
Mathieu functions. 
The definition of the function $u(\xi)$ in Eq.~\eqref{eq_u(xi)} allows establishing a correspondence between the spatial Bloch wavefunction $\psi_{j}(w)=e^{i k w} v_j(w)\leftrightarrow e^{i\beta \xi}v(\xi)=u(\xi)$ with the temporal description of the ion motion. The definition of the function $u(\xi)$ will be useful in the next section, where the quantization of ion motion is described.

\subsection{Quantum Motion}\label{sec_Motion_Q}

Quantum effects in the trapped-ion motion are crucial to explain quantum gates and quantum simulation experiments, as well as laser cooling and non-classical motional states. We account for quantum effects by considering the $p$ and $x$ variables as operators $\hat{p}$ and $\hat{x}$ to quantize the motional Hamiltonian while retaining the time dependence of the potential. As we will show below, it is possible to find the corresponding creation and annihilation operators that have exactly the same properties as their static counterparts~\cite{Leibfried2003, Glauber2006}. The quantized version of the motional Hamiltonian is:
\begin{equation}
\hat{H}=\frac{\hat{p}^2}{2{\rm m}}  + \frac{{\rm m}}{2}\frac{\omega_{\rm rf}^2}{4}\left[a_x + 2 q_x\cos(\omega_{\rm rf} t)\right]\hat{x}^2.
\label{eq_motion_quant}
\end{equation}
It is possible to define a new operator $\hat{C}$ that is a constant of motion~\cite{Leibfried2003} by combining the velocity and position operators weighted by the Mathieu functions in Eq.~\eqref{eq_u(xi)}, $u(t)$ and $u^*(t)$, respectively: 
\begin{equation}
\hat{C}(t)=i\sqrt{\frac{{\rm m}}{2\hbar \omega}} \left[ \dot{\hat{x}}(t) u(t) - \hat{x}(t) \dot{u}(t)  \right]=\hat{C}(0)= \sqrt{\frac{1}{2 {\rm m} \hbar \omega}}\left[ {\rm m}\omega \hat{x}(0) + i \hat{p}(0)  \right] \equiv \hat{a},
\label{eq_C}
\end{equation}
where we have chosen the normalization of $u$ such that $u\dot{u}^*-\dot{u} u^*=2i\omega$. Identifying $\hat{C}$ with the annihilation operator leads to a natural definition of the time-dependent position and momentum operators in the Heisenberg representation:
\begin{eqnarray}
    \hat{x}_{H}(t)&=& \sqrt{\frac{\hbar}{2{\rm m}\omega}}\left[ \hat{a} u^*(t)+ \hat{a}^\dagger u(t)\right],\\
    \hat{p}_{H}(t)&=& \sqrt{\frac{\hbar {\rm m}}{2\omega}}\left[ \hat{a} \dot{u}^*(t)+ \hat{a}^\dagger \dot{u}(t)\right],
\end{eqnarray}
where the time dependence is entirely taken care of by the $u(t)$ functions defined in the previous section from the classical Mathieu equation. This quantized description is necessary to characterize the system at sufficiently low temperature $(k_BT\lesssim\hbar\omega)$, and it will also be used to express the collective modes of the ion crystals in the next section and the laser-ion interaction in Section~\ref{sec_laser-ion_interactions}. Furthermore, since the creation and annihilation operators defined in Eq.~\eqref{eq_C} have exactly the same properties as their static counterparts, it is acceptable to write the Fock motional states as well as coherent and squeezed motional states in the number basis defined by the operator $\hat{n}=\hat{a}^\dag \hat{a}$ (for more details, see Ref.~\cite{Leibfried2003}).





 



\subsection{Normal Modes}
\begin{figure}[t]
    \centering
    \includegraphics[width=0.8\columnwidth]{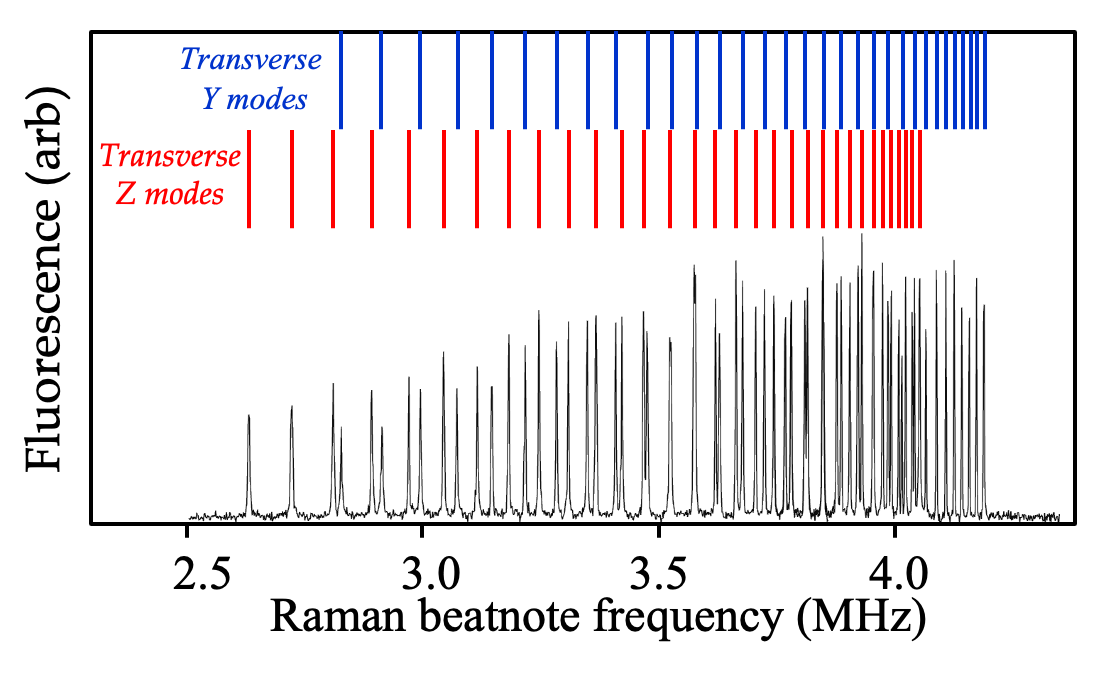}
    \caption{Sideband spectroscopy of radial motional modes of the ion crystal composed of 32 ions. The two sets of transverse modes are distinguishable. Adapted from Ref.~\cite{monroe2021programmable}.
    }
    \label{fig:4}
\end{figure}

A quantum computer or simulator is comprised of many ions, where each charged atom is a quantum information carrier (qubit). 
Multiple ions in the same trap will experience Coulomb repulsion. At sufficiently low temperatures, usually achieved with Doppler cooling~\cite{Wineland1978, Neuhauser1980}, they form a Wigner crystal, where the equilibrium positions are determined by the competition of Coulomb interaction and harmonic confinement. The small vibrations around the equilibrium positions are quantized to be accurately described as collective modes. A crystal composed of $N$ ions will feature $3N$ normal modes along the principal axes of the trapping potentials, forming two sets of transverse modes (see Fig.~\ref{fig:4}) and one set of axial modes. The normal modes of motion are the workhorse of trapped-ion quantum information processing as they offer ideal ancillary degrees of freedom to mediate the interactions among distant ions.


It is possible to fully characterize the equilibrium positions and the eigenfrequencies and eigenvectors of the normal modes by considering the full classical Hamiltonian: 
\begin{equation}
\label{eq_H_motion}
H=\sum_{i=1}^N \sum_{\alpha=x,y,z} \left(  \frac{p_{i\alpha}^2}{2{\rm m}} +\frac{ {\rm m}\omega_\alpha^2 x^2_{i\alpha}}{2} \right) + \frac{1}{2}\frac{e^2}{4\pi\epsilon_0}\sum_{i=1}^N\sum_{ j\neq i} \sqrt{  \frac{1}{\sum_{\alpha=x,y,z}(x_{i\alpha}-x_{j\alpha})^2}}.
\end{equation}
As mentioned previously, the radial confinement in a linear trap is much stronger than the axial confinement. Therefore, we can assume $\omega_x=\omega_y=\omega_z/\sqrt{\beta}$, with $\beta\ll 1$~\cite{James1998}. In this configuration, the ions will crystallize in a linear chain along the axial $z$-direction, where the typical length scale is given by the ratio between the Coulomb energy and the harmonic oscillator energy along the $z$-direction:
\begin{equation}
\ell=\left( \frac{e^2}{4\pi\epsilon_0 {\rm m} \omega_z^2} \right)^{1/3}.
\end{equation}
For typical parameters and atomic masses, $\ell$ is in the order of a few microns, which allows for individual detection and addressing with laser light. We can rewrite the potential energy of the Hamiltonian in Eq.~\eqref{eq_H_motion} as a function of normalized coordinates $u_{i\alpha}=x_{i\alpha}/\ell$:
\begin{equation}
\label{eq_2}
V=\frac{{\rm m}\omega^2_z \ell^2}{2}\left[
\sum_{i=1}^N \left(\frac{u^2_{ix}}{\beta}+\frac{u^2_{iy}}{\beta}+u^2_{iz}\right) + \sum_{i=1}^N\sum_{ j\neq i} \sqrt{ \frac{1}{\sum_{\alpha=x,y,z}(u_{i\alpha}-u_{j\alpha})^2}}
 \right].
\end{equation}
The equilibrium positions are the stationary points in the electrostatic potential given by the condition $\left.\partial V/\partial u_{i\alpha}\right \vert_{u_{i\alpha}=\bar{u}_{i\alpha}}=0$. Given the trapping potential aspect ratio $\beta\ll 1$, the equilibrium positions in the radial directions are simply $\bar{u}_{ix}=\bar{u}_{iy}=0$, and the ones along the axial direction are the solutions of the following system of equations~\cite{James1998}:
\begin{equation}
    \bar{u}_{iz}=\sum_{j=1}^{i-1} \frac{1}{(\bar{u}_{iz} - \bar{u}_{jz} )^2 } - \sum_{j=i+1}^{N} \frac{1}{(\bar{u}_{iz} - \bar{u}_{jz} )^2 }.
    \label{eq_equilibrium}
\end{equation}
For example, in the case of two ions, the problem can be solved analytically to find the equilibrium positions $u_{1z}=-\left({1}/{2}\right)^{2/3}$ and $ u_{2z}=\left(  {1}/{2}\right)^{2/3}$. For $N>2$, the system of equations in Eq.~\eqref{eq_equilibrium} can easily be solved numerically~\cite{James1998}.

In order to characterize the collective normal modes of motion, we assume small vibrations around the equilibrium positions $\bar{u}_{i\alpha}$, namely $u_{i\alpha}=\bar{u}_{i\alpha} + \xi_{i\alpha}$. Rewriting the potential energy as a function of the displacements $\xi_{i\alpha}$, we obtain~\cite{Marquet2003}:
\begin{equation}
    V=V_0 + \frac{1}{2} \sum_{i,j} \sum_{\alpha,\beta} \left.\frac{\partial^2 V}{\partial u_{i\alpha} \partial u_{j\beta} }\right|_{u_{i\alpha}=\bar{u}_{i\alpha}} \xi_{i\alpha} \xi_{j\beta},
\end{equation}
where $V_0$ is the zero-order constant, and the next non-zero term is the Hessian matrix evaluated at the equilibrium positions on the right-hand side of the equation. Therefore, we can define two matrices, one for the axial modes of motion along $z$~\cite{Marquet2003}:
\begin{equation}
A_{ij} = \left\{
 \begin{array}{ll}
 \displaystyle \frac{1}{2}\frac{\partial^2 V}{\partial^2 u_{iz} }=1 + 2\sum_{k\neq i}\frac{1}{|\bar{u}_{iz}- \bar{u}_{kz}|^3}, &i=j \\
\displaystyle \frac{1}{2}\frac{\partial^2 V}{\partial u_{iz} \partial u_{jz} }=- \frac{2}{|\bar{u}_{iz}- \bar{u}_{jz}|^3}, & i\neq j
\end{array}
\right.
\end{equation}
and the other matrix describes the radial modes of motion in the $x$-$y$ plane:
\begin{equation}
B_{ij}=\left\{
    \begin{array}{ll}
      \displaystyle\frac{1}{2}\frac{\partial^2 V}{\partial^2 u_{ix} }=\frac{1}{\beta} - \sum_{k\neq i}\frac{1}{|\bar{u}_{iz}- \bar{u}_{kz}|^3},  & i=j\\ 
    \displaystyle\frac{1}{2}\frac{\partial^2 V}{\partial u_{ix} \partial u_{jx} } = \frac{1}{|\bar{u}_{iz}- \bar{u}_{jz}|^3}, &i\neq j 
    \end{array}
  \right.
\end{equation}
The radial and axial mode matrices are related to each other by the relation:
\begin{equation}
B_{ij} = \left( \frac{1}{2} + \frac{1}{\beta} \right)  \delta_{ij} - \frac{A_{ij}}{2}. 
\label{eq_radial_axial}
\end{equation}
This equation implies that the eigenvectors of the axial and radial modes are exactly the same because the identity matrix is diagonal. Therefore, we can define the eigenvector matrix with $b_{im}$ satisfying both following equations:
\begin{eqnarray}
    \sum_j B_{ij} b_{jm} &=& \gamma_{xm} b_{im}, \quad \omega_{xm} = \sqrt{\gamma_{xm}} \omega_z,\\
    \sum_j A_{ij} b_{jm} &=& \gamma_{zm} b_{im}, \quad \omega_{zm} = \sqrt{\gamma_{zm}} \omega_z.
\end{eqnarray}
The matrix element $b_{im}$ of the eigenvector matrix determines the participation of the $i$-th ion to the $m$-th mode, and it is the same for both families of modes. The implication of the minus sign in Eq.~\eqref{eq_radial_axial} is that the eigenfrequency order is reversed for the axial and radial modes, and the two eigenfrequency families are related by the following relationship:
\begin{equation}
    \gamma_{xm}=\left(\frac{1}{2}+\frac{1}{\beta}\right)-\frac{\gamma_{zm}}{2}.
    \label{eq_eigenfreq}
\end{equation}
This means that the $m=1$ mode corresponding to the symmetric eigenvector $b_{i1}=1/\sqrt{N}$ has the largest (lowest) frequency for the radial (axial) modes. It is possible to demonstrate that for any number of ions $N$, $\gamma_{z2}=3$ for the second axial mode, and the corresponding eigenvector is given by the normalized equilibrium positions, namely $b_{i2}=\bar{u}_{iz}/\sqrt{\sum_k\bar{u}^2_{kz}}$. From Eq.~\eqref{eq_eigenfreq}, it also follows that $\omega_{x2}= \sqrt{\omega_{x1}^2-\omega_{z1}^2}$. Finally, when the last radial eigenfrequency becomes negative, namely when $\beta<\beta_c=2/(\gamma_{zN}-1)$, the matrix $B_{ij}$ becomes non-positive definite, and the ion chain undergoes a structural phase transition to a ``zig-zag'' configuration~\cite{Fishman2008}.


The motion of the ion crystal can be quantized by expressing the displacement of each ion $i$ in the normal mode basis using the eigenvector matrix $b_{im}$. This means that the position of the $i$-th ion along, say, the $x$ principal axis of the trap can be expressed as:
\begin{equation}
    \hat{x}_i = \bar{x}_i + \sum_{m=1}^{N} b_{im}\xi_m^{(0)}(\hat{a}^\dag_m e^{i\omega_m t} + \hat{a}_m e^{-i\omega_m t}),
\end{equation}
where $\xi_m^{(0)} = \sqrt{\hbar/2 {\rm m} \omega_m}$ is the harmonic oscillator position spread of the $m$-th normal mode of motion described by the creation (annihilation) operators $\hat{a}^\dag_m\;(\hat{a}_m)$, and $\bar{x}_i$ is the classical equilibrium position calculated from Eq.~\eqref{eq_equilibrium}.

Since it is experimentally beneficial to have the spacings between the ions uniform across the one-dimensional chain for optimized cooling, manipulation, and detection, one can add high-order, anharmonic terms to the axial potential for longer ion chains to make their spacings quasi-equal. In this case, $|u_{iz}-u_{jz}|\propto|i-j|$, making the diagonal and sub-diagonals of $A_{ij}$ approximately constant and inversely proportional to $|i-j|^3$, respectively. This results in the components of the radial mode vectors $b_{im}$ for $N$ ions forming a discretized pattern with an envelope similar to a sinusoid:
\begin{equation}
    b_{im}\approx \sqrt{\frac{2-\delta_{m1}}{N}}\cos\left({\frac{(2i-1)(m-1)\pi}{2N}}\right),
\end{equation}
where $\delta_{mn}$ is the Kronecker delta. This behavior can be useful for interaction engineering~\cite{Kyprianidis2024interaction}.

\section{Trapped-Ion Qubits: Encoding, Preparation, Manipulation, and Detection}
\label{sec_qubits}

We now turn to the considerations that guide the choice of ion species and qubit encodings for storing and processing quantum information. Alkali atoms, including rubidium, cesium, potassium, and sodium, have been widely studied in neutral-atom
experiments~\cite{Metcalf, Fallani} owing to their relatively simple, hydrogen-like electronic structure. In ion-trapping experiments, atoms can be ionized, leaving electronic configurations that closely resemble alkali atoms. However, as the ionization level increases, spectral transitions get shifted into the ultraviolet (UV) region, where laser sources are not easily accessible. As a result, most trapped-ion experiments use singly ionized species, although highly charged ions have recently attracted interest for metrology applications due to their high sensitivity to effects beyond the standard model~\cite{Kozlov2018Highly, King2022, Rehbehn2023}. 
\begin{wrapfigure}[22]{r}{0.5\textwidth}
    \centering
\includegraphics[width=0.45\columnwidth]{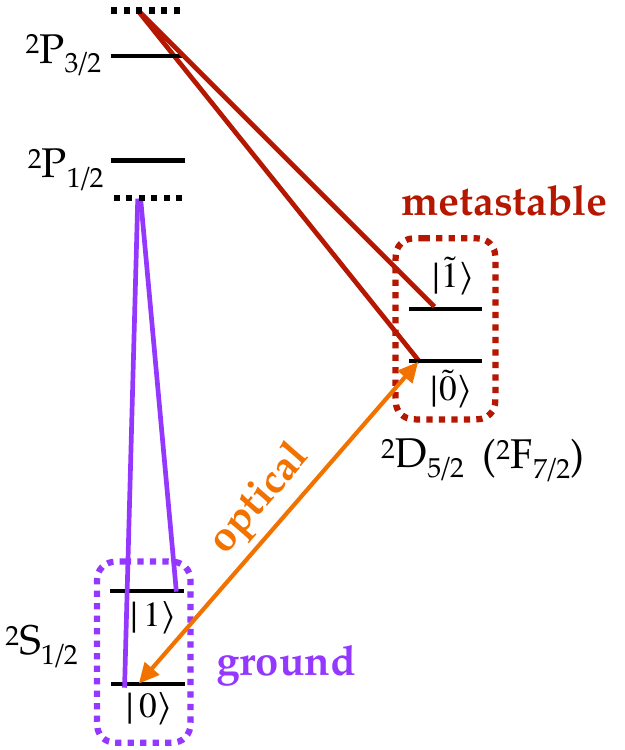}    {\captionsetup{width=0.45\textwidth}\caption{{\bf Trapped-Ion Qubits:} typical atomic structure of most popular ions with the ground-state, optical, and metastable-state qubit encodings.}
\label{fig_ion_structure}
    }
\end{wrapfigure}
The most commonly used ions are singly ionized Group~II and alkaline-earth-like elements. After losing one electron, these ions retain a single valence electron, resulting in an alkali-like atomic structure. These ions typically have $S_{1/2}$ ground state and $^2P_{1/2}$ and $^2P_{3/2}$ as the first dipole-allowed excited states. Except for \Be and \Mg due to their very light masses, there are also low-lying metastable states available in the visible and near UV range, such as $^2D_{3/2}$ and $^2D_{5/2}$, and, in the case of Yb$^+$ ions, also $^2F_{7/2}$ (see Fig.~\ref{fig_ion_structure}). 
Historically, lighter ions such as \Be and \Mg were initially preferred for achieving higher trapping frequencies, with the exception of Al$^+$~\cite{Brewer2019} and In$^+$ \cite{Hausser2025}, that has been used for their $^1S_0\rightarrow{^3P}_0$ clock transitions because of their low sensitivity to quadrupole shifts \cite{Keller2019controlling}. In particular, Al$^+$ currently holds the record for the most accurate atomic clock with $5.5\cdot 10^{-19}$ systematic uncertainty \cite{Marshall2025}. Most experiments now employ Ca$^+$, Ba$^+$, Sr$^+$, and Yb$^+$ which have more favorable cooling and qubit transitions. 
There are three possible ways to encode quantum information in two internal states of an ion:

{\it\textbf{(i)}} {\bf Ground-state qubits:} One approach is to encode the qubit in two ground-state energy levels. This can be either two Zeeman states of the same hyperfine manifold or, in the case of $I\neq0$, two Zeeman states of two distinct hyperfine manifolds that are usually separated by frequencies in the 1-10 GHz range. 

{\it\textbf{(ii)}} {\bf Optical qubits:} The other approach employs metastable states accessed by optical transitions, usually in the visible and UV range (150-800 THz). While they exhibit some advantages in the coherent manipulation, this encoding is fundamentally limited by the finite lifetime $\tau$, which can range widely from about 400\,ms in \Sr to $\sim10^7$\,s in \Yb (see Table~\ref{table_ions}). 

{\it\textbf{(iii)}} {\bf Metastable-state qubits:} Another possible qubit encoding recently emerged consists of using hyperfine states of a metastable manifold~\cite{Allcock2021, quinn2024highfidelityentanglementmetastabletrappedion}. This choice allows for mid-circuit, non-unitary operations, such as sympathetic cooling and measurements, to be carried out with reduced crosstalk~\cite{Yang2022} at the expense of more laser complexity.  Moreover, the metastable-state encoding also enables the conversion of physical leakage errors due to photon scattering events to erasure errors, generally leading to higher thresholds in quantum error correction codes~\cite{Kang2023}. 


\begin{table}[t!]
\centering
\caption{{\bf Trapped-ion qubit encodings:} $\omega_0$ is the ground-state and metastable-state qubit hyperfine splittings, $I$ denotes the nuclear spin, $\lambda$ and $\tau$ are the optical qubit wavelength and lifetime, respectively \cite{Allcock2021}. }
\begin{tabular}{*8c}
Ion Species &  \multicolumn{2}{c}{Ground} & \multicolumn{1}{c}{\quad\,\quad} & \multicolumn{2}{c}{Optical} & \multicolumn{1}{c}{\quad\,\quad} & \multicolumn{1}{c}{Metastable}\\
\hline
  & $I$ & $\omega_0/2\pi$ [GHz]   & & $\lambda$ [nm]    & $\tau$ [s] & & $\omega_0/2\pi$ [MHz] \\
\hline
{ $^9$Be$^+$}   & 3/2 &  1.3        & & -                 & -  &  & -\\
{ $^{25}$Mg$^+$}& 5/2 &  1.8        & & -                 & -  &  & -\\
{$^{40}$Ca$^+$} & 0   &  -          & & 729               & 1.11  & & -  \\
{$^{43}$Ca$^+$} & 7/2 &  3.2        & & 729               & 1.11  & &7, 10, 15, 20, 25  \\
{$^{87}$Sr$^+$} & 9/2 &  5.0        & & 674               & 0.357 & &8.2, 5.2, 2.7, 17, 38 \\
{$^{88}$Sr$^+$} & 0   &  -          & & 674               & 0.357 & & - \\
{$^{133}$Ba$^+$}& 1/2 &  9.9        & & 1762              & 29.8   & &89 \\
{$^{137}$Ba$^+$}& 3/2 &  8.0        & & 1762              & 29.8   & &72, 63, 0.4  \\
{$^{171}$Yb$^+$}& 1/2 &  12.6       & & 467               & $4.98\cdot10^7$    & &3620 \\
\hline
\hline
\label{table_ions}
\end{tabular}\end{table}
Atomic states in ions are subject to decoherence arising from both natural relaxation and dephasing processes. In optical qubits, the main decoherence process is related to the finite lifetime $\tau$ of the metastable state used for encoding (see Table~\ref{table_ions}), which intrinsically limits the $T_1$ and $T_2$ decoherence times. In contrast, natural relaxation in the ground-state qubits is effectively negligible apart from the black-body-induced decay, resulting in significantly longer lifetimes that make them attractive for information storage. In the case of ground- and metastable-state qubits, it is possible to select two states that are insensitive to magnetic fields to the first order, greatly enhancing their coherence properties. For example, $T^*_2=40$ s has been measured in \Caf~\cite{Harty2014}, while $T_2 > 1$ hour has been achieved in \Yb using dynamical decoupling and sympathetic cooling~\cite{Wang2021Single}.

\begin{figure}[t]
    \centering\includegraphics[width=0.9\linewidth]{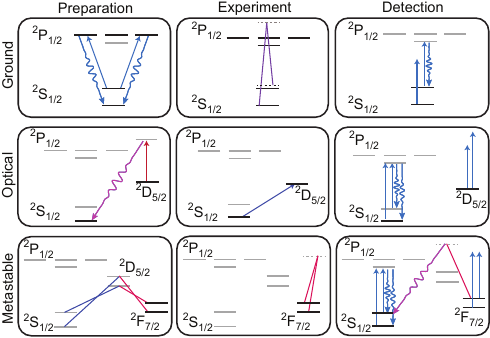}
    \caption{\Yb ion qubit encodings and their respective preparation, detection, and coherent manipulation laser schemes.}
    \label{fig_yb_qubits}
\end{figure}

As for the manipulation of the qubit, every trapped-ion experiment can be divided into three steps: \emph{(a) State preparation, (b) Experiment, (c) State detection}. In Fig.~\ref{fig_yb_qubits}, we show these three steps for ground-state, optical, and metastable-state qubits using \Yb ions as an example.
Errors in state preparation and detection are challenging to decouple, therefore, they are commonly grouped together and referred to as state preparation and measurement (SPAM) errors. State preparation relies on optical pumping, where a certain laser arrangement leads to spontaneous decay into a desired dark state. Detection involves photon scattering: qubits in the ``bright" state cyclically scatter photons, which are collected on a detector, while qubits in the ``dark" state do not scatter photons due to the off-resonant nature of the detection laser (see Fig.~\ref{fig_yb_qubits}). SPAM is an important metric because the current error correction schemes require mid-circuit readouts and resets. Although a correction cycle typically requires more gates than measurement operations, making the gate infidelity the dominant error contributor, SPAM infidelity still cannot be neglected, and, ideally, measurements have to be performed on timescales faster than the qubit decoherence~\cite{An2022High}. 

The ground-state qubit is less favorable for its SPAM fidelity, as the energy levels are spectrally less resolved, and so off-resonant scattering eventually causes bright-to-dark pumping. The best SPAM error achieved in \Yb ground-state qubits without using any auxiliary state is $8\cdot 10^{-4}$ in 100 $\mu$s~\cite{Noek2013}.
Optical qubits, on the contrary, have a very large spectral separation between qubit states and can, therefore, reach very low SPAM errors. The SPAM error of $9\cdot10^{-5}$ was achieved quite early on for optical qubits in \Ca ions~\cite{Myerson2008High}. 

Combining optical and ground-state qubits allows for spectrally resolved ``shelving'', where only one qubit state is transferred to a metastable state, resulting in low SPAM errors. For example, SPAM errors of $3\cdot10^{-4}$ were demonstrated in the $I=1/2$ radioactive isotope of Barium, $^{133}$Ba$^+$~\cite{Christensen2020} and $6 \cdot 10^{-6}$ SPAM was achieved in \Yb ions using the $^2F_{7/2}$ metastable state~\cite{Edmunds2021}.
A combination of optical pumping and shelving has been used to achieve $9.0 \times 10^{-5}$ SPAM errors in even more complicated atoms with large nuclear spin $I>1/2$, such as \Baf $(I=3/2)$~\cite{An2022High}. Furthermore, as mentioned above, the use of metastable-state qubits can further decrease SPAM errors by converting leakage and other errors into erasures that can be efficiently flagged and used for post-selection. Using this technique, Ref.~\cite{Sotirova2024High} achieved SPAM errors of $5 \cdot 10^{-6}$ for the metastable-state qubit, $7 \cdot 10^{-6}$ for the optical qubit, and $8 \cdot 10^{-6}$ for the ground-state qubit. Finally, an interesting scheme to perform partial measurements in trapped-ion chains has been realized in Ref.~\cite{Gaebler2021micromotion}, where the ions are ``hidden'' from the detection light using the micromotion induced by displacing them far from the rf null (see Sec.~\ref{sec_singleion} for details).

Using all qubit encodings at the same time is referred to as \emph{omg} architecture~\cite{Allcock2021}, where different types of qubits carry out separate functions (e.g., register and ancilla). This approach requires more laser complexity to address multiple atomic states in different ions with individual beams as compared to the dual-species architecture, where the register and ancilla functions are carried out by different atomic species with distinct spectral properties, which reduces the optical complexity. However, the metastable-state qubit approach is free from the technical challenges caused by the mass difference. These challenges include time-consuming reordering sequences after background gas collisions and inefficient sympathetic cooling and quantum logic among different species using radial modes~\cite{Sosnova2021}, which is hindered by uneven radial mode participation of the ions in the chain~\cite{Home2013Quantum}.

After having reviewed the preparation and detection schemes for trapped ions, the following section will treat in detail the ``experiment'' step, where the qubit is coherently manipulated by laser fields.

\section{Laser-Ion Interactions: Hamiltonian Simulations and Quantum Gates } \label{sec_laser-ion_interactions}

\subsection{Single Ion Manipulation} \label{sec_singleion}

In the presence of a light field, the total Hamiltonian of a single ion system can be written in terms of its internal-electronic part (Section~\ref{sec_qubits}), its external-motional part (Section~\ref{sec_Motion}), and its interaction with the laser light (in the following sections $\hbar = 1$):
\begin{equation}
\hat{H}=\hat{H}_A + \hat{H}_M + \hat{H}_I.
\end{equation}

Considering a pair of long-lived internal electronic states of the ion, we can represent the internal-electronic part of the Hamiltonian as a two-level qubit system of $\ket{\!\uparrow}$ and $\ket{\!\downarrow}$ states and write it in the following form:

\begin{equation}
\hat{H}_A=\frac{\omega_0}{2}\hat{\sigma}_z,
\label{eq:atom}
\end{equation}
where $\omega_0=\omega_{\uparrow}-\omega_{\downarrow}$ is the energy separation between the two states. The external-motional Hamiltonian of a single ion (see Eq.~\eqref{eq_motion_quant}) along a trap axis ($\alpha = x, y,$ or $z$) can be written as:
\begin{equation}
\label{eq:Hmotion}
\hat{H}_M=\sum_{\alpha}\left\{\frac{\hat{p}_\alpha^2}{2{\rm m}}  + \frac{{\rm m}}{2}\frac{\omega_{\rm rf}^2}{4}\left[a_\alpha + 2 q_\alpha\cos(\omega_{\rm rf} t)\right]\hat{x}_\alpha^2\right\}.
\end{equation}
In this subsection, we simplify the analysis by considering a one-dimensional system and omitting the subscript $\alpha$. However, the approach used in the following can naturally be extended to higher dimensions. It is worth revisiting the characteristics of the ion's motion in a linear Paul trap briefly. In the case of $\alpha = z$, which is the trap axis of the static potential, $q_\alpha = q_z = 0$, and Eq.~\eqref{eq:Hmotion} induces harmonic oscillator quantization with the level spacing of $\omega_z$ corresponding to the axial trap frequency of the ion. On the other hand, as shown in Sec.~\ref{sec_Motion}, when $\alpha = x$ or $y$, $q_\alpha \neq 0$, making the motional quantization from Eq.~\eqref{eq:Hmotion} no longer straightforward. However, in the limit of $|a_{x,y}|, \, q_{x,y}^2 \ll 1$ under which most linear Paul traps operate, micromotion can be neglected, and the motional Hamiltonian can be approximated as harmonic, $\hat{H}_M \approx \omega \hat{a}^\dagger\hat{a}$, where $\omega$ is defined in Eq.~\eqref{eq_omega} as the trap frequency of the ion along either $x$ or $y$ direction, $\hat{a}\; (\hat{a}^\dagger)$ is the annihilation (creation) operator associated with the quantized motion along $x$ or $y$ direction\footnote{Here, we have omitted the constant energy shift of $\frac{1}{2}\omega$.}. 


Since the spread of the electronic wavefunction is generally much less than the wavelength of the light field, the coupling between the near-resonant electronic states and the light field is well described by the lowest-order allowed multipole term of the electromagnetic interaction. For most ions, this usually corresponds to a dipole or quadrupole matrix element. The interaction between a semi-classical laser field and the atomic dipole or quadrupole can be described in terms of Rabi frequency $\Omega$, frequency $\omega_L$, optical phase $\phi$, and wavevector $k_L$~\cite{Steck, Leibfried2003}. Here, the wavevector points along the same trap axis considered in $\hat{H}_M$. The interaction Hamiltonian of a single ion with a laser field can be written as:
\begin{equation}
\hat{H}_I= 
\frac{\Omega}{2}(\hat{\sigma}^+ + \hat{\sigma}^-) \left[ e^{i( k_L \hat{x} - \omega_L t + \phi)} + e^{-i (k_L \hat{x} - \omega_L t + \phi)}  \right],
\label{eq:originInt}
\end{equation}
where $\hat{\sigma}^\pm=\frac{1}{2}\left(\hat{\sigma}_x \pm i\hat{\sigma}_y\right)$. It is important to note that when interacting with an optical qubit, the laser parameters $\Omega,\; \omega_L,\; \phi,$ and $k_L$ correspond to a single light field. On the other hand, for the interaction with a ground-state qubit, they describe the optical beatnote from two laser beams due to the typically required use of a stimulated Raman process (see Fig.~\ref{fig_yb_qubits}). In both cases, these parameters are independently and precisely controllable in experiments. Under the transformation with respect to the free Hamiltonian $\hat{H}_A + \hat{H}_M$, the total Hamiltonian of a single ion system interacting with a light field becomes
\begin{equation}
\hat{H}^{\rm (ord)} = \underbrace{i\left(\frac{\partial }{\partial t}\hat{U}_0^\dagger(t)\right)\hat{U}_0(t)}_{-(\hat{H}_A+ \hat{H}_M)} + \underbrace{\hat{U}_0^\dagger(t) \hat{H} \hat{U}_0(t)}_{\substack{(\hat{H}_A+ \hat{H}_M)\\+(\hat{U}_0^\dagger(t) \hat{H}_I \hat{U}_0(t))}}, \, \hat{U}_0(t)=e^{-i (\hat{H}_A+ \hat{H}_M) t}.
\end{equation}
\par
\begin{figure}[t!]
    \centering    \includegraphics[width=\columnwidth]{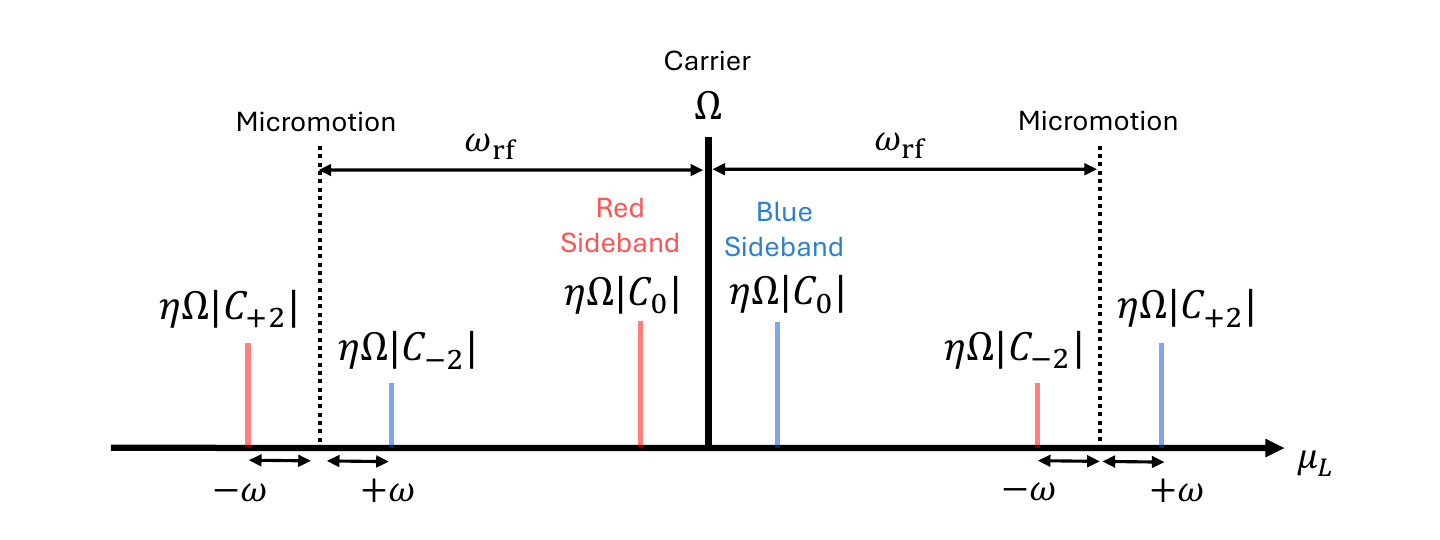}
    \caption{Amplitude suppression of the laser-ion interaction due to secular motion and intrinsic micromotion up to the first order.}
    \label{fig:sideband}
\end{figure}

 By moving into the interaction representation, we essentially rotate away the non-interacting parts of the Hamiltonian. Hence, the dynamics induced by the light field is fully described by the interaction Hamiltonian in the ordinary frame:
\begin{eqnarray}
\hat{H}_I^{\rm (ord)} &=& \hat{U}_0^\dagger(t) \hat{H}_I \hat{U}_0(t) \nonumber \\
&=& \frac{\Omega}{2}e^{i\hat{H}_A t}(\hat{\sigma}^+ + \hat{\sigma}^-)e^{-i\hat{H}_A t} e^{i\hat{H}_M t}\left[ e^{i( k_L \hat{x} - \omega_L t + \phi)} + e^{-i (k_L \hat{x} - \omega_L t + \phi)}  \right]e^{-i\hat{H}_M t} \nonumber \\
&=& \frac{\Omega}{2}(\hat{\sigma}^+ e^{i\omega_0 t} + \hat{\sigma}^- e^{-i\omega_0 t}) e^{i\hat{H}_M t}\left[ e^{i( k_L \hat{x} - \omega_L t + \phi)} + e^{-i (k_L \hat{x} - \omega_L t + \phi)}  \right]e^{-i\hat{H}_M t}, \nonumber\\
\end{eqnarray}
which expands into terms with time-dependent factors of $e^{\pm i(\omega_L \pm \omega_0)t}$. Since the fast rotating terms with the time-dependent factors of $e^{\pm i(\omega_L + \omega_0) t}$ contribute negligibly to the dynamics for $\omega_L + \omega_0$ typically being in the order of 2$\pi\;\times$ 100s of THz, they can be neglected while only the terms that oscillate at frequency $\mu_L = \omega_L - \omega_0 \ll \omega_0$ are retained. This step is generally referred to as \textit{rotating-wave approximation} (RWA)~\cite{Leibfried2003}. It is worth noting that $e^{i\hat{H}_M t}\left[ e^{i k_L \hat{x}} + e^{-i k_L \hat{x}}  \right]e^{-i\hat{H}_M t}$ resembles the similarity transformation of an operator from the Schr\"{o}dinger picture to the Heisenberg picture. Thus, the interaction Hamiltonian can be reduced to:

\begin{eqnarray}
\hat{H}_I^{\rm (ord)}
&=& \frac{\Omega}{2}\hat{\sigma}^+ e^{i[ k_L \hat{x}(t) - \mu_L t + \phi]} + \; \text{h.c.},
\end{eqnarray}
by using the Heisenberg operator from the relation $f(\hat{x}(t),t)=e^{i\hat{H}_M t}f(\hat{x},t)e^{-i\hat{H}_M t}$. Recalling the Heisenberg position operator $\hat{x}(t) = \sqrt{\frac{1}{2{\rm m}\omega}}\left[ \hat{a} u^*(t)+ \hat{a}^\dagger u(t)\right]$ shown in Section~\ref{sec_Motion_Q}, we can further simplify the expression of the interaction Hamiltonian by defining the Lamb-Dicke parameter $\eta \equiv k_L x_0$, where $x_0=\sqrt{\frac{1}{2{\rm m}\omega}}$ describes the ground-state spatial wavefunction spread of the oscillator, and thus get
\begin{eqnarray}
\hat{H}_I^{\rm (ord)}
&=& \frac{\Omega}{2}\hat{\sigma}^+ e^{i( \eta \left[ \hat{a}^\dagger u(t) + \hat{a} u^*(t)\right] - \mu_L t + \phi)} + \; \text{h.c.},\label{H_I_single}
\end{eqnarray}
where we can also write $\eta = 2\pi\frac{x_0}{\lambda_L} = \sqrt{\frac{\omega_\text{rec}}{\omega}}$ for $\lambda_L$ being the optical wavelength of the light field and $\omega_\text{rec}=\frac{k_L^2}{2{\rm m}}$ being the photon recoil frequency.
Performing a Taylor expansion on the exponential in Eq.~\eqref{H_I_single} with respect to its exponent as well as substituting $u(t)$ and $u^*(t)$ with the expression in Eq.~\eqref{eq_u(xi)} leads to:
\begin{eqnarray}
\hat{H}_I^\text{(ord)}&=&\frac{\Omega}{2}\hat{\sigma}^+ \sum_{m=0}^\infty \frac{(i\eta)^m}{m!}\left(\hat{a}e^{-i\beta_x\omega_{\rm rf}t/2}\sum_{n=-\infty}^\infty C_{2n}^* e^{-i n\omega_\text{rf} t}\right. \nonumber \\
 &\quad& + \left. \hat{a}^\dagger e^{i\beta_x\omega_\text{rf}t/2}\sum_{n=-\infty}^\infty C_{2n} e^{i n\omega_\text{rf} t} \right)^m e^{-i(\mu_L t -\phi)} + \text{h.c.},
\end{eqnarray}
where $\beta_x\omega_\text{rf}/2\approx \omega$ for $|a|, \, q^2 \ll 1$.
In typical setups, $\lambda_L \sim 150 - 450$ nm, $\omega/2\pi \sim 2 - 5$ MHz, and thus $x_0 \sim 5 - 10$ nm, which yields small $\eta \sim 0.05 - 0.15$. For $\eta \sqrt{\left<(\hat{a}+\hat{a}^\dagger)^2\right >}\ll1$, it is a good approximation to consider only the zeroth- ($m=0$) and first-order ($m=1$) contributions with respect to $\eta$ in the expansion of the interaction Hamiltonian above, resulting in
\begin{eqnarray}
\hat{H}_I^\text{(ord)}&=&\frac{\Omega}{2}\left( \hat{\sigma}^+\, e^{-i (\mu_L t-\phi)} + \hat{\sigma}^-\, e^{i (\mu_L t-\phi)} \right) \nonumber \\ 
&\quad& + 
\frac{\eta\Omega}{2} 
\left(\sum_{n=-\infty}^\infty i C_{2n} \hat{\sigma}^+ 
\left[ \hat{a}^\dagger e^{i{(\omega+ n\omega_{\rm rf})}t}+ \hat{a} e^{-i{(\omega+ n\omega_{\rm rf})}t} \right] e^{-i(\mu_L t-\phi)} + \rm{h.c.}\right). \nonumber\\
\label{eq:LambDickeApprox}
\end{eqnarray}
This is called \textit{Lamb-Dicke approximation}, which generally holds for high trap frequencies and low motional mode temperatures. The zeroth-order contribution with $\mu_L = 0$ corresponds to the carrier drive used for single-qubit rotation or gate, while all the other terms of the first-order contribution are the combination of the couplings between the qubit and the oscillator motion and between the qubit and the intrinsic micromotion, as shown in Fig.~\ref{fig:sideband}.
The decrease in the coupling strength in the first-order contribution can be intuitively understood in the ion frame as the laser being modulated at $\pm n\omega_\text{rf}$. Interestingly, the carrier Rabi coupling strength is unaffected by the micromotion in the ideal case.

However, this does not hold when the ion is displaced along the relevant trap axis by $u_0$ from the nodal position of the trap's rf field (commonly referred to as \textit{micromotion null}). In this case, the ion experiences excess micromotion from the rf electric field at the shifted position with the amplitude of $A_{\rm mm}=\frac{1}{2}u_0q$~\cite{Berkeland1998, keller2015micromotion}. The interaction Hamiltonian in Eq.~\eqref{eq:originInt} then becomes
\begin{equation}
\hat{H}_I=
\frac{\Omega}{2}(\hat{\sigma}^+ + \hat{\sigma}^-) \left[ e^{i( k_L \hat{x} + k_L A_{\rm mm}\cos{(\omega_{\rm rf}t)} - \omega_L t + \phi)} + \text{h.c.}  \right].
\end{equation}
By introducing the frequency modulation index as $\zeta \equiv k_LA_{\rm mm}$ and using Jacobi-Anger expansion in terms of the Bessel functions with
\begin{equation}
    e^{\pm i\zeta\cos{(\omega_{\rm rf}t)}}=\sum_{\nu=-\infty}^\infty J_\nu(\zeta)e^{\pm i\nu(\omega_{\rm rf}t+\pi/2)},
\end{equation}
the interaction Hamiltonian can be written as:
\begin{equation}
\hat{H}_I=
\frac{\Omega}{2}(\hat{\sigma}^+ + \hat{\sigma}^-)\sum_{\nu=-\infty}^\infty J_\nu(\zeta)\left[ e^{i( k_L \hat{x} + \nu(\omega_{\rm rf}t + \pi/2) - \omega_L t + \phi)} + \text{h.c.}  \right].
\end{equation}
The transformation of this modified expression into the interaction representation, followed by the RWA and Lamb-Dicke approximation, can similarly be carried out as earlier in the subsection, resulting in Eq.~\eqref{eq:LambDickeApprox} with $\hat\sigma^\pm\rightarrow\hat\sigma^\pm\sum_{\nu=-\infty}^\infty J_\nu(\zeta)$ $e^{\pm i\nu(\omega_{\rm rf}t+\pi/2)}$. Therefore, the excess micromotion causes frequency modulation on the ion-light interactions, which induces couplings at frequencies $\pm\nu\omega_{\rm rf}$ ($\nu=1, 2, 3, ...$) and reduces the strength of the overall laser-ion interactions~\cite{Berkeland1998}.

Although it is generally desirable to minimize excess micromotion to enhance laser-ion interaction, micromotion can be used as a feature to decouple the qubit from the laser drive for certain quantum simulation and computation applications, such as mid-circuit measurements and qubit resets. For instance, Ref.~\cite{Gaebler2021micromotion} demonstrates the use of micromotion via the ion displacement from the micromotion null to lower the scattering crosstalk errors from the mid-circuit detection on the nearby unmeasured ions, as described by
\begin{eqnarray}
\frac{\Gamma(\zeta)}{\Gamma_0}= J_0^2(\zeta) + 2\sum_{\nu=1}^\infty \frac{J_\nu(\zeta)^2}{1+ 4\nu^2\omega_{\rm rf}^2/\gamma^2},
\label{eq_BesselGamma}
\end{eqnarray}
where $\Gamma(\zeta)$ is the scattering rate with the frequency modulation index $\zeta$, $\Gamma_0$ is the micromotion-free scattering rate, and $\gamma$ is the atomic transition linewidth.
The decoupling to scattering photons occurs at the zeroes of the Bessel function $J_0(\zeta)$ and in the sideband-resolved limit $\omega_{\rm rf}>\gamma$ where the higher order terms in \eqref{eq_BesselGamma} are also suppressed.

For brevity, we may neglect the effects of the intrinsic ($C_{2n}=\delta_{n0}$) and excess micromotions, which simplifies the interaction Hamiltonian to
\begin{equation}
    \hat{H}_I^{\rm (ord)} = \frac{\Omega}{2}\left[e^{i\eta\left(\hat{a}^\dagger e^{i\omega t} + \hat{a} e^{-i\omega t} \right)}\hat{\sigma}^+\, e^{-i (\mu_L t - \phi)}+\text{h.c.}\right].
    \label{eq:ordHint}
\end{equation}
This Hamiltonian will be used in the derivations of other ion-light interactions from here onward.
In this case, when $\mu_L=0$, the dominating term is the carrier drive, which describes a single-qubit rotation or gate with
\begin{eqnarray}
    \hat{H}_{\rm C} &=& \frac{\Omega}{2}\left(\hat{\sigma}^+\, e^{i\phi} + \hat{\sigma}^-\, e^{-i\phi} \right) \nonumber \\
    &=& \frac{\Omega}{2}\hat{\sigma}_\phi,
\end{eqnarray}
where $\hat{\sigma}_\phi\equiv\hat{\sigma}^+\, e^{i\phi} + \hat{\sigma}^-\, e^{-i\phi}$ can also be written as $\hat{\sigma}_\phi=\hat{\sigma}_x\cos{\phi}-\hat{\sigma}_y\sin{\phi}$, a convenient expression for visualizing the qubit operations on the Bloch sphere. 
\subsection{Spin-dependent Force for $N$ ions} \label{sec_MSgate}
The interaction between a single ion and a light field can be extended to a chain of $N$ ions. For the purpose of generalization, we consider the case where the individual-addressing beams are applied to the ions, indexed by $j$, and have the wavevectors propagate along a direction that corresponds to a set of collective motional modes $m$. The interaction Hamiltonian of the $N$-body system is simply the sum of the interaction Hamiltonians of the single-ion systems:
\begin{equation}
    \hat{H}_I^{\rm (ord)} = \sum_{j=1}^N\frac{\Omega_j}{2}\left[e^{i \sum\limits_{m=1}^N \eta_{jm}\left(\hat{a}_m^\dagger e^{i\omega_m t} + \hat{a}_m e^{-i\omega_m t} \right)}\hat{\sigma}_j^+\, e^{-i (\mu_L^j t - \phi_j)} + \text{h.c.} \right].
\end{equation}
Here, we have ignored the effects from the micromotion and applied the harmonic oscillator quantization to the motion, where the motional Hamiltonian is $\hat{H}_M=\sum_m\omega_m\hat{a}_m^\dagger\hat{a}_m$, and the Lamb-Dicke parameter for ion $j$ on mode $m$ is $\eta_{jm}=b_{jm} \xi^{(0)}_m k_L$ with $\xi^{(0)}_m=\sqrt{\frac{1}{2{\rm m}\omega_m}}$.
To describe the configuration with global beams illuminating all the ions equally, one can drop the index $j$ on $\Omega_j, \mu_L^j,$ and $\phi_j$.

\begin{figure}[t!]
    \centering
    \includegraphics[width=\columnwidth]{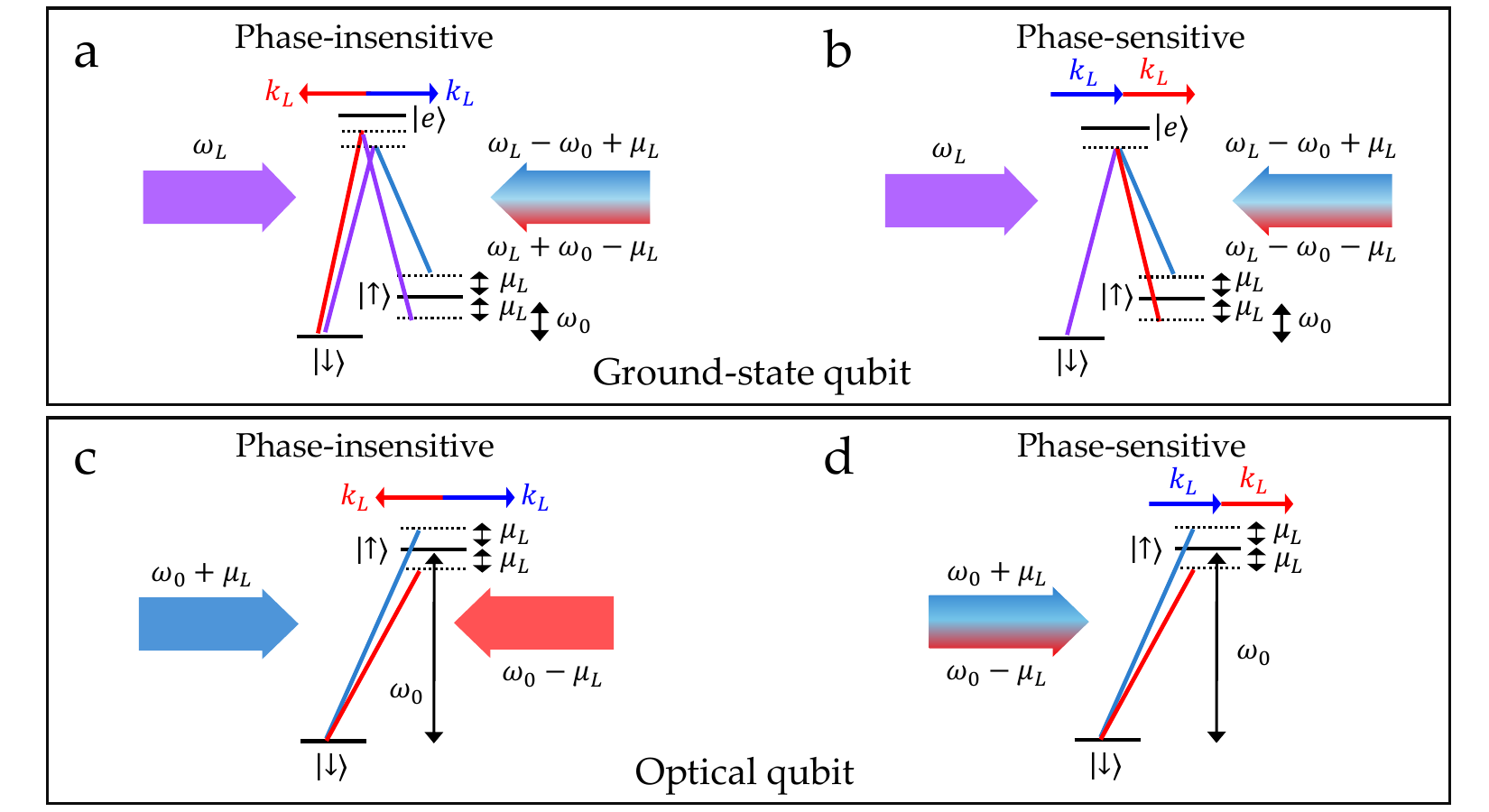}
    \caption{Laser beam geometries that determine the phase sensitivity of the spin-dependent force on a single ion. For a ground-state qubit, configurations (a) and (b) make the interaction sensitive to the motional and spin phases, respectively. Configurations (c) and (d) are the optical qubit's counterparts.}
    \label{fig:phase}
\end{figure}

In the Lamb-Dicke regime, where the excursion of the ion motion is much less than the wavelength of the driving transitions, ($\eta \sqrt{\left <(\hat{a}+\hat{a}^\dagger)^2\right >}\ll1$), the interaction Hamiltonian can be expressed up to the first order as in Sec.~\ref{sec_singleion} obtaining:
\begin{eqnarray}
\label{eq:H_I_int}
\hat{H}_I^{\rm (ord)}&=&\sum_{j=1}^N\frac{\Omega_j}{2}\left( \hat{\sigma}_j^+\, e^{-i (\mu_L^j t - \phi_j)} + \hat{\sigma}_j^-\, e^{i (\mu_L^j t - \phi_j)} \right)    \nonumber \\
&\quad& +  \sum_{j=1}^N\sum_{m=1}^N\frac{i\eta_{jm}\Omega_j}{2} 
\left( \hat{\sigma}_j^+\, e^{-i (\mu_L^j t - \phi_j)} - \hat{\sigma}_j^-\, e^{i (\mu_L^j t - \phi_j)} \right) \nonumber\\ &\quad&\quad\times \left[ \hat{a}_m^\dagger e^{i\omega_m t} + \hat{a}_m e^{-i\omega_m t} \right].
\end{eqnarray}
With the laser parameter choice of $\mu_B^j=\mu$, such that $\delta_{m}=\mu- \omega_m$ with optical phase \textcolor{black}{$\phi_B^j$} and $\mu_R^j=-\mu$ with optical phase \textcolor{black}{$\phi_R^j$}, the interaction Hamiltonian describes the \textcolor{black}{blue} and \textcolor{black}{red} sideband interaction as follows:
\begin{eqnarray}
{\color{black}\hat{H}_B}&=&\sum_{j=1}^N\frac{\Omega_j}{2}\left( \hat{\sigma}_j^+\, e^{-i \mu t + i\phi_B^j} + \hat{\sigma}_j^-\, e^{i \mu t - i\phi_B^j} \right)    \nonumber \\
&\quad&+
{\color{black}\sum_{j=1}^N\sum_{m=1}^N\frac{i\eta_{jm}\Omega_j}{2} 
\left( \hat{\sigma}_j^+  \hat{a}_m^\dagger\, e^{-i \delta_{m} t+i\phi_B^j}  - \hat{\sigma}_j^- \hat{a}_m \, e^{i \delta_{m} t-i\phi_B^j} \right) } \\
&\quad&+ {\color{black}\sum_{j=1}^N\sum_{m=1}^N\frac{i\eta_{jm}\Omega_j}{2} 
\left( \hat{\sigma}_j^+ \hat{a}_m \, e^{-i (\mu+\omega_m) t + i\phi_B^j}  - \hat{\sigma}_j^- \hat{a}_m^\dagger\, e^{i (\mu+\omega_m) t -i\phi_B^j} \right) },\nonumber
\label{eq_BSB}
\end{eqnarray}
or
\begin{eqnarray}
{\color{black}\hat{H}_R}&=&\sum_{j=1}^N\frac{\Omega_j}{2}\left( \hat{\sigma}_j^+\, e^{i \mu t + i\phi_R^j} + \hat{\sigma}_j^-\, e^{-i \mu t - i\phi_R^j} \right)    \nonumber \\
&\quad&+ {\color{black}\sum_{j=1}^N\sum_{m=1}^N\frac{i\eta_{jm}\Omega_j}{2} 
\left( \hat{\sigma}_j^+ \hat{a}_m \, e^{i \delta_{m} t + i\phi_R^j}  - \hat{\sigma}_j^- \hat{a}_m^\dagger\, e^{-i \delta_{m} t - i\phi_R^j} \right) } \\
&\quad&+
{\color{black}\sum_{j=1}^N\sum_{m=1}^N\frac{i\eta_{jm}\Omega_j}{2} 
\left( \hat{\sigma}_j^+  \hat{a}_m^\dagger\, e^{i (\mu+\omega_m) t + i\phi_R^j}  - \hat{\sigma}_j^- \hat{a}_m \, e^{-i (\mu+\omega_m) t - i\phi_R^j} \right) }. \nonumber
\label{eq_RSB}
\end{eqnarray}
By combining the two sideband interactions, the resultant interaction is a spin-dependent force on the ions that is described by:
\begin{eqnarray}
\hat{H}_{\rm SDF} &=& \color{black}{\hat{H}_B} + \color{black}{\hat{H}_R} \nonumber\\ 
&=& \sum_{j=1}^N\Omega_j\underbrace{(\hat{\sigma}_j^+e^{i(\phi_s^j-\pi/2)} + \hat{\sigma}_j^-e^{-i(\phi_s^j-\pi/2)})}_{\hat{\sigma}_x^j\cos{(\phi_s^j-\pi/2)}-\hat{\sigma}_y^j\sin{(\phi_s^j-\pi/2)}}\cos(\mu t - \psi_j) \nonumber \\ &\quad& + \sum_{j=1}^N\sum_{m=1}^N\frac{\eta_{jm}\Omega_j}{2} 
\underbrace{(\hat{\sigma}_j^+e^{i\phi_s^j} + \hat{\sigma}_j^-e^{-i\phi_s^j})}_{\hat{\sigma}_x^j\cos{\phi_s^j}-\hat{\sigma}_y^j\sin{\phi_s^j}}\left[ \hat{a}_m \, e^{i (\delta_{m} t - \psi_j)}  +  \hat{a}_m^\dagger\, e^{-i (\delta_{m} t - \psi_j)}\right] \nonumber\\
&\quad& + \sum_{j=1}^N\sum_{m=1}^N\frac{\eta_{jm}\Omega_j}{2} 
\underbrace{(\hat{\sigma}_j^+e^{i\phi_s^j} + \hat{\sigma}_j^-e^{-i\phi_s^j})}_{\hat{\sigma}_x^j\cos{\phi_s^j}-\hat{\sigma}_y^j\sin{\phi_s^j}}\\
&\quad&\quad\times\left[ \hat{a}_m \, e^{-i [(\mu+\omega_m) t - \psi_j]}  +  \hat{a}_m^\dagger\, e^{i [(\mu+\omega_m) t - \psi_j]}\right], \nonumber
\end{eqnarray}
where $\phi_s^j=\frac{\phi_B^j+\phi_R^j}{2}+\frac{\pi}{2}$ denotes the spin phase that is the angle of the qubit rotation axis in the $x$-$y$ plane of the Bloch sphere for ion $j$, and $\psi_j=\frac{\phi_B^j-\phi_R^j}{2}$ describes the motional phase that is the displacement angle in phase-space for ion $j$. To further simplify the expression, we define the spin operator for ion $j$ as $\hat{\sigma}_{\phi_s^j}\equiv\hat{\sigma}_x^j\cos{\phi_s^j}-\hat{\sigma}_y^j\sin{\phi_s^j}$ and thus obtain
\begin{eqnarray}
\hat{H}_{\rm SDF}&=&\sum_{j=1}^N\Omega_j \hat{\sigma}_{\phi_s^j-\pi/2} \cos(\mu t-\psi_j)  \nonumber \\
&\quad& +
\sum_{j=1}^N\sum_{m=1}^N\frac{\eta_{jm}\Omega_j}{2}  
\hat{\sigma}_{\phi_s^j}\left( \hat{a}_m \, e^{i (\delta_{m} t -\psi_j)}  + \hat{a}_m^\dagger \, e^{-i (\delta_{m} t -\psi_j)} \right) \\
&\quad& +
\sum_{j=1}^N\sum_{m=1}^N\frac{\eta_{jm}\Omega_j}{2}  
\hat{\sigma}_{\phi_s^j}\left( \hat{a}_m \, e^{-i [(\mu+\omega_m) t -\psi_j]}  + \hat{a}_m^\dagger \, e^{i [(\mu+\omega_m) t -\psi_j]} \right), \nonumber
\label{eq:full_SDK}
\end{eqnarray}
where the first term describes the off-resonant carrier drive, which can be neglected for $\Omega_j\ll\mu$, while the second and third terms correspond to spin-motion coupling interactions. The third term is the counter-rotating term that can be neglected within RWA if $\mu+\omega_m\gg\delta_m$. For larger $\Omega_j$, fast-rotating carrier drive leads to errors for the spin-motion coupling interaction because $[\hat{\sigma}_{\phi_s^j-\pi/2},\hat{\sigma}_{\phi_s^j}]\neq 0$. It is worth noting that the $\pi/2$ phase difference comes from the ``$i$" in the imaginary linear term of the Lamb-Dicke expansion in Eq.~\eqref{eq:H_I_int}~\cite{monroe2021programmable}. Retaining both rotating and counter-rotating terms, the spin-dependent force Hamiltonian can be expressed as:
\begin{eqnarray}
\hat{H}_{\rm SDF}&=&\sum_{j=1}^N\Omega_j \hat{\sigma}_{\phi_s^j-\pi/2} \cos(\mu t-\psi_j)  \nonumber \\
&\quad& +
\sum_{j=1}^N\sum_{m=1}^N\eta_{jm}\Omega_j\cos{(\mu t - \psi_j)} 
\hat{\sigma}_{\phi_s^j}\left( \hat{a}_m \, e^{-i\omega_m t}  + \hat{a}_m^\dagger \, e^{i\omega_m t} \right).
\label{eq:MS_SDK}
\end{eqnarray}
In realistic experimental conditions, fluctuations $\delta k \bar{x}_j$ can arise due to interferometric instability of the ions' position with respect to the laser optical phase. This can affect either the spin or motional phase, where the sensitivity depends on the geometry of the laser beams, shown in Fig.~\ref{fig:phase}.

The configuration, where the wavevectors of the blue sideband and red sideband lights co-propagate, is referred to as phase-sensitive. Meanwhile, the opposite case is called phase-insensitive. 
In phase-insensitive geometry, optical path noise affects the motional phase $\psi_j$ while leaving the spin phase common-mode, improving robustness of entangling gates. The phases in the two configurations are summarized as follows~\cite{monroe2021programmable}:







\vspace{-0.5cm}
\begin{equation}
\begin{array}{llllll}
&\, & &\textrm{Spin phase} & &\textrm{Motional phase} \nonumber \\
&\textrm{Phase-sensitive} &
&\phi_s^j =\left(\frac{\phi_{B}+\phi_{R}}{2}\right)+ \frac{\pi}{2} +\delta k \bar{x}_j &
&\psi_j = \left(\frac{\phi_{B}-\phi_{R}}{2}\right).  \label{phasesens}  \\
&\textrm{Phase-insensitive} &
&\phi_s^j =\left(\frac{\phi_{B}+\phi_{R}}{2}\right) &
&\psi_j = \left(\frac{\phi_{B}-\phi_{R}}{2}\right)+ \frac{\pi}{2} +\delta k \bar{x}_j.
\end{array}
\end{equation}

It is worth noting that in the phase-insensitive configuration, $k_B$ and $k_R$ carry opposite signs. This causes the ``$i$" phase difference between the carrier and motional drives to be included in the motional phase, instead of in the spin phase. Thus, Eq.~\eqref{eq:MS_SDK} is modified to:
\begin{eqnarray}
\hat{H}_{\rm SDF}&=&\sum_{j=1}^N\Omega_j \hat{\sigma}_{\phi_s^j} \cos(\mu t-\psi_j+\pi/2)  \nonumber \\
&\quad& +
\sum_{j=1}^N\sum_{m=1}^N\eta_{jm}\Omega_j\cos{(\mu t - \psi_j)} 
\hat{\sigma}_{\phi_s^j}\left( \hat{a}_m \, e^{-i\omega_m t}  + \hat{a}_m^\dagger \, e^{i\omega_m t} \right).
\label{eq:MS_SDK_ph_insensitive}
\end{eqnarray}
In this configuration, the spin operators associated with the carrier and the spin-motion drives commute. Therefore, errors induced by the off-resonant carrier term can be echoed during a fast gate. The phase-insensitive configuration with the addition of a standing wave allows a further suppression of the carrier errors in MS gates, as recently demonstrated in Ref.~\cite{Saner2023}.




\subsection{Mølmer-Sørensen Spin-Spin Interactions}\label{sec_MS}
Here, we review the dynamics of the trapped-ion system undergoing the time-dependent Hamiltonian of the spin-dependent force in Eq.~\eqref{eq:MS_SDK} with $\mu\gg\Omega_j$ and $\delta_m\gg\eta_{jm}\Omega_j$, which leads to an effective Hamiltonian with pure spin-spin interactions that can be used to simulate spin models. 

The dynamics under the time-dependent Hamiltonian in Eq.~\eqref{eq:MS_SDK} can be described by the time-evolution operator:
\begin{equation}
    \hat{U}(\tau) = \hat{\mathcal{T}}\left[e^{-i\int\limits_0^\tau dt\hat{H}_{\rm SDF}(t)}\right],
    \label{eq:time-evolution}
\end{equation}
where $\hat{\mathcal{T}}$ is the time-ordering operator, and $\tau$ is the total interaction time. From the Magnus expansion, the time-evolution operator becomes approximately
\begin{eqnarray}
\hat{U}(\tau) &\approx& \exp \left[ \underbrace{-i\int_0^{\tau} dt\hat{H}_{\rm SDF}(t)}_{\overline{\Omega}_1(\tau)}\underbrace{ - \frac{1}{2}\int_0^{\tau} dt_1 \int_0^{t_1} dt_2 \; [\hat{H}_{\rm SDF}(t_1),\hat{H}_{\rm SDF}(t_2)]}_{\overline{\Omega}_2(\tau)}  \right], \nonumber\\
\label{evolution}
\end{eqnarray}
since the higher-order terms besides the first two terms are negligible at sufficiently large $\mu$.
After evaluating the integral of the first term in the exponent, we obtain
\begin{eqnarray}
\overline{\Omega}_1(\tau) &=& -i\sum_{j=1}^N\frac{\Omega_j}{\mu}(\sin{(\mu \tau - \psi_j)}+\sin{\psi_j})\hat{\sigma}_{\phi_s^j-\pi/2} \nonumber \\
&\quad& + \sum_{j=1}^N\sum_{m=1}^N\hat{\sigma}_{\phi_s^j}(\alpha_{jm}(\tau)\hat{a}_m^\dagger - \alpha_{jm}^*(\tau) \hat{a}_m) \\
&=& -i\sum_{j=1}^N\frac{\Omega_j}{\mu}(\sin{(\mu \tau - \psi_j)}+\sin{\psi_j})\hat{\sigma}_{\phi_s^j-\pi/2} \nonumber \\
&\quad& - \;i\sum_{j=1}^N\sum_{m=1}^N\hat{\sigma}_{\phi_s^j}\left(\text{Re}[\alpha_{jm}]\xi^{(0)}_m\hat{p}_m-\text{Im}[\alpha_{jm}]\frac{1}{\xi^{(0)}_m}\hat{x}_m\right),
\end{eqnarray}
where $\hat{x}_m=\xi^{(0)}_m(\hat{a}_m+\hat{a}_m^\dagger)$ and $\hat{p}_m=-i\frac{1}{\xi^{(0)}_m}(\hat{a}_m-\hat{a}_m^\dagger)$ represent the position and momentum quadratures in phase space. The first term resembles a rotation Hamiltonian, but it can be neglected for $\Omega_j \ll \mu$, analogous to dropping the off-resonant coupling term of the spin-dependent force Hamiltonian in Eq.~\eqref{eq:full_SDK}. On the other hand, the second term describes the spin-dependent coherent displacements of the motional modes through phase space, where the amount of displacement for mode $m$ is:
\begin{eqnarray}
\alpha_{jm}(\tau) &=& \frac{-i\eta_{jm}\Omega_j}{\mu^2-\omega_m^2}\left(e^{i\omega_m \tau}\left[\mu\sin{(\mu \tau - \psi_j)}+i\omega_m\cos{(\mu \tau - \psi_j)}\right]\right. \nonumber \\
&\quad& \quad \quad \quad \quad \quad \left. + \;\mu\sin{\psi_j}-i\omega_m\cos{\psi_j}\right).
\end{eqnarray}
The strength of $\alpha_{jm}$ determines the degree of entanglement between ion $j$ and motional mode $m$, which is greater for smaller $\delta_{m}/\eta_{jm}\Omega_j = (\mu-\omega_m)/\eta_{jm}\Omega_j$. It is worth noting that, even with large $\alpha_{jm}$, the spin-phonon coupling remains bounded over time, therefore it is always possible to determine specific times $t_m$ when the ion $j$ can be disentangled from mode $m$, namely $\alpha_{jm}(t_m)\approx 0$. When $\delta_{m}$ is small with sufficiently large $\mu$, the displacement amplitude approximately becomes
\begin{eqnarray}
    \alpha_{jm}(\tau)=-\frac{\eta_{jm}\Omega_j e^{i\psi_j}}{2\delta_{m}}\left(1-e^{-i\delta_{m}\tau}\right),
\end{eqnarray}
which is equivalent to the result from the integration without considering the counter-rotating term in Eq.~\eqref{eq:full_SDK}. In this case, $t_m = 2\pi \ell/\delta_m$, where $\ell$ is an integer.

The second integral of the exponent of the time-evolution operator in Eq.~\eqref{evolution} can be expanded as:
\begin{eqnarray}
\overline{\Omega}_2(\tau) &=& \underbrace{ - \frac{1}{2}\int_0^{\tau} dt_1 \int_0^{t_1} dt_2 \; [\hat{H}_{\rm SDF}^{\rm C}(t_1),\hat{H}_{\rm SDF}^{\rm C}(t_2)]}_{\overline{\Omega}_2^{\rm CC}(\tau)} \nonumber \\
&\quad& \underbrace{ - \frac{1}{2}\int_0^{\tau} dt_1 \int_0^{t_1} dt_2 \; [\hat{H}_{\rm SDF}^{\rm C}(t_1),\hat{H}_{\rm SDF}^{\rm SM}(t_2)]}_{\overline{\Omega}_2^{\rm CS}(\tau)} \\
&\quad& \underbrace{ - \frac{1}{2}\int_0^{\tau} dt_1 \int_0^{t_1} dt_2 \; [\hat{H}_{\rm SDF}^{\rm SM}(t_1),\hat{H}_{\rm SDF}^{\rm C}(t_2)]}_{\overline{\Omega}_2^{\rm SC}(\tau)} \nonumber \\
&\quad& \underbrace{ - \frac{1}{2}\int_0^{\tau} dt_1 \int_0^{t_1} dt_2 \; [\hat{H}_{\rm SDF}^{\rm SM}(t_1),\hat{H}_{\rm SDF}^{\rm SM}(t_2)]}_{\overline{\Omega}_2^{\rm SS}(\tau)}, \nonumber
\end{eqnarray}
where $\hat{H}_{\rm SDF}^{\rm C}(t)$ and $\hat{H}_{\rm SDF}^{\rm SM}(t)$ are the off-resonant carrier and spin-motion coupling drives in the spin-dependent force Hamiltonian in Eq.~\eqref{eq:MS_SDK}, respectively. For clarity, we shall evaluate these integrals separately (see results in Appendix~\ref{app_2ndMagnus}). Due to the commutation relation of the Pauli operators, $\overline{\Omega}_2^{\rm CC}(\tau) = 0$. 

The contributions from $\overline{\Omega}_2^{\rm CS}(\tau)$ and $\overline{\Omega}_2^{\rm SC}(\tau)$ are negligible under the conditions $\mu\gg\Omega_j$ and $\delta_m\gg\eta_{jm}\Omega_j$, which are usually met when operating in the quantum simulation regime, also known as dispersive regime. Therefore, among the four integrals, only $\overline{\Omega}_2^{\rm SS}(\tau)$ determines the dynamics of the system. However, as shown in its explicit form in Appendix~\ref{app_2ndMagnus}, $\overline{\Omega}_2^{\rm SS}(\tau)$ consists of a linear term and time-bounded terms. At long time $\tau$, the integral is dominated by the linear term and can be approximated as:

\begin{figure}[t!]
    \centering
    \includegraphics[width=0.8\columnwidth]{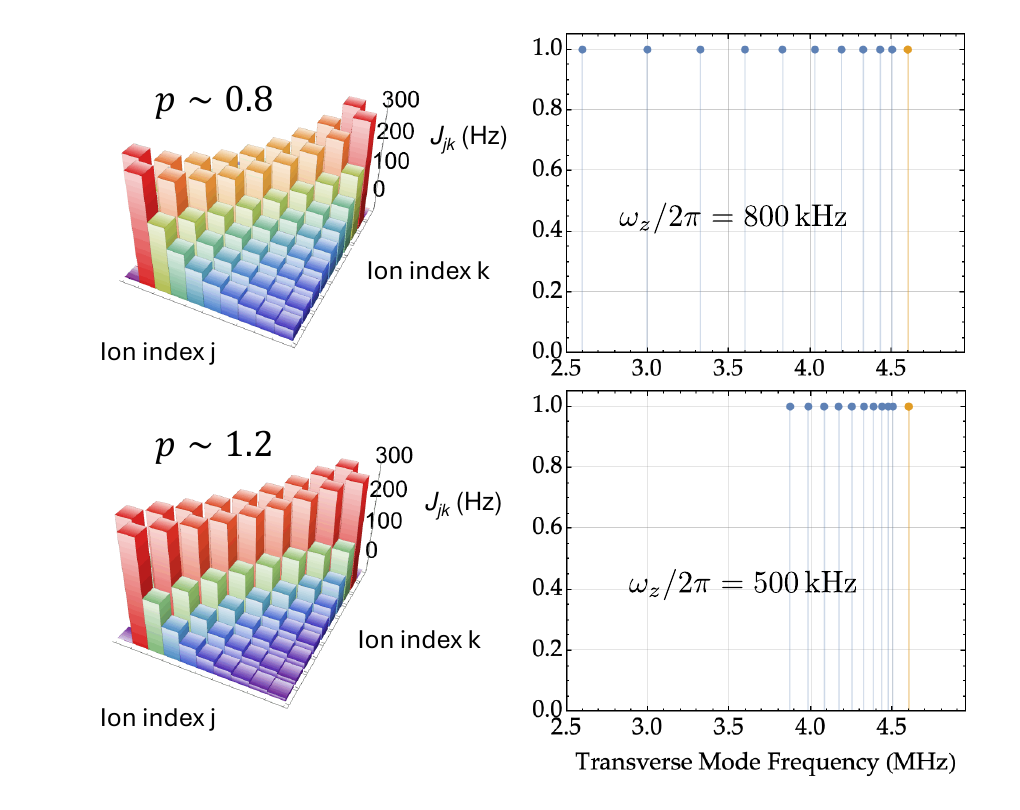}
    \caption{The power-law spin-spin interaction strengths $J_{jk}=\frac{J_0}{|j-k|^p}$ with two different exponents $p=0.8$ and $1.2$ corresponding to the radial mode spectra given by the axial center-of-mass frequencies $\omega_z/2\pi = 0.800$ MHz and $0.500$ MHz, respectively.}
    \label{fig:powerlaw}
\end{figure}

\begin{eqnarray}
    \overline{\Omega}_2^{\rm SS}(\tau) &\approx& -i\sum_{j<k}\left(\sum_{m=1}^N\frac{\eta_{jm}\eta_{km}\Omega_j\Omega_k}{\mu^2-\omega_m^2}\hat{\sigma}_{\phi_s^j}\hat{\sigma}_{\phi_s^k}\cos{(\psi_j-\psi_k)}\omega_m\right) \tau \nonumber \\
    &\equiv& -i\sum_{j<k}J_{jk}\hat{\sigma}_{\phi_s^j}\hat{\sigma}_{\phi_s^k}\tau,
    \label{eq_Magnus_result2}
\end{eqnarray}
where $\sum_{j=1}^N\sum_{k=1}^N\frac{1}{2}\equiv\sum_{j<k}$. This results in the effective Hamiltonian:
\begin{eqnarray}
    \hat{H}_{\rm eff} = \sum_{j<k}J_{jk}\hat{\sigma}_{\phi_s^j}\hat{\sigma}_{\phi_s^k},
    \label{eq_HJij_ions}
\end{eqnarray}
with the spin-spin coupling matrix elements:
\begin{equation}
    J_{jk}=\sum_{m}\frac{\eta_{jm}\eta_{km}\Omega_j\Omega_k}{\mu^2-\omega_m^2}\omega_m\cos{(\psi_j-\psi_k)},
\end{equation}
that can also be written as:
\begin{equation}
    J_{jk}=\Omega_j \Omega_k \omega_{\rm rec} \sum_{m} \frac{b_{jm} b_{km}}{\mu^2 - \omega_m^2}\cos(\psi_{j}-\psi_{k}).
\end{equation}
with $\omega_{\rm rec}=k_L^2/2{\rm m}$, being the recoil angular frequency. 

This interaction describes spin-spin couplings with full connectivity across the chain. In the case of the bichromatic laser drive with a uniform Rabi frequency $\Omega_j = \Omega$, common optical phase such that $\phi_{s}^j=\phi_s$ and $\psi_j=\psi$, and beatnote frequencies $\pm\mu$ such that $\delta_m>0,\;\forall m$, the spin-spin coupling matrix elements can be approximated by a power-law decay with respect to the ion-ion distance, which is described by:
\begin{eqnarray}
    J_{jk}=\frac{J_0}{|j-k|^p},
\end{eqnarray}
where $J_0$ is the nearest-neighbor coupling strength, typically controlled by the laser power. On the other hand, the inverse power law exponent $p$ is experimentally adjusted with either the beatnote frequency $\mu$, which determines the contribution of the motional modes $m$ to $J_{jk}$, or the axial trap frequency $\omega_z$, which sets the frequency spacings between the radial modes, and in turn, the values of the radial mode contribution factors ${b_{jm} b_{km}/(\mu^2 - \omega_m^2)}$, as shown in Fig.~\ref{fig:powerlaw}. Theoretically, the range $0<p<3$ is accessible, where $p\rightarrow 0$ when $\mu\rightarrow 0$, and $p\rightarrow 3$ when $\mu\rightarrow\infty$. In practice, $0.5\lesssim p \lesssim1.5$ is more comfortably achievable with a state-of-the-art trapped ion quantum simulator. Similarly, a beatnote frequency $\mu$ placed between the motional modes $\omega_m$ will give rise to other forms of $J_{jk}$ interaction matrix. By using multiple bichromatic beatnote frequency tones $\mu_l$ ($l>1$), more complex patterns of $J_{jk}$ interaction matrix from the combination of characteristic interaction matrices can be realized, as proposed in Refs.~\cite{Korenblit2012, Davoudi2020towards, Kyprianidis2024interaction}.

In order to apply a precisely tunable transverse field, there are two options: $(i)$ applying a carrier beatnote with a spin phase containing a $\pi/2$ offset with respect to the spin-motion coupling terms; $(ii)$ applying an asymmetric detuning to the bichromatic drive for the spin-dependent force, namely ${\mu_B^j=\omega_m+\delta_{m}-\Delta=\mu-\Delta}$ and ${\mu_R^j=-\omega_m-\delta_{m}-\Delta}$ ${=-\mu-\Delta}$~\cite{lee2016thesis}. Here, we analyze the latter case in more detail: by rotating to a shifted frame with the transformation $\hat{U}_0(t)=e^{-i (\hat{H}_A+ \hat{H}_M-{\hat{\sigma}_z^j \Delta/2}) t}$, Eq.~\eqref{eq:full_SDK} is modified to:
\begin{eqnarray}
    \hat{H}_{\rm SDF}^{\rm (shift)}&=&\sum_{j=1}^N\frac{\Delta}{2} \hat{\sigma}_{z}^j + \sum_{j=1}^N\Omega_j \hat{\sigma}_{\phi_s^j-\pi/2} \cos(\mu t-\psi_j)  \nonumber \\
    &\quad& +
    \sum_{j=1}^N\sum_{m=1}^N\frac{\eta_{jm}\Omega_j}{2}  
    \hat{\sigma}_{\phi_s^j}\left( \hat{a}_m \, e^{i (\delta_{m} t -\psi_j)}  + \hat{a}_m^\dagger \, e^{-i (\delta_{m} t -\psi_j)} \right) \\
    &\quad& +
    \sum_{j=1}^N\sum_{m=1}^N\frac{\eta_{jm}\Omega_j}{2}  
    \hat{\sigma}_{\phi_s^j}\left( \hat{a}_m \, e^{-i [(\mu+\omega_m) t -\psi_j]}  + \hat{a}_m^\dagger \, e^{i [(\mu+\omega_m) t -\psi_j]} \right), \nonumber    \label{eq:HSDKshift}
\end{eqnarray}
with $\Delta\ll\delta_m,\;\forall m$. This can be understood as transforming the Hamiltonian to the frame rotating with the frequency $-\Delta$ away from the bare qubit frequency $\omega_0$ in Eq.~\eqref{eq:atom}. It is worth noting that in this case, Eq.~\eqref{eq:ordHint} becomes
\begin{equation}
    \hat{H}_I^{\rm (shift)} = \frac{\Omega}{2}e^{i\eta\left(\hat{a}^\dagger e^{i\omega t} + \hat{a} e^{-i\omega t} \right)}\left(\hat{\sigma}^+\, e^{-i ((\mu_L+\Delta) t - \phi)} + \hat{\sigma}^-\, e^{i ((\mu_L+\Delta) t - \phi)} \right),
    \label{eq:ordHintshift}
\end{equation}
Hence, in this frame, a single qubit gate or carrier rotation is achieved with the laser beatnote frequency of $\mu_L=-\Delta$ ($\omega_L=\omega_0-\Delta$) instead of $\mu_L=0$  ($\omega_L=\omega_0)$.

By imposing an additional condition of $\Delta/2\ll\eta_{jm}\Omega_j$ on top of the familiar restrictions of $\mu\gg\Omega_j$ and $\delta_m\gg\eta_{jm}\Omega_j$, the effective Hamiltonian in the time-evolution operator can be expressed as:
\begin{eqnarray}
    \hat{H}_{\rm eff} = \sum_{j<k}J_{jk}\hat{\sigma}_{\phi_s^j}\hat{\sigma}_{\phi_s^k} + \sum_{j=1}^N\frac{\Delta}{2} \hat{\sigma}_{z}^j.
\end{eqnarray}
Similar to the two conditions also used in the ordinary frame derivation, this new constraint exists to suppress the non-commuting contributions between the first and other terms in Eq.~\eqref{eq:HSDKshift} when performing the Magnus expansion. The effective Hamiltonian essentially describes the spin-spin interactions with a global transverse field.

In the case of $\hat{\sigma}_{\phi_s^j}\hat{\sigma}_{\phi_s^k} = \hat{\sigma}_x^j\hat{\sigma}_x^k$, we can expand the relation $\hat{\sigma}_x^j=\hat{\sigma}_+^j+\hat{\sigma}_-^j$ and shift the reference frame again by $\sum_j\frac{\Delta}{2}\hat{\sigma}_z^j$ to obtain the following effective Hamiltonian
\begin{eqnarray}
    \hat{H}_{\rm eff} = \sum_{j<k}J_{jk}\left(\hat{\sigma}_+^j\hat{\sigma}_+^k e^{i\Delta t} + \hat{\sigma}_+^j\hat{\sigma}_-^k + \hat{\sigma}_-^j\hat{\sigma}_+^k + \hat{\sigma}_-^j\hat{\sigma}_-^k e^{-i\Delta t}\right),
\end{eqnarray}
In the limit of $\Delta/2 \gg J_{jk}$, the Hamiltonian is further simplified by suppressing the double excitation processes to
\begin{eqnarray}
    \hat{H}_{\rm eff} = \sum_{j<k}J_{jk}\left(\hat{\sigma}_+^j\hat{\sigma}_-^k + \hat{\sigma}_-^j\hat{\sigma}_+^k\right),
\end{eqnarray}
describing spin-exchange interactions that conserve the total number of spin excitations along the transverse field $\sum_j\frac{\Delta}{2}\hat{\sigma}_z^j$. Moreover, we can rewrite this Hamiltonian as an XY Hamiltonian by replacing $\hat{\sigma}_-^j$ and $\hat{\sigma}_+^j$ with their expressions in terms of $\hat{\sigma}_x^j$ and $\hat{\sigma}_y^j$, which results in:
\begin{eqnarray}
    \hat{H}_{\rm eff} = \sum_{j<k}\frac{J_{jk}}{2}\left(\hat{\sigma}_x^j\hat{\sigma}_x^k + \hat{\sigma}_y^j\hat{\sigma}_y^k\right).
\end{eqnarray}

We emphasize that the transverse field $(\Delta/2)\hat{\sigma}_z^j$ shown in the derivation is not necessarily induced by the asymmetric detuning of the bichromatic drive, but it can alternatively be engineered using the carrier method: applying a $\hat{\sigma}_{\phi_s^j+\pi/2}$ drive sandwiched by appropriate $\pi/2$ pulses to map the $\hat{\sigma}_z^j$-basis onto the $\hat{\sigma}_{\phi_s^j+\pi/2}$-basis.

\subsection{Mølmer-Sørensen Gate}\label{sec_MS_gate}

In this section, we will adapt the quantum simulation scheme described above to derive the Mølmer–Sørensen (MS) gate, which forms a digital logic primitive in trapped-ion quantum computing~\cite{Molmer1999, Molmer2000}. To perform the gate, we apply the spin-motion drive with small detunings $\delta_m$ for specific gate times such that the motional modes are quasi-resonantly excited and de-excited.
This means that the conditions used in the quantum simulation or dispersive regime ($\mu\gg\Omega_i$ and $\delta_m\gg\eta_{im}\Omega_i$) do not apply in this case.
We will see that in the simplest two-ion-one-mode case, the obvious choice is $t_0= 2\pi \ell/\delta$ with integer $\ell$, while more complex pulse schemes are needed for a multi-ion chain case, where many motional modes are involved in the dynamics and, therefore, the conditions $\alpha_{jm}(t_{\rm gate})=0$ need to be simultaneously satisfied at the end of the gate.

\begin{wrapfigure}[27]{r}{0.57\textwidth}
    \centering
    \vspace{-0.5cm}
\includegraphics[width=0.55\columnwidth]{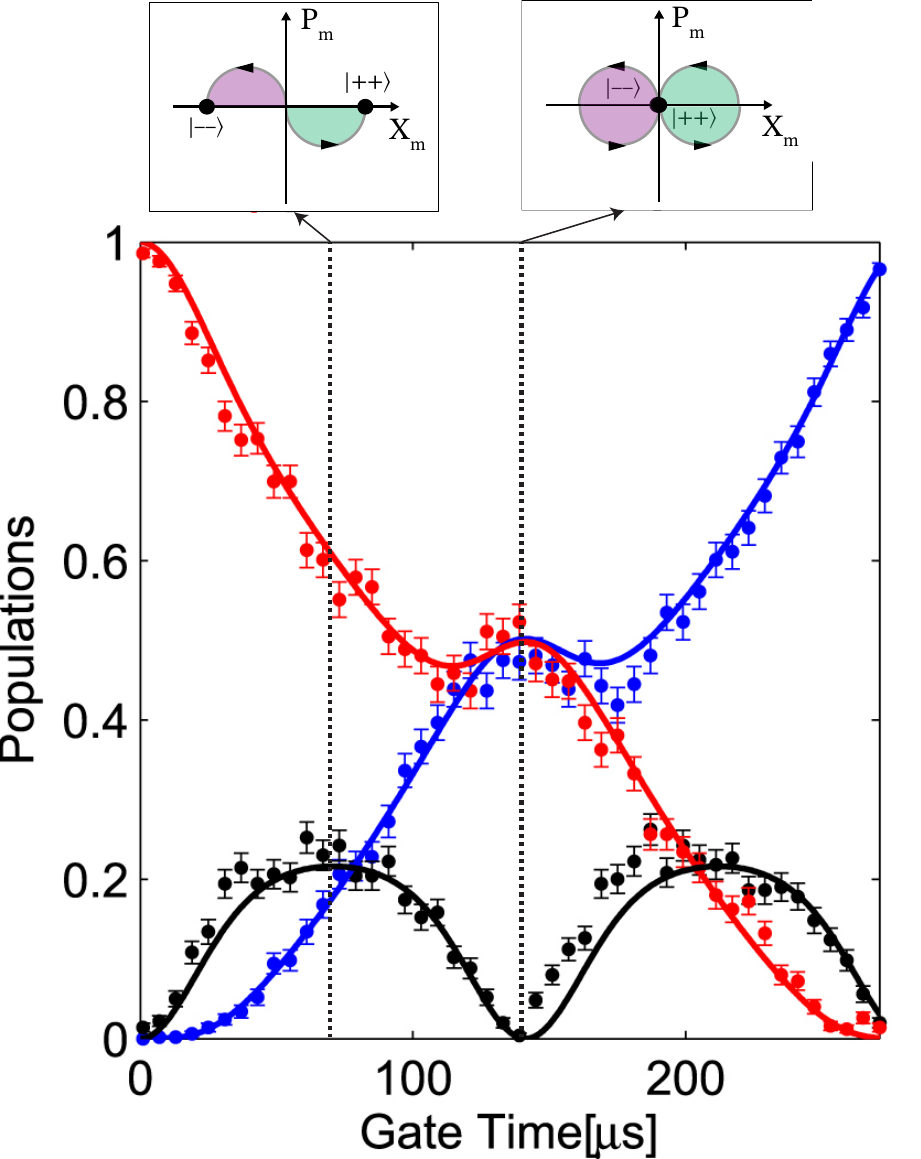}
\vspace{-0.2cm}
    \caption{Time-dependent population dynamics during the MS gate. The ions are initially prepared in the $\left|\downarrow\downarrow\right>$ state. A spin-dependent force pulse of variable gate time is applied. The populations $P_{\uparrow\uparrow}$ (blue), $P_{\uparrow\downarrow}+P_{\downarrow\uparrow}$ (red), and $P_{\downarrow\downarrow}$ (black) in the main plot are measured as a function of gate time. Insets: motional state evolution in the ($X_m$, $P_m$) space for internal states $\left|--\right>$ and $\left|++\right>$ at times marked by the vertical dashed lines. Adapted from Ref.~\cite{Akerman2015}.}
    \label{fig:msgate}
\end{wrapfigure}

As in previous sections, we begin with the interaction Hamiltonian for a chain of $N$ ions in the Lamb–Dicke regime. For the MS gate, two laser tones are applied such that they are symmetrically detuned around the carrier transition with $\pm(\omega_m+\delta_m)$.  Since small detunings $\delta_m$ are applied here (also referred to as resonant regime), it is possible to neglect the carrier and counter-rotating terms in Eqs.~\eqref{eq_BSB} and~\eqref{eq_RSB}, leading to the simplified expressions:
\begin{eqnarray}
   \hat{H}_B &=& \sum_{j=1}^N \sum_{m=1}^N \frac{i \eta_{j m} \Omega_j}{2}\left(\hat{\sigma}_j^{+} \hat{a}_m^{\dagger} e^{-i \delta_m t+i \phi_B^j}-\hat{\sigma}_j^{-} \hat{a}_m e^{i \delta_m t-i \phi_B^j}\right), \\
    \hat{H}_R &=& \sum_{j=1}^N \sum_{m=1}^N \frac{i \eta_{j m} \Omega_j}{2}\left(\hat{\sigma}_j^{+} \hat{a}_m e^{i \delta_m t+i \phi_R^j}-\hat{\sigma}_j^{-} \hat{a}_m^{\dagger} e^{-i \delta_m t-i \phi_R^j}\right).
\end{eqnarray}
Combining these two tones leads to the spin-dependent force defined below:
\begin{eqnarray}
\hat{H}_{\mathrm{SDF}}
&=& \sum_{j=1}^N \sum_{m=1}^N \frac{\eta_{j m} \Omega_j}{2} \hat{\sigma}_{\phi_s^j}\left(\hat{a}_m e^{i\left(\delta_m t-\psi_j\right)}+\hat{a}_m^{\dagger} e^{-i\left(\delta_m t-\psi_j\right)}\right).
\label{eq_SDF_simple}
\end{eqnarray}
where we can write $\hat{\sigma}_{\phi_s^j} = \hat{\sigma}_j^{+} e^{i \phi_s^j}+\hat{\sigma}_j^{-} e^{-i \phi_s^j}$. Similarly to section~\ref{sec_MS}, the evolution under the time-dependent Hamiltonian~\eqref{eq_SDF_simple} can be described through a Magnus expansion:
\begin{eqnarray}
   \hat{U}(\tau) &=& \exp \left[\underbrace{-i \int_0^\tau d t \hat{H}_{\operatorname{SDF}}(t)}_{\bar{\Omega}_1(\tau)} - \underbrace{\frac{1}{2} \int_0^\tau d t_1 \int_0^{t_1} d t_2\left[\hat{H}_{\operatorname{SDF}}\left(t_1\right), \hat{H}_{\mathrm{SDF}}\left(t_2\right)\right]}_{\bar{\Omega}_2(\tau)}\right]. \nonumber\\
\end{eqnarray}
The closed expression in the second-order term above comes from the commutator $[\hat{a}_m,\hat{a}_n^\dag]=\delta_{mn}$ being a c-number and $[\hat{\sigma}_{\phi_s^j}, \hat{\sigma}_{\phi_s^k}]=0$, which guarantees that any subsequent commutator in the Magnus expansion will also be zero.
Evaluating the commutator $[\hat{H}_{\rm SDF}(t_1), \hat{H}_{\rm SDF}(t_2)]$:
\begin{eqnarray}
    [\hat{H}_{\rm SDF}(t_1), \hat{H}_{\rm SDF}(t_2)] &=& \frac{i}{2} \sum_{j,k=1}^N \sum_{m=1}^N \eta_{j m} \eta_{k m} \Omega_j \Omega_k \nonumber \\ &\quad&\quad\quad\times \sin \left[\delta_m\left(t_1-t_2\right)-\left(\psi_j-\psi_k\right)\right] \hat{\sigma}_{\phi_s^j} \hat{\sigma}_{\phi_s^k},
\end{eqnarray}
leading to a term proportional to $\hat{\sigma}_{\phi_s^j} \hat{\sigma}_{\phi_s^k}$. Calculating both first and second-order terms gives: 
\begin{eqnarray}
\overline{\Omega}_1(\tau) &=& -i \int_0^\tau d t \hat{H}_{\operatorname{SDF}}(t) \nonumber\\
&=& \sum_{j=1}^N\sum_{m=1}^N\frac{\eta_{jm}\Omega_j}{2}  
\hat{\sigma}_{\phi_s^j}\left[ \left( -i \int_0^\tau d t e^{-i (\delta_{m} t -\psi_j)}\right)\hat{a}_m^\dagger+ \left(-i \int_0^\tau d t e^{i (\delta_{m} t -\psi_j)}\right) \hat{a}_m\right]\nonumber\\
&=& \sum_{j=1}^N\sum_{m=1}^N\frac{-\eta_{jm}\Omega_j}{2}  
\hat{\sigma}_{\phi_s^j}\left(\frac{ e^{i \psi_j}\left(1-e^{-i \delta_m \tau}\right)}{\delta_m}\hat{a}_m^\dagger  -\frac{e^{-i \psi_j}\left(1-e^{i \delta_m \tau}\right)}{\delta_m}\hat{a}_m\right)\nonumber\\
&=& \sum_{j=1}^N\sum_{m=1}^N\hat{\sigma}_{\phi_s^j}(\alpha_{jm}(\tau)\hat{a}_m^\dagger - \alpha_{jm}^*(\tau) \hat{a}_m) \nonumber\\
&=& \sum_{j=1}^N \hat{\zeta}_j(\tau) \hat{\sigma}_{\phi_s^j},
\end{eqnarray}
where $\alpha_{jm}(\tau)=-\frac{\eta_{j m} \Omega_j e^{i \psi_j}}{2 \delta_m}\left(1-e^{-i \delta_m \tau}\right)$, and $\hat{\zeta}_j(\tau)=\sum_{m=1}^N\left[\alpha_{jm}(\tau) \hat{a}_m^{\dagger}-\alpha_{jm}^*(\tau) \hat{a}_m\right]$. Similarly, evaluating the second term leads to:

\begin{eqnarray}
\overline{\Omega}_2(\tau) &=& -\frac{1}{2} \int_0^\tau d t_1 \int_0^{t_1} d t_2\left[\hat{H}_{\operatorname{SDF}}\left(t_1\right), \hat{H}_{\mathrm{SDF}}\left(t_2\right)\right]\nonumber\\
&=& -\frac{1}{2} \int_0^\tau d t_1 \int_0^{t_1} d t_2 \frac{i}{2} \sum_{j, k=1}^N \sum_{m=1}^N \eta_{j m} \eta_{k m} \Omega_j \Omega_k \nonumber \\
&\quad&\quad\times\;\sin \left[\delta_m\left(t_1-t_2\right)-\left(\psi_j-\psi_k\right)\right] \hat{\sigma}_{\phi_s^j} \hat{\sigma}_{\phi_s^k} \nonumber\\
&=& -\frac{i}{2} \sum_{j, k =1}^N \chi_{jk}(\tau)\hat{\sigma}_{\phi_s^j}\hat{\sigma}_{\phi_s^k},
\end{eqnarray}
where:
\begin{eqnarray}
\chi_{jk}(\tau) &=& \frac{\Omega_j \Omega_k}{2}\sum_m \eta_{jm} \eta_{km}  \left( \int_0^\tau dt_1 \int_0^{t_1} dt_2  \sin[\delta_m(t_1-t_2)-(\psi_j-\psi_k)] \right)\nonumber\\
&=& \frac{\Omega_j \Omega_k}{2} \sum_m \frac{\eta_{j m} \eta_{k m}}{\delta_m}\left[\tau \cos \left(\psi_j-\psi_k\right)\right.\nonumber\\
&\quad&\quad\quad\quad\left.-\;\frac{1}{\delta_m}\left(\sin \left(\delta_m \tau-\left(\psi_j-\psi_k\right)\right)+\sin \left(\psi_j-\psi_k\right)\right)\right].
\end{eqnarray}
Since $[\hat{\sigma}_{\phi_s^j}, \hat{\sigma}_{\phi_s^j}] = [\hat{\sigma}_{\phi_s^j}, \hat{\sigma}_{\phi_s^k}] = 0$, it is possible to use the Baker-Campbell-Hausdorff formula to expand the unitary operator as follows:
\begin{eqnarray}
\hat{U}(\tau) &=& \exp \left[\sum_{j=1}^N \sum_{m=1}^N\left[\alpha_{jm}(\tau) \hat{a}_m^{\dagger}-\alpha_{jm}^*(\tau) \hat{a}_m\right] \hat{\sigma}_{\phi_s^j} - \frac{i}{2} \sum_{j, k =1}^N \chi_{jk}(\tau)\hat{\sigma}_{\phi_s^j}\hat{\sigma}_{\phi_s^k}\right]\nonumber\\
&=&\left(\prod_m\hat{\mathcal{D}}_m\left[\sum_{j=1}^N \alpha_{jm}(\tau)\hat{\sigma}_{\phi_s^j}\right]\right)\exp\left[ -\frac{i}{2} \sum_{j, k =1}^N \chi_{jk}(\tau)\hat{\sigma}_{\phi_s^j}\hat{\sigma}_{\phi_s^k}\right],
\end{eqnarray}
where $\hat{\mathcal{D}}_m(\alpha) = \exp\left[ \alpha \hat{a}^\dagger_m - \alpha^*\hat{a}_m \right]$ is a displacement operator. Importantly, the time evolution of the ion system can be decomposed into the application of two unitaries. In the first unitary, the spin degree of freedom is coupled with the motional degree of freedom, resulting in a spin-dependent displacement. In the second unitary, a geometric phase $\chi_{jk}(\tau)$ is accrued, which is proportional to the area enclosed in phase space by the ions during the gate (see Fig.~\ref{fig:msgate}). To implement a high-fidelity gate, it is necessary to suppress the residual entanglement between spin and motion at the end of the gate evolution. As mentioned previously, this is achieved by choosing the gate time $\tau$ such that all displacement amplitudes $\alpha_{jm}(\tau)$ vanish, ensuring that the motional modes return to their original state, realizing a pure spin–spin unitary on the two qubits.

{ MS Gate between two ions} - As an example, let us consider a system consisting of two ions with spin phases set to zero. In this case, the time evolution operator is simplified to:

\begin{equation}
\hat{U}(\tau) =\left(\prod_m\hat{\mathcal{D}}_m\left(\alpha_{1m}(\tau)\hat{\sigma}_x^{(1)}+\alpha_{2m}(\tau)\hat{\sigma}_x^{(2)}\right)\right)\exp\left[ -i \chi_{12}(\tau)\hat{\sigma}_x^{(1)}\hat{\sigma}_x^{(2)}\right].
\end{equation}

Expressed in the $x$-basis $\{ \ket{++}, \ket{+-}, \ket{-+}, \ket{-–} \}$, the unitary becomes:
\begin{eqnarray}
U_{i j}(\tau)&=& e^{-i \chi_{1 2}(\tau)}\left|++\right\rangle\left\langle++\right| \prod_m \hat{\mathcal{D}}_m\left[\alpha_{1 m}(\tau)+\alpha_{2 m}(\tau)\right] \nonumber\\
&\quad&+\; e^{-i \chi_{1 2}(\tau)}\left|--\right\rangle\left\langle--\right| \prod_m \hat{\mathcal{D}}_m\left[-\alpha_{1 m}(\tau)-\alpha_{2 m}(\tau)\right] \nonumber\\
&\quad&+\; e^{i \chi_{1 2}(\tau)}\left|+-\right\rangle\left\langle+-\right| \prod_m \hat{\mathcal{D}}_m\left[\alpha_{1 m}(\tau)-\alpha_{2 m}(\tau)\right] \\
&\quad&+\; e^{i \chi_{1 2}(\tau)}\left|-+\right\rangle\left\langle-+\right| \prod_m \hat{\mathcal{D}}_m\left[-\alpha_{1 m}(\tau)+\alpha_{2 m}(\tau)\right].\nonumber
\end{eqnarray}
To decouple the spin and motional degrees of freedom (i.e., to close the motional loops), it is necessary to choose the gate evolution such that $\delta_m \tau = 0\mod 2\pi$, ensuring that all the displacement operators vanish (see Fig.~\ref{fig:msgate}). Furthermore, by tuning the Rabi frequencies, we can engineer the accumulated phases $\chi_{ij}(\tau)$ to be $\pi/4$. 
Under these conditions, the MS gate takes the following matrix form in the $x$-basis $\{ \ket{++}, \ket{+-}, \ket{-+}, \ket{--} \}$:
\begin{equation}
\hat{U}_{\mathrm{MS}}=e^{i\pi/4}\left(\begin{array}{cccc}
-i & 0 & 0 & 0 \\
0 & 1 & 0 & 0 \\
0 & 0 & 1 & 0 \\
0 & 0 & 0 & -i
\end{array}\right).
\end{equation}
Expressed in the $z$-basis $\{ \ket{\uparrow\uparrow}$, $ \ket{\uparrow\downarrow}$, $ \ket{\downarrow\uparrow}$, $ \ket{\downarrow\downarrow} \}$, the unitary becomes:
\begin{equation}
\hat{U}_{\mathrm{MS}}=\frac{1}{\sqrt{2}} \left(\begin{array}{cccc}
1 & 0 & 0 & -i \\
0 & 1 & -i & 0 \\
0 & -i & 1 & 0 \\
-i & 0 & 0 & 1
\end{array}\right).
\end{equation}
This gate produces a maximally entangling operation when acting on separable input states.

{\it Optimal Control} -  In practice, the fidelity of quantum gates is limited by various sources of error, including motional dephasing, motional heating, spin dephasing, and laser intensity noise~\cite{Ballance2016}. Depending on the specific characteristics of the noise, it can be advantageous to shape the phase and intensity of the laser control pulses to break the symmetries of how noise influences the results of the gate~\cite{Debnath2016, Shapira2018}. It is important to note, however, that mitigating one type of error often worsens others.

A major challenge in implementing high-fidelity MS gates lies in minimizing the residual entanglement between the motional and spin degrees of freedom at the end of the gate operation. 
This requires closing all the motional-state ``loops" with the gate time $\tau_g$, such that $\Sigma_{j=1}^N \alpha_{jm}(\tau_g)=0$, while simultaneously acquiring the desired geometric phase. This becomes increasingly difficult in long chains of ions, where the motional modes are closely spaced. One way to achieve closing all the motional-state ``loops" is by setting a very small detuning $\delta_0$ from a single motional mode and using $\tau_{g}=2\pi/\delta_0$, which gives a long gate time for small $\delta_0$. An alternative strategy involves operating in a regime where multiple motional modes are coupled to the qubits simultaneously. This leads to shorter gate times at the expense of increasingly complex dynamics with more parameters to calibrate. For instance, to implement a two-qubit $XX$ gate, all motional trajectories must be closed (resulting in $2N$ constraints), and the correct phase accumulation between the two ions must be ensured (leading to an additional constraint), yielding a total of $2N+1$ constraints. To meet these conditions, Ref.~\cite{Choi2014Optimal} proposed dividing a constant pulse into $2N+1$ segments, each with tunable motional phase and/or Rabi frequency. This reduces the problem into a system of linear equations with a guaranteed solution. In this multi-mode regime, the carrier term in the spin-dependent force also plays a key role. It is possible to suppress the influence of this carrier term by structuring the light field. For instance, by placing an ion in the anti-node of a standing wave on the quadrupole transition, the carrier component is suppressed~\cite{Saner2023, Ricci2023}, and the sidebands are still efficiently driven. This can subsequently reduce gate times. Additionally, pulse shaping can also be used to design pulses that are robust against motional frequency fluctuations~\cite{Leung2018Robust}, enable simultaneous entangling gates across multiple qubit pairs~\cite{Figgatt2019Parallel}, and reduce gate durations below 1 $\mu$s~\cite{Schaefer2018Fast}.

\subsection{Light Shift Gate} 
As outlined above, the phase-insensitive scheme features several advantages, such as faster gate implementation and robustness to qubit dephasing error. For these reasons, several groups use light shift gates ($\propto \hat{\sigma}_z\hat{\sigma}_z$) generated by the spin-dependent force in the $z$-basis for both quantum simulations~\cite{Britton2012, Guo2024} and quantum gates~\cite{Leibfried2003, Schaefer2018Fast, Baldwin2021}. In this case, a pair of Raman laser beams is used with a frequency difference such that there is no direct coupling between the qubit states ($\ket{\!\uparrow}$ and $\ket{\!\downarrow}$). Instead, the laser drive independently imprints a state-dependent AC Stark shift on the qubit states by coupling to an intermediate state ($\equiv\ket{e}$). In addition, the two Raman beams are configured with wavevectors $k_L$ pointing along the same principal trap axis (say, $x$), frequencies $\mu_L$ from the excited state $\ket{e}$, and optical phases $\phi_L$, where $L=B,\;R$ are the labels for the Raman beams with higher and lower frequencies, respectively. Suppose $\Delta k \hat{x} = k_B\hat{x}-k_R\hat{x}$, $\Delta\phi=\phi_B-\phi_R$, and $\mu_B-\mu_R = \omega + \delta$ for $\omega$ being the center-of-mass secular frequencies along $x$, the interaction Hamiltonian of a single ion is ($\hbar=1$)
\begin{equation}
\hat{H}_I=
\sum_{s=\uparrow,\downarrow}\sum_{L=B,R}\frac{\Omega_L^s}{2}(\ket{e}\bra{s} + \ket{s}\bra{e}) \left[ e^{i( k_L \hat{x} - (\omega_s+\mu_L) t + \phi_L)} + e^{-i (k_L \hat{x} - (\omega_s+\mu_L) t + \phi_L)}  \right],
\end{equation}
where $\Omega_{L}^s$ is the Rabi coupling strength between states $\ket{s}$ and $\ket{e}$ from Raman beam $L$, and their frequency difference is $\omega_s\gg\mu_L$, similar to Eq.~\eqref{eq:originInt}. By moving into the interaction representation and applying RWAs, the Hamiltonian becomes
\begin{eqnarray}
\hat{H}_I^{\rm (ord)}
&=& \sum_{s=\uparrow,\downarrow}\sum_{L=B,R}\frac{\Omega_L^s}{2}\ket{e}\bra{s}e^{i( k_L \hat{x}(t) - \mu_L t + \phi_L)} + \; \text{h.c.},
\end{eqnarray}

It has been shown in Ref.~\cite{James2007} that the rapidly oscillating Hamiltonian of the form $\hat{H}_I(t)=\sum_n\hat{h}_n\exp{(-i\omega_n t)}+ \hat{h}_n^\dagger\exp{(i\omega_n t)}$ can be transformed into the time-averaged effective Hamiltonian $\hat{H}_{\rm eff}(t)=\sum_{n,m}\frac{1}{2}\left(\frac{1}{\omega_m}+\frac{1}{\omega_n}\right)\left[\hat{h}_m^\dagger,\hat{h}_n\right]$ $\exp{(i[\omega_m-\omega_n]t)}$, which is referred as \textit{James-Jerke approximation}. By applying this approximation and neglecting the terms associated with the constant light shifts and energy shift on $\ket{e}$~\cite{Schaefer2018Fast,schafer2018thesis}, we obtain the effective Hamiltonian as follows:
\begin{eqnarray}
\hat{H}_{\rm eff}
&=& \sum_{s=\uparrow,\downarrow}\frac{\Omega_B^s\Omega_R^s}{4\Delta}\ket{s}\bra{s}\left(e^{i( \Delta k \hat{x}(t) - (\omega+\delta) t + \Delta\phi)} + \; \text{h.c.}\right),
\end{eqnarray}
where $\Delta$ is the harmonic average of $\mu_B$ and $\mu_R$ with $\frac{1}{\Delta}=\frac{1}{2}\left(\frac{1}{\mu_B}+\frac{1}{\mu_R}\right)$. We can rewrite the effective Hamiltonian as
\begin{eqnarray}
\hat{H}_{\rm eff}
&=& \frac{\Omega_{\rm eff}}{2}\hat{\sigma}_z\left(e^{i( \Delta k \hat{x}(t) - (\omega+\delta) t + \Delta\phi)} + \; \text{h.c.}\right),
\end{eqnarray}
where we defined the effective Rabi frequency $\Omega_{\rm eff}\equiv\frac{\Omega_B^\uparrow\Omega_R^\uparrow-\Omega_B^\downarrow\Omega_R^\downarrow}{4\Delta}$ and ignored the common energy shift terms. As in Sec.~\ref{sec_MSgate}, we then apply the Lamb-Dicke approximation while neglecting the contributions from the micromotion and obtain:
\begin{eqnarray}
\hat{H}_{\rm eff}&=&\frac{\Omega_{\rm eff}}{2}\hat{\sigma}_z\left(e^{-i ((\omega+\delta) t - \Delta\phi)} + e^{i ((\omega+\delta) t - \Delta\phi)} \right)    \nonumber \\
&\quad& + \frac{i\eta\Omega_{\rm eff}}{2} \hat{\sigma}_z\left(e^{-i ((\omega+\delta) t - \Delta\phi)} - e^{i ((\omega+\delta) t - \Delta\phi)} \right)  \left[ \hat{a}^\dagger e^{i\omega t} + \hat{a} e^{-i\omega t} \right].
\end{eqnarray}
The first term describes a light shift on the qubit states, which can be ignored for $\omega+\delta\gg\Omega_{\rm eff}$. This is analogous to the off-resonant carrier term in the MS case. In addition, since it is commuting with the spin-dependent force, it can effectively be echoed out. By expanding the second term and dropping the fast-rotating terms at $\omega$, the Hamiltonian is simplified to
\begin{eqnarray}
\hat{H}_{\rm eff}&=&\frac{i\eta\Omega_{\rm eff}}{2} \hat{\sigma}_z\left(\hat{a}^\dagger e^{-i (\delta t - \Delta\phi)} - \hat{a}e^{i (\delta t - \Delta\phi)} \right) \nonumber \\
&=&\frac{\eta\Omega_{\rm eff}}{2} \hat{\sigma}_z\left(\hat{a}e^{i (\delta t - \Delta\phi - \pi/2)} + \hat{a}^\dagger e^{-i (\delta t - \Delta\phi - \pi/2)} \right) \\
&=& \frac{\eta\Omega_{\rm eff}}{2} \hat{\sigma}_z\left(\hat{a}e^{i (\delta t - \psi)} + \hat{a}^\dagger e^{-i (\delta t - \psi)} \right), \nonumber
\end{eqnarray}
which describes the spin-dependent force in phase space with the motional phase $\psi=\Delta\phi+\pi/2$, similar to the one derived in Sec.~\ref{sec_MSgate} but in the $z$-basis instead of on the $x$-$y$ plane of the Bloch sphere. Analogously to the MS case, the time-evolution operator of this interaction gives rise to the effective Hamiltonian with $\hat{\sigma}_z^j\hat{\sigma}_z^k$ used for the light shift gate.


\subsection{Spin-Boson Model}
The Hamiltonian of a trapped-ion system with $N$ qubits interacting with light fields, $\hat{H}=\hat{H}_{A} +\hat{H}_M+\hat{H}_I$, can also be mapped to a spin-boson model by moving to a different resonant frame via the following transformation~\cite{Schneider2012}:
\begin{eqnarray}
\hat{H}^{\rm (res)} &=& \underbrace{i\left(\frac{\partial }{\partial t}\hat{U}^\dagger(t)\right)\hat{U}(t)}_{-(\hat{H}_A+ \hat{H}_M+\hat{H}_{\delta})} \; + \underbrace{\hat{U}^\dagger(t) \hat{H} \hat{U}(t)}_{\substack{(\hat{H}_A+ \hat{H}_M+\hat{H}_{\delta})\\+(\hat{U}^\dagger(t) \hat{H}_I \hat{U}(t)-\hat{H}_{\delta})}},
\end{eqnarray}
where $\hat{U}(t)=e^{-i (\hat{H}_A+ \hat{H}_M+\hat{H}_{\delta}) t}$ with $\hat{H}_{\delta}\equiv\sum_m \delta_m \hat{a}_m^\dag \hat{a}_m$, and $\hat{H}$ can be rewritten as $\hat{H}= (\hat{H}_{A} +\hat{H}_M+\hat{H}_{\delta})
+(\hat{H}_{I}-\hat{H}_{\delta})$. The time-independent term $-\hat{H}_\delta$ corresponds to the bosonic Hamiltonian of the spin-boson model.

Following the same derivations in Secs.~\ref{sec_singleion} and~\ref{sec_MSgate}, the interaction term of the resonant Hamiltonian becomes
\begin{eqnarray}
\label{eq:}
\hat{H}_I^{\rm (res)} &=& \hat{U}^\dagger(t) \hat{H}_I \hat{U}(t)\nonumber\\
&=&\sum_{j=1}^N\frac{\Omega_{j}}{2}\left( \hat{\sigma}_j^+\, e^{-i (\mu_L^j t - \phi_j)} + \hat{\sigma}_j^-\, e^{i (\mu_L^j t - \phi_j)} \right) \\
&+&  \sum_{j=1}^N\sum_{m=1}^N\frac{i\eta_{jm}\Omega_j}{2} 
\left( \hat{\sigma}_j^+\, e^{-i (\mu_L^j t - \phi_j)} - \hat{\sigma}_j^-\, e^{i (\mu_L^j t - \phi_j)} \right) \nonumber \\ &\times&\left[ \hat{a}_m^\dagger e^{i(\omega_m+\delta_m) t} + \hat{a}_m e^{-i(\omega_m+\delta_m) t} \right]. \nonumber
\end{eqnarray}
To induce the spin-dependent force that maps to the spin-boson coupling term of the spin-boson model, we simultaneously apply two symmetric sideband beams with ${\color{black}\mu_B^j=\mu}$ and optical phase \textcolor{black}{$\phi_B^j$} and ${\color{black}\mu_R^j=-\mu}$ and optical phase \textcolor{black}{$\phi_R^j$} to each ion $j$, as for the generation of the MS gate in Sec.~\ref{sec_MSgate}. The resultant interaction Hamiltonian is described by:
\begin{eqnarray}
\label{eq:spin-boson_SDK}
\hat{H}_{\rm SDF}^{\rm (res)}&=& \sum_{j=1}^N\Omega_j \hat{\sigma}_{\phi_s^j-\pi/2}^j \cos(\mu t-\psi_j)+\sum_{j=1}^N \sum_{m=1}^N\frac{\eta_{jm}\Omega_j}{2} \hat{\sigma}^j_{\phi_s^j} \left( \hat{a}_m \, e^{-i\psi_j} + \hat{a}_m^\dagger\, e^{i\psi_j} \right) \nonumber\\
&\quad&+
\sum_{j=1}^N\sum_{m=1}^N\frac{\eta_{jm}\Omega_j}{2}  
\hat{\sigma}_{\phi_s^j}\left( \hat{a}_m \, e^{-i (2\mu t -\psi_j)}  + \hat{a}_m^\dagger \, e^{i (2\mu t -\psi_j)} \right),\label{eq_HSDF_res}
\end{eqnarray}
where the off-resonant carrier drive in the first term can be neglected for $\Omega_j\ll\mu$. In this case, differently from the spin-spin MS interaction case analyzed in Section~\ref{sec_MS}, we impose the conditions $\mu+\omega_m\gg|\mu-\omega_m|=\delta_m$ and $\delta_m \sim \eta_{jm}\Omega_j$ such that applying the RWA to the third term in Eq.~\eqref{eq_HSDF_res} that rotates at $\sim2\mu$ is also justified. This makes the time-independent term the only dominant contribution to the Hamiltonian. The final expression of the total Hamiltonian is then given by:
\begin{eqnarray}
\label{eq:spin-boson_interaction}
\hat{H}^{\rm (res)}&=& \hat{H}_{\rm SDF}^{\rm (res)} -\hat{H}_\delta \nonumber \\
&=&\sum_{j=1}^N \sum_{m=1}^N\frac{\eta_{jm}\Omega_j}{2} \hat{\sigma}^j_{\phi_s^j} \left( \hat{a}_m \, e^{-i\psi_j} + \hat{a}_m^\dagger\, e^{i\psi_j} \right) - \sum_m \delta_m \hat{a}_m^\dag \hat{a}_m,
\end{eqnarray}
which corresponds to the desired spin-boson model.

In practice, the condition for the RWA does not hold for all motional mode $m$ with a particular choice of $\mu$. Therefore, it is beneficial to only transform the trapped-ion Hamiltonian into the resonant frame with respect to the motional modes $q$ that satisfy the RWA criteria; particularly, $\hat{H}_\delta = \sum_q \delta_q \hat{a}_q^\dagger \hat{a}_q$. In this frame, Eq.~\eqref{eq:spin-boson_interaction} is modified to
\begin{eqnarray}
\label{eq:spin-boson_SDK_full}
\hat{H}^{\rm (res)} &=& \sum_{j=1}^N \sum_{q}\frac{\eta_{jq}\Omega_j}{2} \hat{\sigma}^j_{\phi_s^j} \left( \hat{a}_q \, e^{-i\psi_j} + \hat{a}_q^\dagger\, e^{i\psi_j} \right) - \sum_q \delta_q \hat{a}_q^\dag \hat{a}_q \nonumber\\
&\quad& + \underbrace{\sum_{j=1}^N \sum_{m\neq q}\eta_{jm}\Omega_j \cos(\mu t-\psi_j) \hat{\sigma}^j_{\phi_s^j} \left( \hat{a}_m \, e^{-i\omega_m t} + \hat{a}_m^\dagger\, e^{i\omega_m t} \right)}_{\hat{H}_{jk}^{\rm cr}},
\end{eqnarray}
where the second term resembles the off-resonant spin-dependent force in the ordinary frame. Since $\delta_m \gg \eta_{jm}\Omega_j, \forall m \neq q$ in this frame, it effectively gives rise to unwanted two-body spin coupling from the Magnus expansion, $\hat{H}^{\rm cr}_{jk}=\sum_{j<k} J^{\rm cr}_{jk} \hat{\sigma}_{\phi_s^j}^j \hat{\sigma}_{\phi_s^k}^k$, with the coupling strength:
\begin{equation}
    J^\text{cr}_{jk}=\Omega_j \Omega_k \omega_{\rm rec} \sum_{m\neq q} \frac{b_{jm} b_{km}}{\mu^2 - \omega_m^2}\cos(\psi_{j}-\psi_{k}),
    \label{eq_Jcr}
\end{equation}
that can lead to errors in the dynamics of the spin-boson system of interest. Ideally, the contribution of this term may be reduced by an optimal choice of $\mu$, phase, and amplitude optimization, or a counteracting MS interaction mediated by another set of normal modes.



\section{Trapped-Ion Quantum Simulation: a Focus on Chemical Dynamics}\label{sec_progress}

This section reviews the latest developments in trapped-ion quantum simulation using the tools explained in Section~\ref{sec_laser-ion_interactions}. Most of the quantum simulations with trapped ions performed in the last 15 years are pertinent to studying the many-body physics of spin models, such as the Ising~\cite{Kim_2009,kim10}, the XY~\cite{Jurcevic2014}, and the Heisenberg~\cite{Kranzl2023observation} models. In this context, the flexibility of trapped-ion simulators enabled the simulation of magnetic ordering of adiabatically loaded equilibrium ground states~\cite{Islam2011, Islam2013, Richerme2013}, spectroscopy of low-lying excited states~\cite{Senko2014}, and a plethora of non-equilibrium dynamical phenomena, such as the violation of Lieb-Robinson bounds in long-range interacting systems~\cite{Richerme2014}, dynamical phase transitions of quantum magnetism~\cite{Zhang2017a, Jurcevic2017}, time crystals~\cite{Zhang2017, Kyprianidis2021}, quantum thermalization~\cite{Neyenhuis2017, Kranzl2023experimental}, and lack thereof~\cite{Smith2016, Hess20170107, Morong2021}, hydrodynamics~\cite{joshi2022observing} and spin-1 models~\cite{Senko2015, edmunds2024constructingspin1haldanephase}. 

Beyond simulating spin models pertinent to condensed matter physics, the tunable long-range interactions in trapped-ion systems also allow for the simulation of lattice-discretized, strongly coupled quantum field theories relevant to high-energy particle and nuclear physics~\cite{preskill2018simulating, Banuls2020, Bauer2023, Bauer2023Quantum}. A large variety of phenomena and models have been simulated, including the Dirac equation in 1+1 dimensions~\cite{Gerritsma2010dirac}, confinement of meson-like quasiparticles~\cite{tan2021domain,de2024observationstringbreakingdynamicsquantum}, scattering amplitudes~\cite{Gustafson2021, davoudi2024scattering}, baryon $\beta$-decay~\cite{Farrell2023}, and dense neutrino gases dynamics~\cite{Pehlivan2011, Amitrano2023}. There has also been significant progress in the simulation of real-time dynamics in lattice gauge theories such as the 1+1D quantum electrodynamics (QED), investigating both the non-equilibrium dynamics~\cite{Muschik_2017, Nguyen2022digital, Mueller2023quantum} and the variational ground states~\cite{Kokail2019}. Recently, the use of qudits has further enabled the simulation of 2D+1 QED~\cite{Meth2025}. 

For more comprehensive reviews on trapped-ion simulation of spin physics and of models related to nuclear and high-energy physics, we refer to the reviews~\cite{monroe2021programmable, defenu2023} and~\cite{Foss-Feig2024}, respectively, and to the references therein. Conversely, here, we will focus exclusively on a new emerging research direction: the trapped-ion analog simulation of chemical dynamics using programmable open quantum systems.

\subsection{Analog Simulation of Chemical Dynamics in Open Quantum Systems}

In recent years, the use of digital quantum computation for chemistry problems on trapped-ion platforms has shown success in studying the characteristics of molecular systems, such as their ground-state energy surfaces~\cite{shen2017electronicstructure, Hempel2018chemistry}, molecular structures~\cite{Seetharam2023nmr}, and coherent dynamics~\cite{Richerme2023hydrogen}. At the same time, analog simulation and analog-digital approaches have also made significant progress in demonstrating various chemical dynamics, including the observations of vibrationally assisted energy transport (VAET)~\cite{Gorman2018}, geometric-phase interference around a conical intersection~\cite{Whitlow2023, Valahu2023}, molecular vibronic spectra~\cite{MacDonell2023timedomain}, and polarized light-induced electron transfer (PLET)~\cite{Ke2023}. These studies have been enabled by the natural correspondence between the native spin and bosonic degrees of freedom and their programmable interactions in trapped-ion systems with quantum chemistry models of excitation-energy transfer and vibronic (electronic-vibrational) coupling~\cite{MacDonell2021quantumchemistry}. Upon scaling the system size, state-of-the-art trapped-ion simulators may lead to a possible quantum advantage over classical-digital algorithms in certain coupling regimes, even with the current technical limitations~\cite{Kang2024}. Another important feature of trapped-ion analog quantum simulators is the possibility to directly engineer reservoirs that account for the effect of the environment on the chemical reactions or systems~\cite{Lemmer2018structure,schlawin2021continuously,olayaagudelo2024simulatingopensystemmoleculardynamics, wang2024spinboson}, which has been shown to be crucial in models of energy transfer in biomolecules~\cite{Fassioli2014} and organic semiconductors~\cite{Sneyd2022}. Therefore, in this review, we focus exclusively on results related to the analog quantum simulation of open quantum system dynamics relevant to chemical processes using trapped-ion technologies. 

\begin{figure}[t!]
    \centering
    \includegraphics[width=\columnwidth]{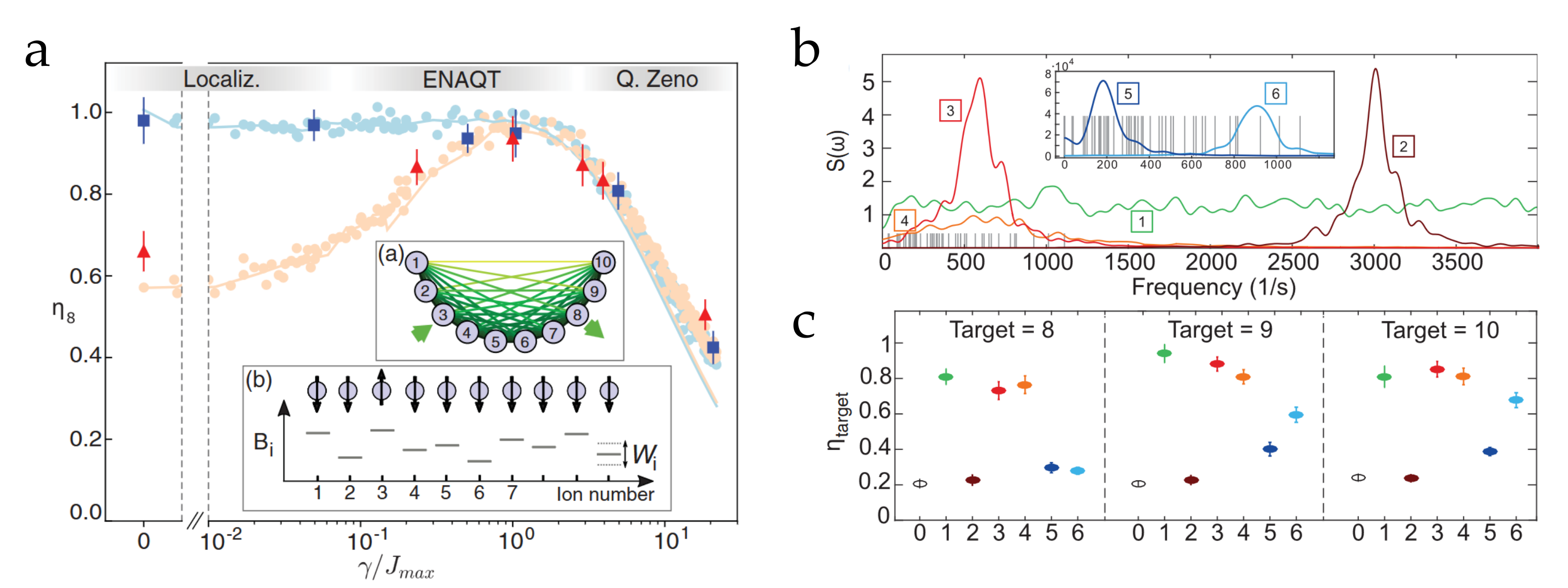}
    \caption{(a) Transport efficiency $\eta_8$ to target ion $i=8$ under different static disorder strengths (blue: $B_{\rm max} = 0.5J_{\rm max}$; red: $B_{\rm max} = 2.5J_{\rm max}$) as a function of the Markovian-like dephasing rate $\gamma $. (b) Frequency differences between the eigenenergies of the disordered system (vertical black lines) and spectral density functions for different noise models (identified by number and color). (c) Corresponding transport efficiency to the noise profile (matched by number and color). Adapted from Ref.~\cite{Maier2019ENAQT}.}
    \label{fig:enaqt}
\end{figure}

One of the first trapped-ion studies of the effect of environmental noise on energy transport was performed using laser-ion interactions on a chain of 10 ions at the University of Innsbruck~\cite{Maier2019ENAQT}. The system was described by long-range hopping $J_{ij}\approx J_{\rm max}/|i-j|^\alpha$ among the excitation sites (each being represented by the spin state of the individual ion) and local and time-varying magnetic fields, namely:
\begin{equation}
    \begin{split}
    \hat{H} = \sum_{i\neq j}J_{ij}\left(\hat{\sigma}_i^+\hat{\sigma}_j^- + \text{h.c.}\right)+\sum_i[B_i+W_i(t)]\hat{\sigma}_i^z.
    \end{split}
    \label{eq_Henaqt}
\end{equation}
Here, the on-site energies $B_i$, which were randomly selected from a uniform distribution ranging from $-B_{\rm max}$ and $B_{\rm max}$, create static disorder in the system, while the temporally modulated on-site energies $W_i(t)$, were engineered with mean value $\langle\langle W_i(t)=0 \rangle\rangle$ and tunable spectral power $S(\omega)=\lim\limits_{T\rightarrow \infty}\frac{1}{T}\int_0^T\int_0^T\langle\langle W_i(t)W_i(t')\rangle\rangle$ $e^{i\omega(t-t')}dt'dt$, where $\langle\langle \cdot \rangle\rangle$ represents averaging over noise trajectories. This setup allowed for the realization of environment-assisted quantum transport (ENAQT) through dynamical dephasing noise under different configurations of static disorder, where the time-varying terms in the Hamiltonian induce dephasing on the excitation states.
 
The simulation evolved the single-excitation initial state $\ket{\downarrow\downarrow\uparrow\downarrow\downarrow\downarrow\downarrow\downarrow\downarrow\downarrow}$ with the Hamiltonian in Eq.~\eqref{eq_Henaqt} (see Inset (b) of Fig.~\ref{fig:enaqt}a), which conserves the number of excitations throughout the evolution of the system. The authors measured the dynamics of the probability of finding the excitation at a particular site, $p_{i}(t) = \left(\langle\hat{\sigma}_{i}^z(t)\rangle+1\right)/2$, which allows for the extraction of its transport efficiency, defined by $\eta_{i}\equiv\int p_{i}(t)dt$. First, they investigated the regime of Markovian (white) dephasing noise, described by a constant $S(\omega)\approx W_{\rm max}^2/\lambda$, realized by varying $W_i(t)$ between two values, $-W_{\rm max}/2$ and $W_{\rm max}/2$, with equal probabilities at the rate of $\lambda$ throughout the simulation time span. This method works as long as the noise sampling rate $\lambda$ is much larger than the maximum spin coupling $J_{\rm max}$, resulting in an effective dephasing rate $\gamma = W_{\rm max}^2/\lambda$. As shown in Fig.~\ref{fig:enaqt}a, optimal Markovian noise ($\gamma\approx J_{\rm max}$) can increase transport in the case of strong disorder ($B_{\rm max}>J_{\rm max}$) by lifting the destructive interference associated with Anderson localization, which otherwise suppresses transport under weak noise ($\gamma < J_{\rm max}$). When dephasing dominates the dynamics ($\gamma>J_{\rm max}$), a quantum Zeno effect is observed, with reduced transport efficiency for both weak ($B_{\rm max}<J_{\rm max}$) and strong disorders.

Applying non-Markovian dephasing noise on the dynamics of the strongly disordered system $(B_{\rm max}>J_{\rm max})$ led to different results. The non-Markovian noise was realized by applying $W_i(t)$ corresponding to a Lorentzian-shaped $S(\omega)$, which could be controlled to spectrally overlap with the frequency differences among the eigenenergies of the disordered system (see Fig.~\ref{fig:enaqt}b). This overlap was the crucial feature in determining the levels of local transport enhancement, which allowed for increased efficiency on a selected site $i$, as shown in Fig.~\ref{fig:enaqt}c. Interestingly, the narrow-band noise structure of non-Markovian dephasing maintains better coherences than its white noise counterpart despite enabling similar transport enhancements at half the energy cost.


This experiment displays the capabilities of a trapped-ion simulator in precisely mimicking the Hamiltonian of a complex system and engineering a structured noise profile of the environment over a wide range of parameters. This approach enables the study of time-resolved dynamics in regimes that individuate the independent roles of coherences and decoherences in environment-critical processes, such as ENAQT.

Inspired by recent theoretical works~\cite{Lemmer2018structure,schlawin2021continuously}, the electron transfer (ET) process between two molecular sites coupled to an environmentally dissipative vibrational mode was recently simulated using a $^{171}$Yb$^+$--$^{172}$Yb$^+$ ion crystal at Rice University~\cite{so2024ETsim}. The donor and acceptor electronic sites are encoded in the ground-state $^{171}$Yb$^+$ qubit, while the vibrational mode of the system, which defines the reaction coordinate, is encoded in a shared motional mode of the dual-species crystal. For the analog simulation, the Hamiltonian that governs the unitary evolution of the vibronic system is composed of terms that are directly mapped to the native light-ion interactions, described in Sec.~\ref{sec_laser-ion_interactions}:
\begin{equation}
    \begin{split}
    \hat{H} = V_x\hat{\sigma}_x + \frac{\Delta E}{2}\hat{\sigma}_z + \frac{g}{2}\hat{\sigma}_z\left(\hat{a}+\hat{a}^\dag\right)+\omega \hat{a}^\dag \hat{a},
    \end{split}
    \label{eq:HET}
\end{equation}
where $V_x$ and $g$ determine the electronic and vibronic coupling strengths, respectively. Similarly, $\Delta E$ represents the donor-acceptor energy difference, while $\omega$ is the vibrational energy of the system. Moreover, the reorganization energy $\lambda\equiv g^2/\omega$ defines the energy required to relocate the vibronic wave packet from the donor site to the acceptor site without changing its electronic configuration.

By simultaneously cooling the collective bosonic mode via the narrow-linewidth optical transition of the $^{172}$Yb$^+$ coolant ion, the vibronic system evolves according to the Lindblad's master equation:
\begin{eqnarray}
    &&\frac{\partial\hat{\rho}}{\partial t}=-i[\hat{H},\hat{\rho}] + \gamma (\bar{n}+1)\hat{\mathcal{L}}_{\hat{a}}[\hat{\rho}] + \gamma \bar{n} \hat{\mathcal{L}}_{\hat{a}^\dagger}[\hat{\rho}],\nonumber\\
    &&\hat{\mathcal{L}}_{\hat{c}}[\hat{\rho}]=
    \hat{c}\hat{\rho} \hat{c}^\dagger - \frac{1}{2}\{\hat{c}^\dagger \hat{c},\hat{\rho}\}, 
\end{eqnarray}
where $\gamma$ is the rate at which the vibrational energy is removed from the system, and $\bar{n}\sim0.1-0.3$ is the average thermal phonon number at frequency $\omega$ that describes the bath temperature, $k_B T \approx \omega/\log(1+1/\bar{n})$. When $\gamma \ll \omega$ and $\gamma\ll k_B T$, the dynamics of the system under the master equation is equivalent to that of the donor-acceptor vibronic system being linearly coupled to a continuous harmonic-oscillator environment with an Ohmic spectral density ($J(\omega)\sim\omega$), a popular choice for molecule-environment interactions in ET models.

\begin{figure}[t!]
    \centering
    \includegraphics[width=\columnwidth]{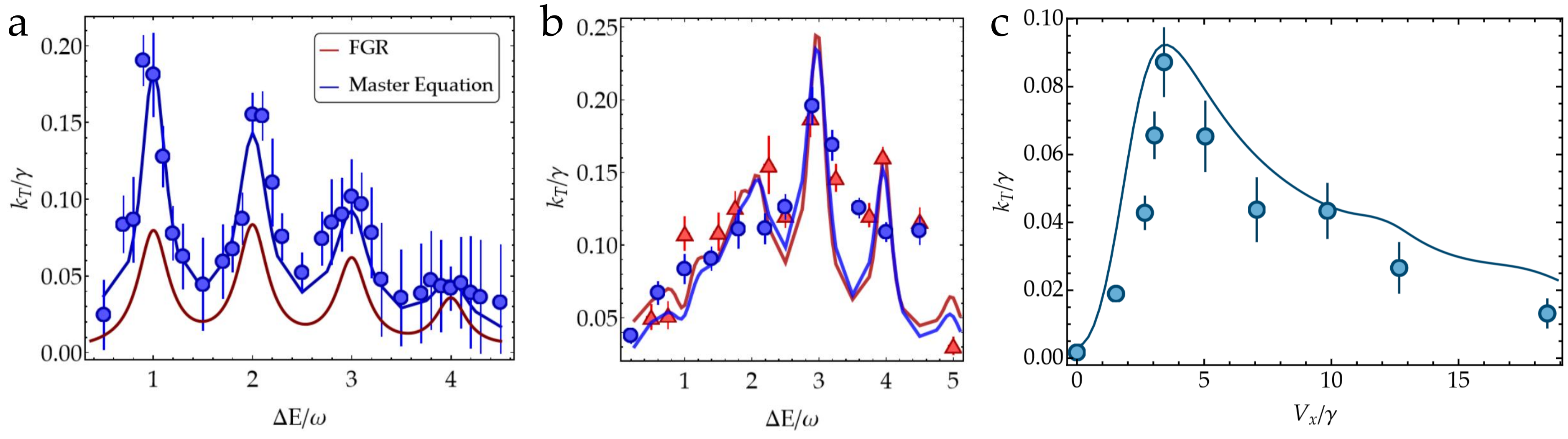}
    \caption{Transfer rate $k_T$ in units of the relaxation rate $\gamma$ as a function of the donor-acceptor energy gap $\Delta E$ in the (a) nonadiabatic and (b) adiabatic regimes and as a function of (c) the electronic coupling strength $V_x/\gamma$ showing an optimal transfer regime. Adapted from Ref.~\cite{so2024ETsim}.}
    \label{fig:et}
\end{figure}

In this experiment, the time-resolved donor population, $P_D(t)=(\langle\hat{\sigma}_z(t)\rangle + 1)/2$, was measured, from which the transfer rate of the ET process was extracted according to $k_T=\int P_D(t)dt/\int t P_D(t)dt$. The initial motional wave packet of the system was centered on the uncoupled donor energy surface, initialized with a temperature similar to the bath.
From the transfer-rate spectra with respect to the energy difference $\Delta E$, unique characteristics of different ET regimes of adiabaticity were identified in Figs.~\ref{fig:et}a and b. In the nonadiabatic regime ($V_x\lesssim\gamma, V_x<\lambda/4$), the transfer-rate spectrum resembles the vibrational-mode structure as resonances are observed at $\Delta E = l\omega$ with $l$ being a non-zero integer, and the dissipation rate $\gamma$ determines their widths. This can be explained by the large overlaps between the nonadiabatic eigenstates and that of the uncoupled ($V_x=0$) donor-acceptor system: $\ket{\uparrow}_z\otimes\hat{\mathcal{D}}(- g/2\omega)\ket{n_D}$ and $\ket{\downarrow}_z\otimes\hat{\mathcal{D}}(+ g/2\omega)\ket{n_A}$, where the motional states are the displaced Fock states associated with the donor and acceptor population.

On the other hand, the electronic coupling can mix the donor-acceptor sites, forming the upper and lower hybridized states in the strongly adiabatic regime ($V_x>\gamma,V_x\sim\lambda/4$). Hence, the transfer-rate characteristics associated with this regime feature proportionality to the dissipation rate $\gamma$ and the absence of the vibrational-mode structure at low $\Delta E$ and the recovery of the distinct resonant peaks from the release of the initially trapped population in the upper delocalized energy surface at high $\Delta E$ despite the general decrease in the transfer rates with increasing $\Delta E$. By sweeping $V_x/\gamma$ at $\Delta E = l\omega$ ($l=2$), it was possible to identify an optimal transfer regime in Fig.~\ref{fig:et}c, which highlights the importance of decoherence in excitation transfer, as previously reported in solid-state, biomolecular, atomic systems, and in Ref.~\cite{Maier2019ENAQT}.

Another analog quantum simulation of open-system molecular dynamics was recently demonstrated by Navickas et al.~\cite{Navickas2025} to study the nonadiabatic photochemical dynamics of a pyrazine molecule interacting with an effectively infinite-temperature bath. Without coupling to the environment, the closed molecular system can be modeled as:
\begin{equation}
    \begin{split}
    \hat{H} = \frac{\Delta E}{2}\hat{\sigma}_z + \frac{g_1}{2}\hat{\sigma}_z\left(\hat{a}_1+\hat{a}_1^\dag\right)+\frac{g_2}{2}\hat{\sigma}_x\left(\hat{a}_2+\hat{a}_2^\dag\right)+\sum_{i=1}^2\omega_i \hat{a}_i^\dag \hat{a}_i,
    \end{split}
\end{equation}
where the states $\ket{\uparrow}$ and $\ket{\downarrow}$, defining the Pauli operators $\hat{\sigma}_{x,z}$, represent two electronic states, specifically the bright (large transition dipole moment) $\pi\pi^*$ and dark $n\pi*$ states. In addition, parameters $g_1$ and $g_2$ determine the interaction strengths of the electronic states to the tuning ($\omega_1$) and coupling ($\omega_2$) vibrational modes, respectively. The vibronic and harmonic terms together induce a conical intersection, where the potential energy surfaces coincide, leading to the breakdown of the Born-Oppenheimer approximation, which enables fast population transfer between the electronic states. Meanwhile, $\Delta E$ introduces a slope to the energy landscape of the system.

\begin{figure}[t!]
    \centering
    \includegraphics[width=0.8\columnwidth]{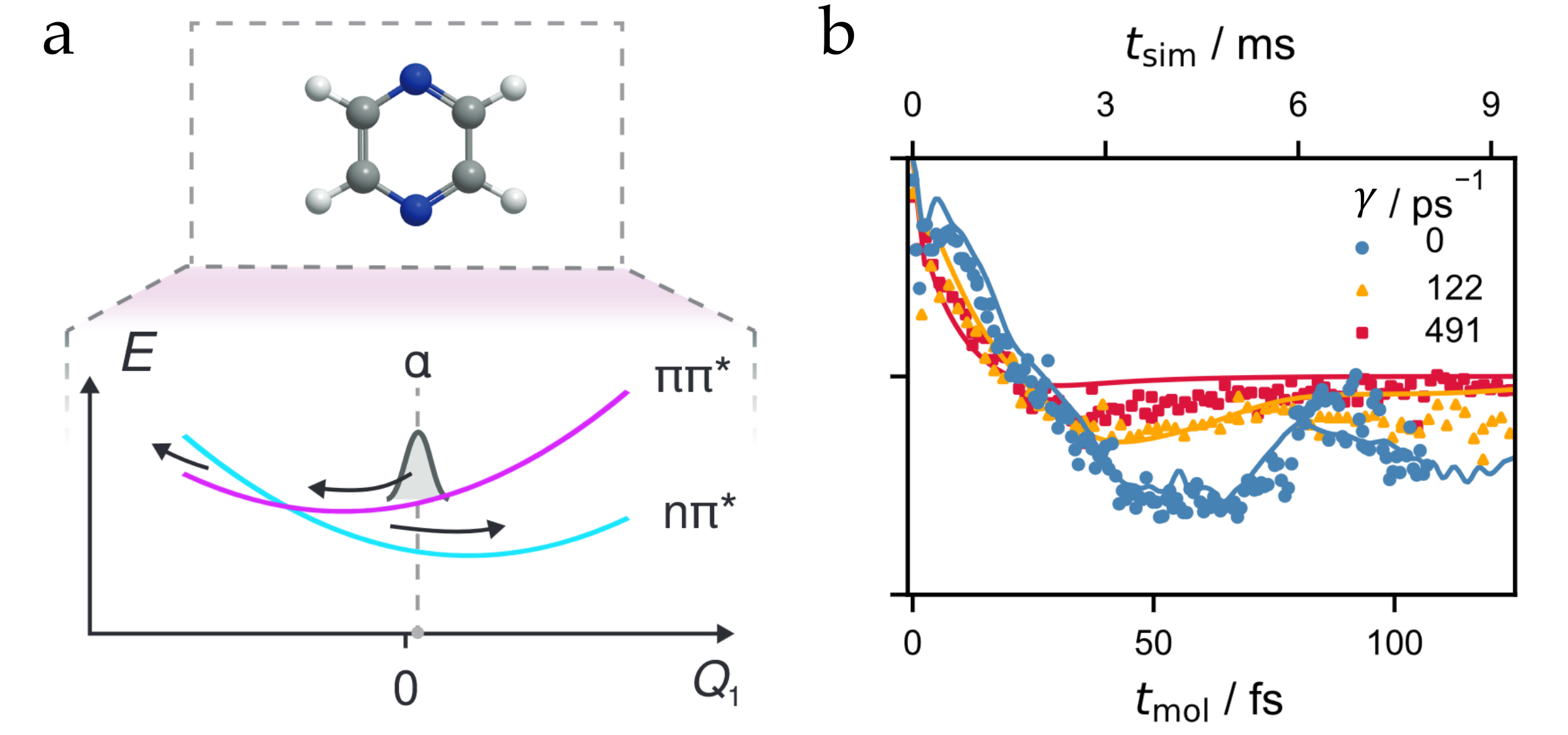}
    \caption{(a) One-dimensional adiabatic potential energy surfaces of photoexcited pyrazine (slice at $Q_2 = 0$) with the wavepacket initialized by displacement to $Q_1=\alpha$ and a sloped conical intersection. (b) Time dynamics of the pyrazine with varying $\gamma$. Adapted from Ref.~\cite{Navickas2025}.}
    \label{fig:pyrazine}
\end{figure}

In contrast to the low-temperature thermal bath in Ref.~\cite{so2024ETsim} to which the system effectively dissipates its vibrational energy, the authors considered an environment at effectively infinite temperature in this work, which continuously inputs energy to the molecular vibrations of the system. Thus, the dynamics of the open system can be described by the Lindblad's master equation:
\begin{eqnarray}
    &&\frac{\partial\hat{\rho}}{\partial t}=-i[\hat{H},\hat{\rho}] + \sum_{i=1}^2\gamma_i\left(\hat{\mathcal{L}}_{\hat{a}_i}[\hat{\rho}] + \hat{\mathcal{L}}_{\hat{a}_i^\dagger}[\hat{\rho}]\right).
\end{eqnarray}
Here, the effect of the bath is engineered by injecting a noisy electric field with frequency components that include the secular frequencies of the motional modes encoding the molecular vibrations ($\omega_i$). Their corresponding amplitudes are the control knobs for adjusting $\gamma_i$. Since this method does not require an ancilla qubit for reservoir engineering, the experiment was performed with a single trapped $^{171}$Yb$^+$ ion and two motional modes.

Prior to the simulation, the electronic state and vibrational wave packet of the system were prepared in $\ket{\uparrow}$ and $\hat{\mathcal{D}}_1(\alpha)\ket{0}_1\otimes\ket{0}_2$, respectively.
For the scaled system parameters ({$\omega_1,\;\omega_2,\;\Delta E,\;g_1,\;g_2,\;\alpha$}) associated with a photoexcited pyrazine, most of the population in the bright $\pi\pi^*$ state decays to the dark $n\pi^*$ state, realizing the non-radiative decay of the wave packet through a sloped conical intersection, as shown in Fig.~\ref{fig:pyrazine}a. When coupled to the environment, the coherence of the dynamics is quickly destroyed with increasing $\gamma_1=\gamma_2=\gamma$, which leads to a steady state population of 0.5, as expected from a Boltzmann distribution at high temperature (see Fig.~\ref{fig:pyrazine}b). Similar to the aforementioned works~\cite{Maier2019ENAQT, so2024ETsim}, this experiment showcases the programmability, resource efficiency, and open-system simulation capabilities of trapped-ion technologies by realizing the nonadiabatic photochemical dynamics of a real molecule interacting with an environment.

\begin{figure}[t!]
    \centering
    \includegraphics[width=\columnwidth]{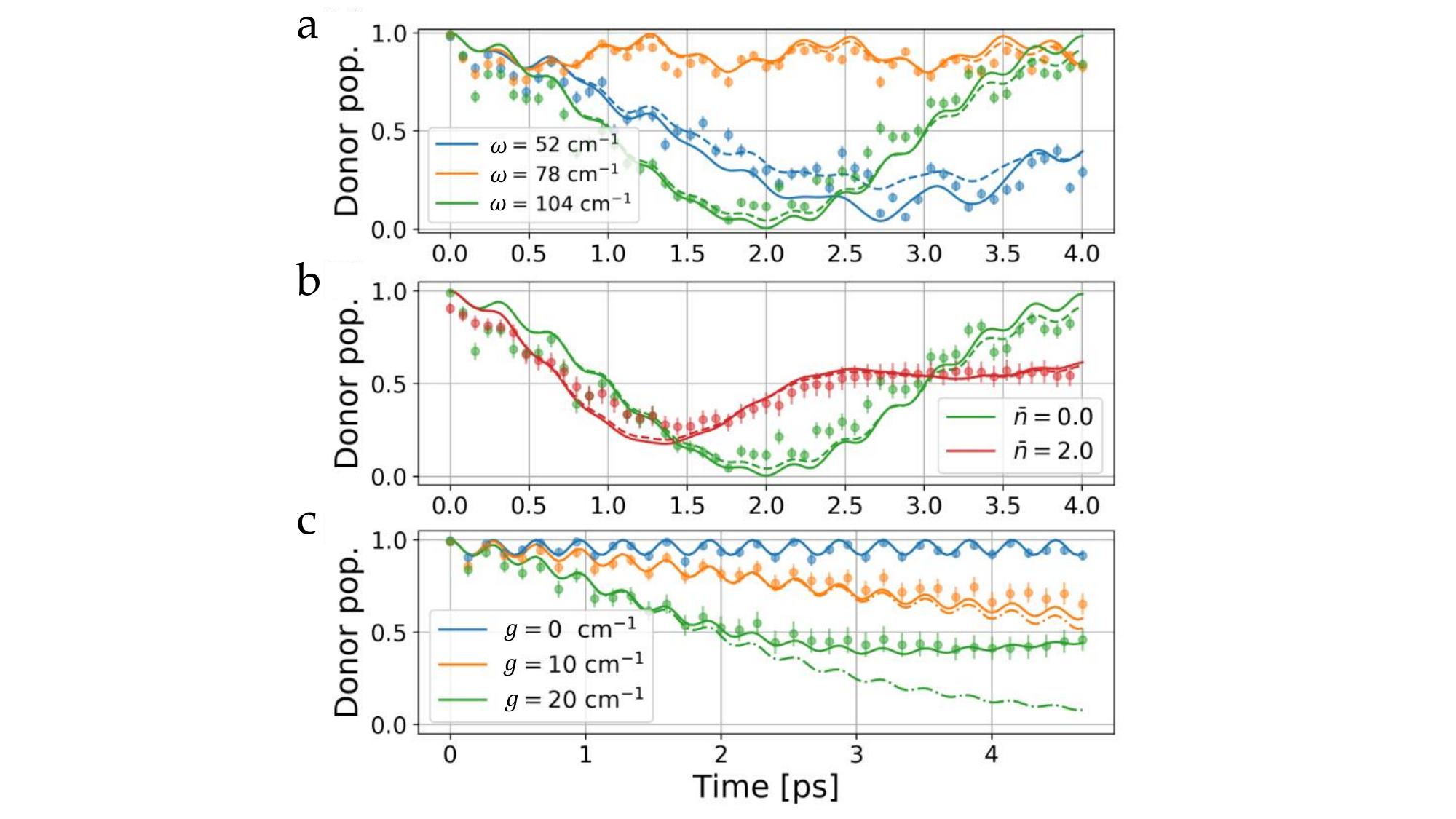}
    \caption{Time evolution of the donor population in the VAET model with $\Delta E=100\; {\rm cm}^{-1}$ and $V_x=15\; {\rm cm}^{-1}$ for (a) different $\omega$ with $g = 30\;{\rm cm}^{-1}$ and $\bar{n}=0$, (b) different $\bar{n}$ with $g = 30\;{\rm cm}^{-1}$ and $\omega=104\;{\rm cm}^{-1}$, and (c) different $g$ with $\omega=104\;{\rm cm}^{-1}$, $\bar{n}=0$, and $\gamma = 10\;{\rm cm}^{-1}$. Adapted from Ref.~\cite{sun2024quantumsimulationspinbosonmodels}.}
    \label{fig:vaet}
\end{figure}

To conclude our discussion on simulating open-system chemical dynamics with trapped ions, we present the results from Ref.~\cite{sun2024quantumsimulationspinbosonmodels}, which demonstrates a new method to simulate open quantum systems consisting of a spin coupled to a bath of harmonic oscillators subject to dephasing. This can be achieved by adding random frequency offsets $\delta_m(t)$ to the laser tones used to generate the spin-boson interaction and averaging the results over many trials. The value of $\delta_m(t)$ at time $t$ is randomly sampled from a normal distribution with the mean of 0 and standard deviation $\delta_{m,\text{std}}$. The dynamics of the total system can be described by:
\begin{eqnarray}
    &&\frac{\partial\hat{\rho}}{\partial t}=-i[\hat{H},\hat{\rho}] + \sum_{i}\gamma_i\hat{\mathcal{L}}_{\hat{a}_i^\dagger\hat{a}_i}[\hat{\rho}],
\end{eqnarray}
where the dephasing rate of the $i$-th oscillator $\gamma_i=\delta_{m,\text{std},i}^2\tau$ with $\tau$ being the duration of the discretized time step of the evolution, and the spin-boson Hamiltonian is written as follows:
\begin{equation}
    \begin{split}
    \hat{H} = V_x\hat{\sigma}_x + \frac{\Delta E}{2}\hat{\sigma}_z + \sum_i\frac{g_i}{2}\hat{\sigma}_z\left(\hat{a}_i+\hat{a}_i^\dag\right)+\sum_i\omega_i \hat{a}^\dag_i \hat{a}_i,
    \end{split}
\end{equation}
where the first two terms represent the Hamiltonian acting on the spin, while the last term describes the bath Hamiltonian containing discrete harmonic oscillators, each coupled to the spin at the rate $g_i$. 
Unlike a dissipative bath in~\cite{so2024ETsim} or an infinite-temperature bath in~\cite{Navickas2025}, the dephasing applied to the bath oscillators is energy conserving: in the absence of coupling to the spin, it does not change their energy. Therefore, the initial temperatures of the bath oscillators can play a crucial role in determining the spin-boson dynamics. To prepare a thermal state with the average phonon number $\bar{n}(\omega_i)$, a controlled amount of resonant spin-dependent force operations is applied on the associated mode $i$ near its motional ground state with a randomized motional phase $\psi_{m,i}(t)$ at time $t$ of each operation. When averaging the results over many evolutions, this technique yields the final temperature of $\bar{n}(\omega_i)=\bar{n}_0(\omega_i)+N_{i, \rm SDF}\Omega_{i, \rm SDF}^2\tau_{i, \rm SDF}^2/4$, where $\bar{n}_0(\omega_i)\approx0$ is the near ground-state phonon number, $N_{i, \rm SDF}$ is the number of operations, $\Omega_{i, \rm SDF}$ is the sideband Rabi frequency, and $\tau_{i, \rm SDF}$ is the duration of each time step.

Under the conditions of $\gamma_i<\omega_i/2, \;\gamma_i/2 \ll 2\pi k_BT,$ and weak $g_i$ (or short evolution time), the spin dynamics of the dephased spin-boson model can be approximated to that of the spin-boson model with the following spectral density:
\begin{equation}
    \begin{split}
    J(\omega)=\sum_i g_i^2\left[\frac{\gamma_i}{(\gamma_i/2)^2+(\omega-\omega_i)^2}-\frac{\gamma_i}{(\gamma_i/2)^2+(\omega+\omega_i)^2}\right],
    \end{split}
\end{equation}
where the first (second) term for each oscillator has a Lorentzian profile with the peak frequency at $+\omega_i\;(-\omega_i)$ and full width at half maximum (FWHM) of $\gamma_i$. 
It is worth noting that various forms of spectral density $J(\omega)$ can be effectively engineered by combining the interactions between the spin and multiple motional modes ($i$), each with a stochastically varying detuning ($\gamma_i$) and independent parameters ($g_i$ and $\omega_i$). 
As a minimal example, the authors investigated the vibrationally assisted energy transport (VAET) dynamics with a dephased vibrational mode, whose coherent evolution is described by the same Hamiltonian in Eq.~\eqref{eq:HET}, where the donor and acceptor sites encoded by $\ket{\uparrow}$ and $\ket{\downarrow}$, respectively, are coupled to the vibration in the environment with frequency $\omega$ at the rate $g$.

The key differences between the coherent dynamics in this example and the ones studied in Ref.~\cite{so2024ETsim} include the choice for the initial state of the spin-boson system, the vibronic coupling strength, and the system-environment boundary. In Ref.~\cite{so2024ETsim}, the initial state was chosen to be a thermal distribution of displaced Fock states, ensuring that the wave packet starts at the center of the donor potential energy surface, which is displaced by $|g/\omega|$ with $|g|\gtrsim|\omega|$ from that of the acceptor along the reaction coordinate. Moreover, the concepts of state-dependent potential energy surfaces and a reaction coordinate were made possible by the choice to consider the vibronic coupling and harmonic terms in the ET Hamiltonian as parts of the donor-acceptor vibronic system. On the other hand, by viewing the oscillator-related terms as the environment in the weak vibronic coupling regime ($|g|\ll|\omega|$), the authors in Ref.~\cite{sun2024quantumsimulationspinbosonmodels} observed the manifestation of energy transfer when the resonance condition ($l\omega=\sqrt{(2V)^2+\Delta E^2}$ with $l$ being an integer) of coherent VAET ($\gamma=0$) is met in Fig.~\ref{fig:vaet}a. This occurs when integer units of phonon energy are exchanged between the sites and the environment. As shown in Fig.~\ref{fig:vaet}b, the transfer becomes faster at higher $\bar{n}$, which can be explained by the involvement of more vibrational quanta. When dephasing is introduced ($\gamma\neq0$) in Fig.~\ref{fig:vaet}c, the coherence of the resonant dynamics is damped even at $\bar{n}=0$, similar to the expected evolution of the VAET model subject to vibrational damping. However, the dynamics given by the dephased and damped spin-boson models deviate in the strong-$g$ regime or long evolution time~\cite{Tamascelli2018}. Despite being limited to simulating the dynamics in the weak coupling regime, this technique provides a unique way to engineer a complex bath structure that can be useful for studying unexplored biochemical models and can also be combined with the reservoir engineering technique based on sympathetic cooling used in Ref.~\cite{so2024ETsim}. We note that the VAET dynamics of systems with two damped bosonic modes was also realized with a dual-species ion chain in Ref.~\cite{So2025}, revealing the effects of constructive interference between VAET pathways in the excitation transfer. By combining cooling with tunable heating applied to the motional modes, thereby enabling independent control of the bath temperatures, Ref.~\cite{so2025finiteT} investigated temperature effects in the single-mode ET and two-mode VAET models, showing that redistribution of the steady-state phonon populations reshapes the transfer-rate spectrum as a function of the donor-acceptor energy gap.

\section{Outlook}\label{sec_outlook}

All the experimental works reviewed in the previous section highlight the potential of trapped-ion systems for open-system analog simulation of fundamental processes in quantum chemistry, where the effect of the environment cannot be neglected. Trapped-ion technology natively gives access to the intermediate coupling regime in energy flow problems~\cite{Fassioli2014}, where the reorganization energy $\lambda$ is comparable or larger than the electronic coupling $J$ between different sites. Therefore, although the current demonstrations are limited to small systems still within the reach of classical computation, the rapid pace of recent advancements shows promising capabilities for investigating regimes that pose significant computational challenges for classical methods, such as tensor networks, especially when long-range interactions are considered~\cite{Kang2024, Somoza2019}.

One exciting prospect is the quantum simulation of vibrationally assisted exciton transfer in chemical and biological complexes, where the role of quantum coherence and delocalization in facilitating the transfer is still debated~\cite{Fassioli2014, Sneyd2022, Engel2007, Plenio2013, Monahan2015, Romero2014, Arsenault2020, Sneyd2021}. To study the role of entanglement and delocalization, hybrid digital-analog protocols can be implemented in trapped ions to prepare initially entangled/delocalized excitations and study the transfer in a tunable open quantum system, as proposed in Ref.~\cite{padilla2025vibrationallyassistedexcitontransfer}. A hybrid digital-analog approach is also advantageous for the quantum simulation of 2D spectroscopy using a probe qubit~\cite{guimarães2024acceleratingtwodimensionalelectronicspectroscopy}, which reduces the measurement overhead compared to standard digital protocols. A promising avenue in the quantum simulation of charge transfer is the use of multiple atomic states (\emph{qudits}) in an open quantum system setting that will allow the simulation of multiple electronic states with a limited ion overhead~\cite{MacDonell2021quantumchemistry} and, in the long term, of triplet-triplet annihilation and singlet fission physics~\cite{campaioli2024optimisationultrafastsingletfission}.


Trapped-ion programmable open quantum simulators with independent tunability of both the unitary and dissipative parts of the evolution can be used for state reservoir engineering. Dissipatively engineered Bell pairs have been realized in trapped-ion systems~\cite{barreiro2011open, lin2013dissipative, Malinowski2022, cole2022resource}, and there are several proposals to create $N$-body spin and boson entangled states~\cite{Morigi2015, reiter2016scalable, Cole2021dissipative, Zhu2025}. 
These capabilities are complemented by the possibility of natively implementing bosonic codes in trapped-ion systems encoding logical qubits in the harmonic oscillator states~\cite{Fluhmann2019encoding, Fluhmann2020, Brennan2022, rojkov2024stabilizationcatstatemanifoldsusing, Valahu2024}, which can be prepared with both reservoir engineering~\cite{Kienzler2015} or coherent pulses~\cite{Matsos2024}.

Trapped-ion research continues to expand rapidly, driven by international efforts in academia and industry. One of the outstanding challenges to performing quantum simulations and computations in a regime that is intractable for classical numerical methods is scaling up the trapped-ion systems by increasing the number of ions while maintaining fine control over them. One possible approach relies on increasing the number of ions in a single crystal. Combining cryogenic setups and new trap designs, single ion crystals with more than 100 ions have been achieved in both 1D~\cite{Pagano2018, Yao2022} and 2D~\cite{Kiesenhofer2023, Guo2024}. 

However, the number of ions that can be individually controlled in a single RF or Penning trap has practical limits due to normal mode crowding that hinders the high-fidelity gates and axial mode softening that, in large chains, leads to high heating rates and axially delocalized ions that cannot be efficiently individually addressed~\cite{Cetina2022}.
For these reasons, various strategies have emerged in recent years to realize scalable trapped-ion architectures. One prominent solution is the Quantum Charge Coupled Device (QCCD) approach, pioneered at NIST~\cite{Wineland1998,kielpinski2012}. By printing electrodes on a 2D substrate, it becomes possible to create multiple trapping zones and dynamically shuttle ions among them for high-fidelity interactions~\cite{H1_data, Gaebler2021micromotion}. Although large-scale QCCD demonstrations have mostly used 1D or quasi-1D geometries~\cite{Pino2021, Moses2023}, recent progress in reliably transporting multi-species ion crystals through 2D junctions~\cite{Burton2023, delaney2024scalable} suggests that fully 2D QCCD devices may soon be within reach. In tandem, integrated photonics (waveguides, splitters, grating couplers) offers a promising path to reduce free-space optical complexity. Recently integrated waveguides for the delivery of laser beams for cooling and quantum logic operations have been demonstrated~\cite{Niffenegger2020, Mehta2020, Ivory2021} as well as the ability to independently address ions in multiple trap zones using integrated photonics~\cite{Kwon2024, Mordini2025}.

Another path to scalable 2D arrays relies on Penning traps~\cite{Jain2020}, which combine static quadrupole potentials with large magnetic fields, eliminating the need for large RF voltages. This configuration has already demonstrated low motional heating rates ($\lesssim$ 1 quanta/s~\cite{Jain2024}) and straightforward 2D transport, thereby bypassing several of the known obstacles in standard RF-based architectures.

A third approach, modular quantum computing with optical interconnects~\cite{Duan2001, Monroe2013}, links smaller trapped-ion nodes (tens of qubits each) into larger networks via photonic entanglement. Two ions in separate modules emit entangled photons that interfere on a beam splitter, producing a heralded remote entangled state~\cite{Stephenson2020}. The entangling rate can be expressed as the product~\cite{Moehring2007,Hucul2015} $R_{\rm ent}=\frac{1}{2} \left(P_{\rm gen} P_{\rm coll}\right)^2 R$, where $P_{\rm gen}$ and $P_{\rm coll}$ are the photon generation and collection probabilities, and $R$ is the experimental attempt rate. Although photon loss currently limits entangling rates to a few 100s of entangling events per second~\cite{OReilly2024}, in-vacuum optics~\cite{Carter2024, Saha2025} and optical cavities~\cite{Schupp2021} have the potential to significantly boost photon collection efficiencies, making this scheme increasingly viable for large-scale quantum computation.

Collectively, these technological developments hold the promise to scale and improve trapped-ion quantum systems beyond fault tolerance, heralding exciting opportunities to explore physics and computations that remain intractable on classical hardware.

\begin{acknowledgement}
The authors acknowledge Christian Roos, Ting Rei Tan, Kenneth Brown, and Jungsang Kim for granting permission to reuse figures from their works. 
G.P. acknowledges the support of the Welch Foundation Award C-2154, the Office of Naval Research Young Investigator Program (Grant No. N00014-22-1-2282), the NSF CAREER Award (Grant No. PHY-2144910), the Army Research Office (W911NF22C0012), and the Office of Naval Research (Grants No. N00014-23-1-2665 and N00014-24-12593). We acknowledge that this material is based on work supported by the U.S Department of Energy, Office of Science, Office of Nuclear Physics under the Early Career Award No. DE-SC0023806.
\end{acknowledgement}
\ethics{Competing Interests}{
G.P. is a co-founder of TAMOS Inc.
W.A. and V.S. have no conflicts of interest to declare that are relevant to the content of this chapter.}


\section*{Appendix}
\addcontentsline{toc}{section}{Appendix}

\subsection{Second term of Magnus Series for Mølmer-Sørensen Spin-Spin Interactions} \label{app_2ndMagnus}

Here, we evaluate the individual integrals that make up the second term of the Magnus expansion of the time-evolution operator in Eq.~\eqref{evolution}, where $\overline{\Omega}_2(\tau)=\overline{\Omega}_2^{\rm CC}(\tau)+\overline{\Omega}_2^{\rm CS}(\tau)+\overline{\Omega}_2^{\rm SC}(\tau)+\overline{\Omega}_2^{\rm SS}(\tau)$. Since $[\hat{\sigma}_{\phi_s^j},\hat{\sigma}_{\phi_s^k}]=0 \;\forall j,k$, we get $[\hat{H}_{\rm SDF}^{\rm C}(t_1),\hat{H}_{\rm SDF}^{\rm C}(t_2)]=0$, and thus the first integral yields $\overline{\Omega}_2^{\rm CC}(\tau) = 0$.
Similarly, it can easily be shown that $[\hat{\sigma}_{\phi_s^j-\pi/2},\hat{\sigma}_{\phi_s^j}]=2i\hat{\sigma}^j_z$. Hence, we get
\begin{eqnarray}
    \overline{\Omega}_2^{\rm CS}(\tau) &=& -i\sum_{j,m=1}^N\eta_{jm}\Omega_j^2\hat{\sigma}_{z}^j \int_0^{\tau} dt_1 \int_0^{t_1} dt_2 \left(\hat{a}_me^{-i\omega_m t_2}+\hat{a}_m^\dagger e^{i\omega_m t_2}\right) \nonumber\\
    &\quad& \quad \quad \quad \quad \quad \quad \quad \quad \cos{(\mu t_1 - \psi_j)}\cos{(\mu t_2 - \psi_j)}, \\
    \overline{\Omega}_2^{\rm SC}(\tau) &=& i\sum_{j,m=1}^N\eta_{jm}\Omega_j^2\hat{\sigma}_{z}^j \int_0^{\tau} dt_1 \int_0^{t_1} dt_2 \left(\hat{a}_me^{-i\omega_m t_1}+\hat{a}_m^\dagger e^{i\omega_m t_1}\right)\nonumber\\
    &\quad& \quad \quad \quad \quad \quad \quad \quad \quad  \cos{(\mu t_1 - \psi_j)}\cos{(\mu t_2 - \psi_j)},
\end{eqnarray}
\begin{eqnarray}
        \overline{\Omega}_2^{\rm CS}(\tau)+\overline{\Omega}_2^{\rm SC}(\tau) &=& -i\sum_{j,m=1}^N\eta_{jm}\Omega_j^2\hat{\sigma}_{z}^j\int_0^{\tau} dt_1 \int_0^{t_1} dt_2 \left(\hat{a}_m\left[e^{-i\omega_m t_2}-e^{-i\omega_m t_1}\right]\right. \nonumber \\
    &\quad& \left.+\;\hat{a}_m^\dagger \left[e^{i\omega_m t_2}-e^{i\omega_m t_1}\right]\right) \cos{(\mu t_1 - \psi_j)}\cos{(\mu t_2 - \psi_j)} \nonumber \\
    &=& -i\sum_{j=1}^N\sum_{m=1}^N \left(X_{jm}^{(1)}(\tau)\hat{a}_m+X_{jm}^{(2)}(\tau)\hat{a}_m^\dagger \right)\hat{\sigma}_z^j,
\end{eqnarray}
with the time-dependent amplitudes
\begin{eqnarray}
    X_{jm}^{(1)}(\tau) &=& \frac{-\eta_{jm}\Omega_j^2}{2\delta_m\mu(\mu+\omega_m)(\mu+\delta_m)(2\mu+\omega_m)}\left(4\mu^3-\mu\omega_m^2-3\mu\omega_m^2\cos{(2\psi_j)}\right. \nonumber \\
    &\quad& + \; e^{-i\omega_m\tau}\left[ 3\mu\omega_m^2\cos{(2\mu\tau-2\psi_j)} +i\omega_m(2\mu^2+\omega_m^2)\sin{(2\mu\tau-2\psi_j)} \right. \nonumber \\
    &\quad&-\;(4\mu^2-\omega_m^2)\left(\mu+2ie^{i\omega_m \tau}\omega_m\cos{\psi_j}\sin{(\mu\tau-\psi_j)}\right. \nonumber\\
    &\quad&+\;2i\omega_m\cos{(\mu\tau-\psi_j)}\sin{\psi_j} \nonumber\\
    &\quad&\left.\left.-\;2\mu\sin{(\mu\tau-\psi_j)\sin{\psi_j}} +2e^{i\omega_m\tau}\mu\sin{(\mu\tau-\psi_j)}\sin{\psi_j}\right)\right] \nonumber \\
    &\quad& \left.+\;i\omega_m(2\mu^2+\omega_m^2)\sin{(2\psi_j)}\right), \\
    X_{jm}^{(2)}(\tau) &=& \frac{\eta_{jm}\Omega_j^2}{2\delta_m\mu(\mu+\omega_m)(\mu+\delta_m)(2\mu+\omega_m)}\left(\mu(\omega_m^2-4\mu^2)\right. \nonumber \\
    &\quad&-\;3e^{i\omega_m\tau}\mu\omega_m^2\cos{(2\mu\tau-2\psi_j)} \nonumber\\
    &\quad& +\;3\mu\omega_m^2\cos{(2\psi_j)}+ie^{i\omega_m\tau}\omega_m(2\mu^2+\omega_m^2)\sin{(2\mu\tau-2\psi_j)} \nonumber \\
    &\quad& -\;(4\mu^2-\omega_m^2)\left[2i\omega_m\cos{\psi_j}\sin{(\mu\tau-\psi_j)}\right. \nonumber\\
    &\quad& +\;2ie^{i\omega_m\tau}\omega_m\cos{(\mu\tau-\psi_j)}\sin{\psi_j} \nonumber\\
    &\quad& \left.+\;\mu(2e^{i\omega_m\tau}\sin{(\mu\tau-\psi_j)}\sin{\psi_j}-2\sin{(\mu\tau-\psi_j)}\sin{\psi_j}-e^{i\omega_m\tau}) \right]\nonumber \\
    &\quad&\left.+\;i\omega_m(2\mu^2+\omega_m^2)\sin{(2\psi_j)}\right).
\end{eqnarray}
Here, the denominators are in the order of $\delta_m\mu^4$, while the numerators are at most in the order of $\eta_{jm}\Omega_j^2\mu^3$. Hence, it is acceptable to discard these terms at sufficiently large $\mu$ and $\delta_m$, where $\mu\gg\Omega_j$ and $\delta_m\gg\eta_{jm}\Omega_j$. This is similar to completely neglecting the off-resonant carrier drive of the spin-dependent force Hamiltonian in the evaluation of the time-evolution operator in Eq.~\eqref{eq:time-evolution} from the beginning. Lastly, we can simplify $\overline{\Omega}_2^{\rm SS}(\tau)$ before performing the integration by invoking the commutation relation $[\hat{a}_m,\hat{a}_n^\dagger]=\delta_{mn}$ and obtain
\begin{eqnarray}
    \overline{\Omega}_2^{\rm SS}(\tau) &=& -\frac{1}{2}\sum_{j=1}^N\sum_{k=1}^N\sum_{m=1}^N\eta_{jm}\eta_{km}\Omega_j\Omega_k\hat{\sigma}_{\phi_s^j}\hat{\sigma}_{\phi_s^k} \nonumber \\
    &\quad& \int_0^{\tau} dt_1 \int_0^{t_1} dt_2 \left(e^{-i\omega_m(t_1-t_2)}-e^{i\omega_m(t_1-t_2)}\right) \nonumber \\
    &\quad&\quad \times \cos{(\mu t_1 - \psi_j)}\cos{(\mu t_2 - \psi_k)} \nonumber\\
    &=& -i\sum_{j<k}\sum_{m=1}^N\frac{\eta_{jm}\eta_{km}\Omega_j\Omega_k}{\mu^2-\omega_m^2}\hat{\sigma}_{\phi_s^j}\hat{\sigma}_{\phi_s^k}\left[\cos{(\psi_j-\psi_k)\omega_m \tau}\right. \nonumber \\
    &\quad& +\;\frac{\omega_m((5\mu^2-\omega_m^2)\cos{\psi_j}\sin{\psi_k}-(3\mu^2+\omega_m^2)\sin{\psi_j}\cos{\psi_k})}{2\mu(\mu-\omega_m)(\mu+\omega_m)} \nonumber \\
    &\quad& +\; \omega_m\cos{\psi_k}\left(\frac{\sin{(2\mu \tau -\psi_j)}}{2\mu}-\frac{\sin{((\mu-\omega_m)\tau-\psi_j)}}{\mu-\omega_m}\right. \nonumber\\
    &\quad& \left.-\;\frac{\sin{((\mu+\omega_m)\tau-\psi_j)}}{\mu+\omega_m}\right) + \sin{\psi_k}\left(-\frac{\omega_m\cos{(2\mu \tau -\psi_j)}}{2\mu}\right. \nonumber \\
    &\quad& \left.\left.-\;\frac{\mu\cos{((\mu-\omega_m)\tau-\psi_j)}}{\mu-\omega_m}+ \frac{\mu\cos{((\mu+\omega_m)\tau-\psi_j)}}{\mu+\omega_m}\right)\right],
    \label{eq:spin-spin_full}
\end{eqnarray}
where $\sum_{j=1}^N\sum_{k=1}^N\frac{1}{2}\equiv\sum_{j<k}$ with the factor of 1/2 accounting for double counting of indices $j$ and $k$. The only term in the expression above that is linear in time is the first term, while the other terms are bounded in time. Hence, at a long time $\tau$, we can ignore the fast-oscillating terms, and the integral becomes
\begin{eqnarray}
    \overline{\Omega}_2^{\rm SS}(\tau) &\approx& -i\sum_{j<k}\left(\sum_{m=1}^N\frac{\eta_{jm}\eta_{km}\Omega_j\Omega_k}{\mu^2-\omega_m^2}\hat{\sigma}_{\phi_s^j}\hat{\sigma}_{\phi_s^k}\cos{(\psi_j-\psi_k)}\omega_m\right) \tau.
    \label{eq:spin-spin_approx}
\end{eqnarray}

\bibliographystyle{spphys}
\bibliography{varenna}

\end{document}